\documentclass[aps,onecolumn,preprint,superscriptaddress,nofootinbib,floats,tightenlines,12pt]{revtex4} 
\pdfoutput=1

\usepackage{amsmath,amssymb,color,mathrsfs, graphicx,verbatim,epsfig, bbm, wasysym}
\usepackage[hyperfootnotes=false]{hyperref}
\usepackage{slashed}
\allowdisplaybreaks

\usepackage{slashed} 
\usepackage{color, multirow} 
\usepackage{ulem}

\definecolor{darkblue}{cmyk}{1,0.3,0,0.2}
\definecolor{violet}{cmyk}{0,1,0,0.2}
\hypersetup{colorlinks, bookmarksnumbered, citecolor=darkblue, linkcolor=darkblue, pdfstartview=FitH, urlcolor=darkblue, linktocpage}

\begin{document} 

\baselineskip=18pt


\thispagestyle{empty}
\vspace{20pt}
\font\cmss=cmss10 \font\cmsss=cmss10 at 7pt

\begin{flushright}
\small 
\end{flushright}

\hfill
\vspace{20pt}

\begin{center}
{\Large \textbf
{
Anomalous triple gauge couplings 
\\[0.15cm]
in electroweak dilepton tails at the LHC
\\[0.15cm] 
and interference resurrection
}}
\end{center}

\vspace{15pt}
\begin{center}
{Haeyun Hwang$^{\, a}$, Ui Min$^{\, b}$, Junghyeon Park$^{\, b}$,  Minho Son$^{\, b}$ and Jae Hyeok Yoo$^{\, a}$}
\vspace{30pt}

$^{a}$ {\small \it Department of Physics, Korea University, Seoul, Republic of Korea
}
\vskip 3pt
$^{b}$ {\small \it Department of Physics, Korea Advanced Institute of Science and Technology, \\ 291 Daehak-ro, Yuseong-gu, Daejeon 34141, Republic of Korea
}

\end{center}

\vspace{20pt}
\begin{center}
\textbf{Abstract}
\end{center}
\vspace{5pt} {\small
We study the electroweak dilepton production with two forward jets at the LHC, aiming to measure the anomalous triple gauge couplings in the Effective Field Theory (EFT) approach. This process exhibits a distinctive feature, namely, the interference between Standard Model (SM) and beyond the SM is resurrected in the inclusive cross section of the full amplitude, including two forward jets. As a concrete illustration, we perform the detailed analytic and numerical study of the interference using a simpler toy process, and discuss the subtlety of the effective $W$ approximation. We propose a new kinematic variable, VBFhardness, that controls the amount of energy flowing into the dilepton subprocess. We show that an appropriate cut on VBFhardness makes the interference resurrection manifest.
Finally, we use the invariant mass of the dilepton system as well as the transverse momentum, as done in the literature, to derive the sensitivity to anomalous triple gauge couplings at the LHC and the high luminosity LHC. Our result is compared with the existing limits from the experiments.
}

\vfill\eject
\noindent

\tableofcontents
\newpage

\section{Introduction}
\label{sec:intro}

Although the LHC has been performing great including the discovery of the Higgs boson~\cite{ATLAS:2012yve,CMS:2012qbp}, it continuously shows no evidence for the new physics, or beyond the Standard Model (BSM), only confirming the Standard Model (SM) to a  better precision. It indicates that either new particles, if they exist, are very weakly coupled to the SM or they may be hidden in the energy scale beyond the LHC reach, especially, if a new physics has a sizeable coupling to the SM. Given the strong indication for the mass gap between the electroweak and new physics scales, the effective field theory approach makes sense to parametrize the possible new physics effects encoded in the higher-dimensional operators. Deviating from the SM with the Higgs doublet under the SM gauge symmetry, the effective Lagrangian, known as the SM Effective Field Theory (SMEFT), below the cutoff $\Lambda$ is written as\\
\begin{equation}\label{eq:EFT}
 \mathcal{L} = \mathcal{L}_{\rm SM} + \sum_i \frac{c_i^{(6)}}{\Lambda^2} \mathcal{O}_i^{(6)} + \sum_i \frac{c_i^{(8)}}{\Lambda^4} \mathcal{O}_i^{(8)} + \cdots~,
\end{equation}
where the lepton number conservation was assumed and $c_i^{(d)}$ is the Wilson coefficient for the dimension-$d$ operator $\mathcal{O}_i^{(d)}$.  The non-vanishing effect from the new physics on the Wilson coefficients of higher-dimensional operators will cause a  deviation of couplings among SM particles from the SM prediction. 

In this work, we focus on the precision measurements of the cubic interaction of the gauge bosons at the LHC.
Taking into account the property of the SMEFT up to dimension-6 operators, the deviation of the triple gauge couplings from the SM can be parametrized in terms of three anomalous Triple Gauge Couplings (aTGC) as
\begin{equation} \label{eq:atgc}
\begin{split}
 \mathcal{L}_{\rm tgc}  
 &=i e  \left ( W_{\mu \nu}^+ W_\mu^-  -  W_{\mu \nu}^- W_\mu^+ \right ) A_\nu  +  i  e {c_\theta \over s_\theta} 
 \left (1 + \delta g_{1,z} \right )   \left ( W_{\mu \nu}^+ W_\mu^-  -  W_{\mu \nu}^- W_\mu^+ \right ) Z_\nu
 \\
&+ i e (1 + \delta \kappa_\gamma)  A_{\mu\nu}\,W_\mu^+W_\nu^-   +   i  e {c_\theta \over s_\theta}  \left (1 +  \delta \kappa_z \right ) Z_{\mu\nu}\,W_\mu^+W_\nu^-   
\\
 &+  i   { \lambda_z  e  \over m_W^2 } \left [   W_{\mu \nu}^+W_{\nu \rho}^- A_{\rho \mu}  +  {c_\theta \over s_\theta} W_{\mu \nu}^+W_{\nu \rho}^- Z_{\rho \mu}   \right]~,
\end{split}
\end{equation} 
where $c_\theta = \sqrt{1 - s_\theta^2}$ and $\delta \kappa_z  = \delta g_{1,z} -{s_\theta^2 \over c_\theta^2} \delta \kappa_\gamma$. 
Considering only amplitudes with a single insertion of aTGC, the cross section is in general a quadratic function of aTGC and it can be parametrized as
\begin{equation}\label{eq:EFT:xsec}
 \sigma = \sigma_{\rm SM} + C_i \sigma^i_{\rm SM\times BSM} + C_{i}C_{j} \sigma^{ij}_{\rm BSM\times BSM}~,
\end{equation}
where the index $i$ runs over three aTGCs, $C_i \equiv \{ \lambda_z, \, \delta g_{1,z},\, \delta \kappa_z \}$.

Typically, measurements of aTGC at the LHC have been performed by using diboson processes such as  $WW$, $WZ$, and $W\gamma$ in the lepton-enriched final state channels~\cite{ATLAS:2021ohb,CMS:2021icx,CMS:2021cxr}.  Unlike the precision measurement in LEP from $WW$ production process with the fixed center of mass energy around the electroweak scale, the sensitivity on aTGC from the LHC relies on the accessibility to the higher energy as long as it does not violate the validity of the EFT~\cite{Contino:2016jqw}, or one should not use the data at the energy $E$ above the cutoff $\Lambda$, or $E/\Lambda \lesssim 1$.
While the leptonic channel is clean and thus provides good sensitivity, the accompanying neutrinos can make it difficult to experimentally extract the exact scale of the hard process, especially in the dileptonic $WW$ process. The neutrino reconstruction is rather straightforward in the fully leptonic $WZ$ and $W\gamma$ processes~\cite{CMS:2021cxr} . When one can not impose an appropriate cut on the scale of the hard process to ensure $E/\Lambda \lesssim 1$, one can only set a conservative bound in this situation~\cite{Falkowski:2016cxu,Azatov:2017kzw}.

An issue in the diboson process has been the noninterference between the SM and BSM amplitudes which was found to be dictated by the helicity structure of the amplitudes~\cite{Cheung:2015aba,Azatov:2016sqh}. Including only dimension-6 operators, in the absence of interference, the leading BSM contribution to the total cross section scales $\mathcal{O}(\Lambda^{-4})$, and it may invalidate the EFT expansion in terms of $\Lambda$. This also makes the translation of the data to the SMEFT sensitive to the dimension-8 operators as the leading contribution is in the same order of the interference between dimension-8 operators and the SM.
There have been many attempts to resurrect the interference in the diboson process. While $2\rightarrow 2$ diboson processes are subject to the noninterference, unstable vector gauge bosons must decay. Once the $2\rightarrow 2$ diboson amplitude is extended to $2\rightarrow 3,\, 4$ by gluing with the three point amplitude(s) for a gauge boson decay into two fermions, the total helicity of both amplitudes of the dimension-6 and the SM can match and thus interfere. The authors in~\cite{Panico:2017frx,Azatov:2017kzw,Azatov:2019xxn} suggested to look into differential angular distributions in the leptonic decay channels to resurrect the interference. See~\cite{Franceschini:2017xkh,Aoude:2019cmc} for a related discussion. The authors in~\cite{Azatov:2017kzw,Azatov:2019xxn} pointed out the partial resurrection of the interference due to the QCD next-to-leading order (NLO) effect.  The role of off-shellness of the vector gauge bosons in the diboson process on the interference has been studied in~\cite{Helset:2017mlf}. 

In this work, we newly add the dilepton production process with two associated forward jets in the vector boson fusion (VBF) to the list regarding the interference resurrection. 
This process at $\sqrt{s} = 13$ TeV, using the integrated luminosity of 35.9 fb$^{-1}$, has been analyzed by the CMS collaboration~\cite{CMS:2017dmo}.
Although the signal rate of the VBF process is smaller than the diboson production from the QCD process, it may not be practically irrelevant compared to the diboson process. Besides, it has its own theoretical interest: the interference between the amplitudes with dimension-6 operators and those from the SM is resurrected in the inclusive cross section of the $2\rightarrow 4$ process, and it reveals a nontrivial phase space of the process.
While the electroweak (EW) $\ell\ell$ + jets process which is our main interest in this work may be considered as the EW Drell-Yan process, we aim to measure aTGCs via the tree-level process whereas QCD Drell-Yan process can access them via one loop effect~\cite{Dawson:2018dxp}. One can see~\cite{Farina:2016rws,Greljo:2017vvb,Panico:2021vav,Allwicher:2022gkm,Allwicher:2022mcg,Greljo:2022jac} (and~\cite{CMS:2021ctt,ATLAS:2020zms} for the experiment) for the precision study at the high energy tail of the QCD $\ell\ell$ process focusing on the tree-level four-fermion interactions.

A confusion arises due to the usual effective W approximation (EWA)~\cite{Cahn:1983ip,Dawson:1984gx,Kane:1984bb,Cahn:1984tx,Lindfors:1985yp,Gunion:1986gm,Kleiss:1986xp,Altarelli:1987ue,Johnson:1987tj,Kunszt:1987tk,Kuss:1995yv,Accomando:2006mc,Alboteanu:2008my,Borel:2012by} which factorizes the gauge boson radiated off the quark line and the $WW$ initiated subprocess. If this is the case, the interference will be suppressed again as total helicities of the SM and BSM amplitudes of $WW\rightarrow \ell\ell$ do not match in the massless limit~\cite{Azatov:2016sqh}.  For better understanding, an analytic study of the process that takes the full effect of the forward quark current would be highly beneficial. To this end, we carry out the full analytic calculation for a simpler process $u\gamma \rightarrow d\nu e^+$ that has only one forward quark current and one intermediate gauge boson as a toy process. As will be discussed below in detail, we find that the interference cross section of $u\gamma \rightarrow d\nu e^+$ with respect to the SM counterpart does resurrect the energy growing behavior, interestingly, in the inclusive cross section, that would have been lost in the typical EWA limit. The resurrected energy growing interference in the inclusive cross section appears in the full $u\gamma \rightarrow d\nu e^+$ process when enlarging the phase space to cover beyond the relevant regime for the EWA, and thus provides a counter-example to the usual EWA assumption (see~\cite{Borel:2012by} for a related discussion).  Our simpler toy process provides the proof of concept example for the resurrected interference in the inclusive cross section, and the intuition from it greatly helps for a better qualitative understanding of our EW $\ell\ell$ production process with two associated jets.

It turns out to be crucial that an enough energy must flow into the $\ell\ell$ hard subprocess to resurrect the energy growing interference in the inclusive cross section of the full $2\rightarrow 4$ process. Unlike the QCD Drell-Yan process where $m_{\ell\ell}$ directly controls the fraction of the energy that goes into the dilepton system, it becomes ambiguous in our EW $\ell\ell$ with two forward jets process because some fraction of energy goes to the scattered quarks. In this work, we propose a new variable, what we call {\it VBFhardness}, that allows to control the fraction of energy carried by the $\ell\ell$ subsystem. We demonstrate that the energy growing interference with respect to the SM is clearly resurrected with an appropriate cut on VBFhardness.

In Section~\ref{sec:vfb:twoleptons:inter} we briefly sketch the (non)interference of the dilepton production with two associated jets. In Section~\ref{sec:vbf:toy} we provide the analytic result of a simpler $2\rightarrow 3$ (instead of our $2\rightarrow 4$) toy process as this simpler example can be analytically calculated to capture the full effect of the forward jet from the viewpoint of the interference resurrection and the validity of the EWA. In Section~\ref{sec:vbf:numeric} we perform the numerical simulation of the EW $\ell\ell$ production with two associated jets as our main process of interest. In particular, we validate our simulation against the CMS cut-and-count analysis. We carry out the multivariate analysis using the Boosted Desicion Tree (BDT). We finally derive the sensitivity of aTGC at the LHC and high luminosity LHC (HL-LHC). Our results are compared to the existing limits from various diboson processes.

\section{EW dilepton production with two associated jets}
\label{sec:vfb:twoleptons:inter}

%
%
\begin{figure}[tph]
\begin{center}
\includegraphics[width=0.32\textwidth]{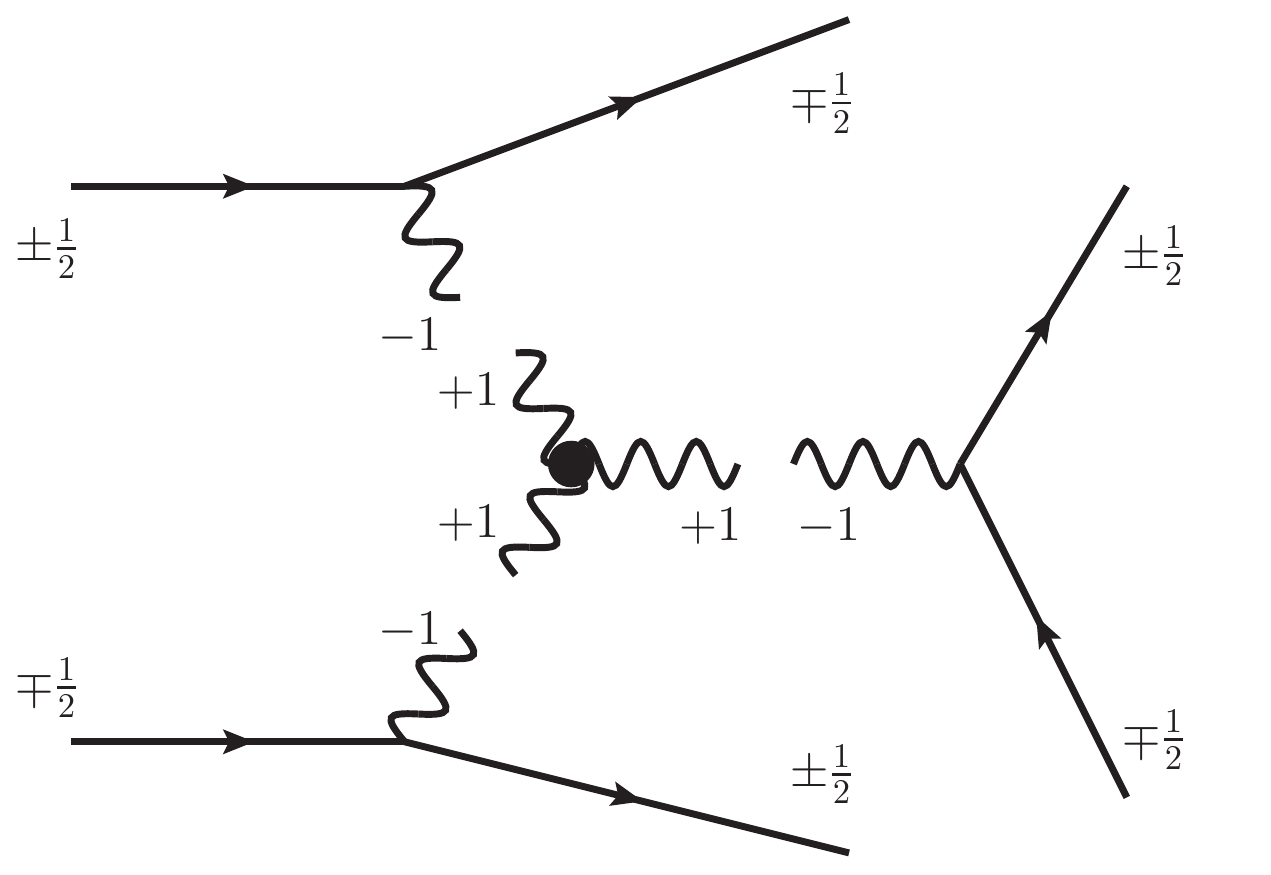}
\includegraphics[width=0.32\textwidth]{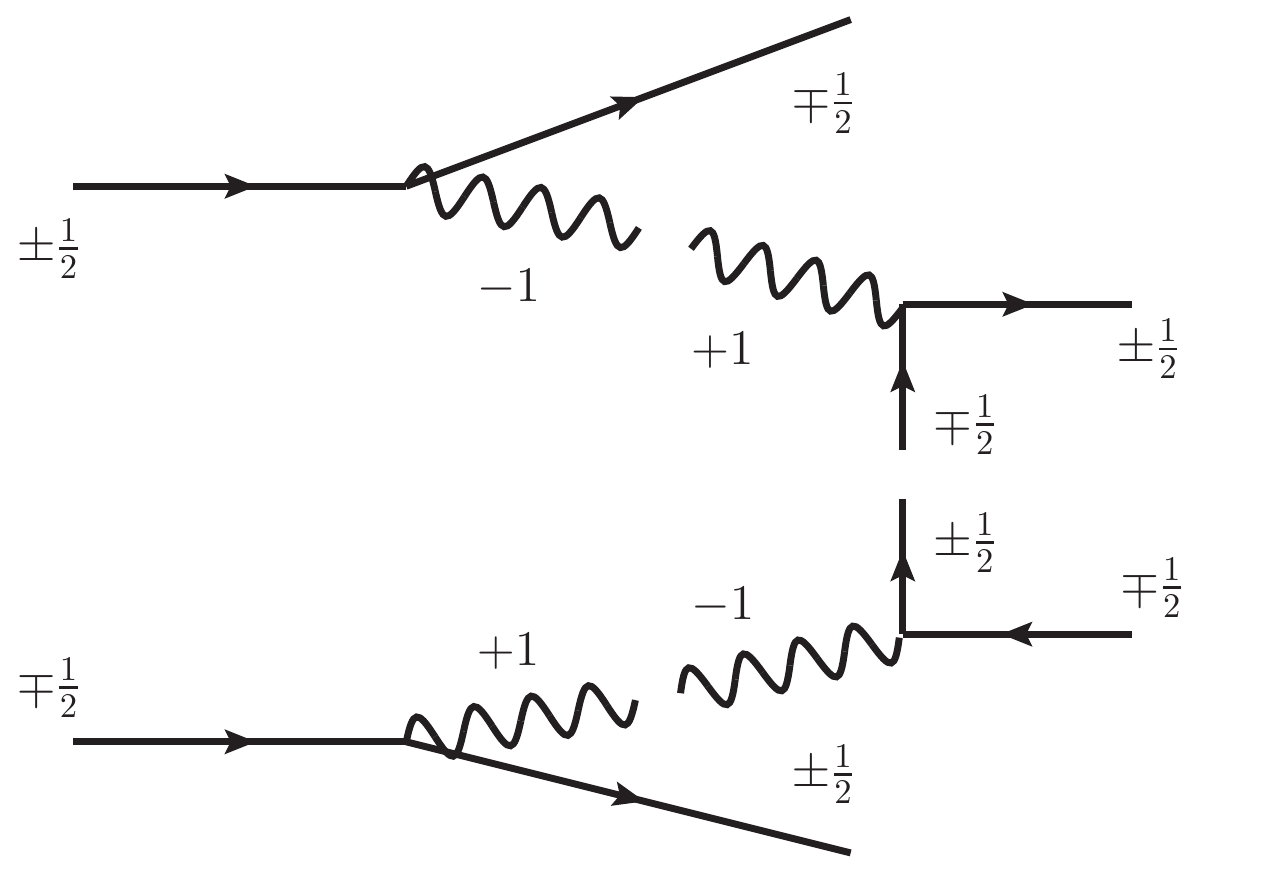}
\includegraphics[width=0.32\textwidth]{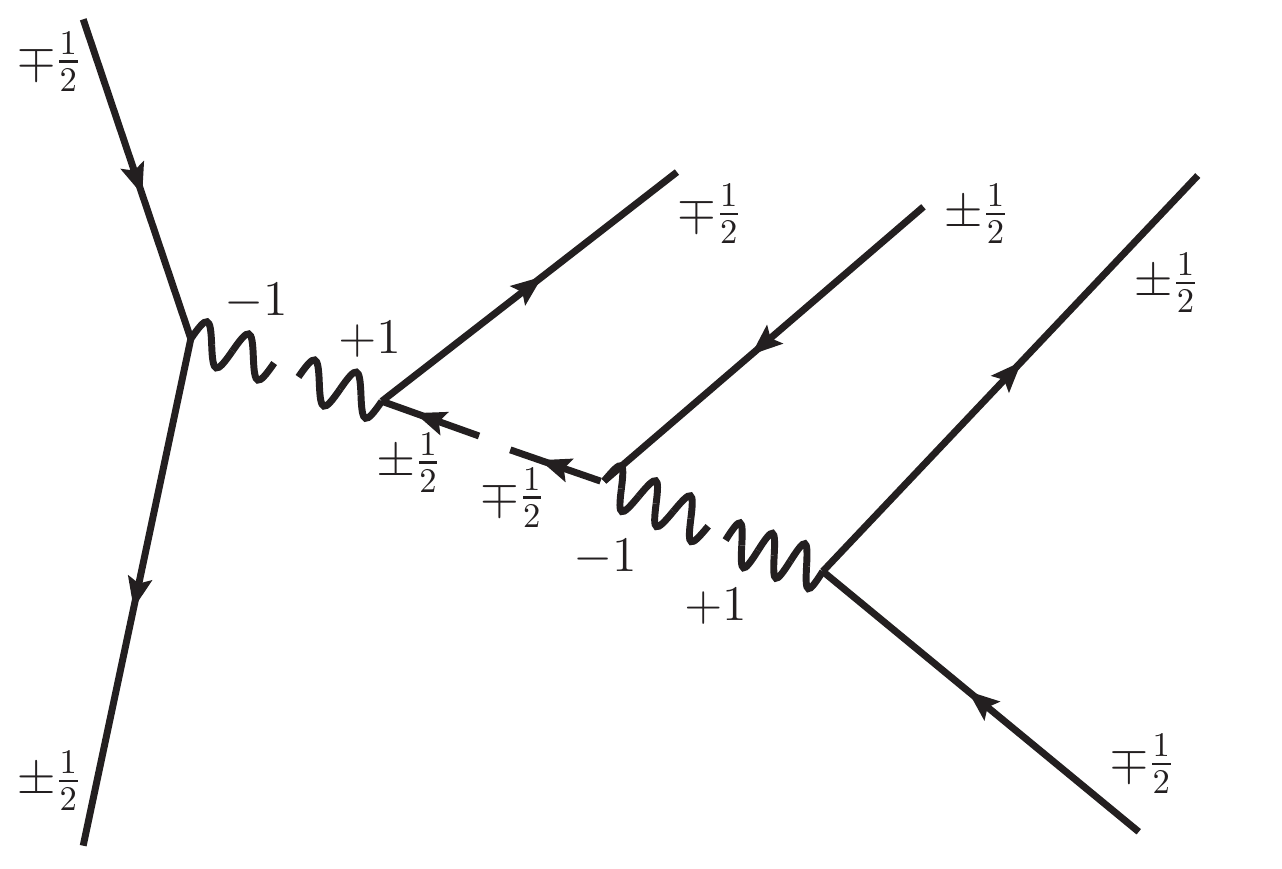}
\caption{\small Interference between BSM and SM diagrams in the massless limit where only two types of SM diagrams are shown. The blob denotes the insertion of the dimension-6 operator ${\rm tr}(W_{\mu\nu}^3)$. The helicity assignment is displayed as an example.}
\label{fig:aTGC:2to4}
\end{center}
\end{figure}

Fig.~\ref{fig:aTGC:2to4} illustrates the subset of diagrams for the $2\rightarrow 4$ amplitudes, leading to the dilepton with two associated jets, and possible helicity assignments which allow the interference between SM and BSM amplitudes. 
There is no similar diagram to the first one in Fig.~\ref{fig:aTGC:2to4} with the SM triple gauge couplings of the transverse modes as the helicity can not be correctly assigned. Its non-vanishing diagram can arise via helicity flips along with the Higgs VEV insertions and it will be suppressed by $\mathcal{O}(m^2_W/E^2)$.

The virtuality of $W$ emitted off the initial quark current induces the energy uncertainty of $W$ whose inverse sets the time uncertainty $\Delta t \sim E/V^2$ where $V$ is the virtuality of $W$ and $E$ is the scale of the hard process. As long as $\Delta t$ is much longer than the typical interaction time $t \sim 1/E$, one can not distinguish the virtual $W$ from on-shell one. In this situation, one can typically compute the partonic cross section of the hard subprocess whose leading contribution is approximated by those, effectively treating $W$ as on-shell gauge boson, and convolutes it with the probability distribution function of the $W$ gauge boson~\cite{Borel:2012by}. This factorization is known as the effective $W$ approximation (EWA).
%
\begin{figure}[tph]
\begin{center}
\includegraphics[width=0.32\textwidth]{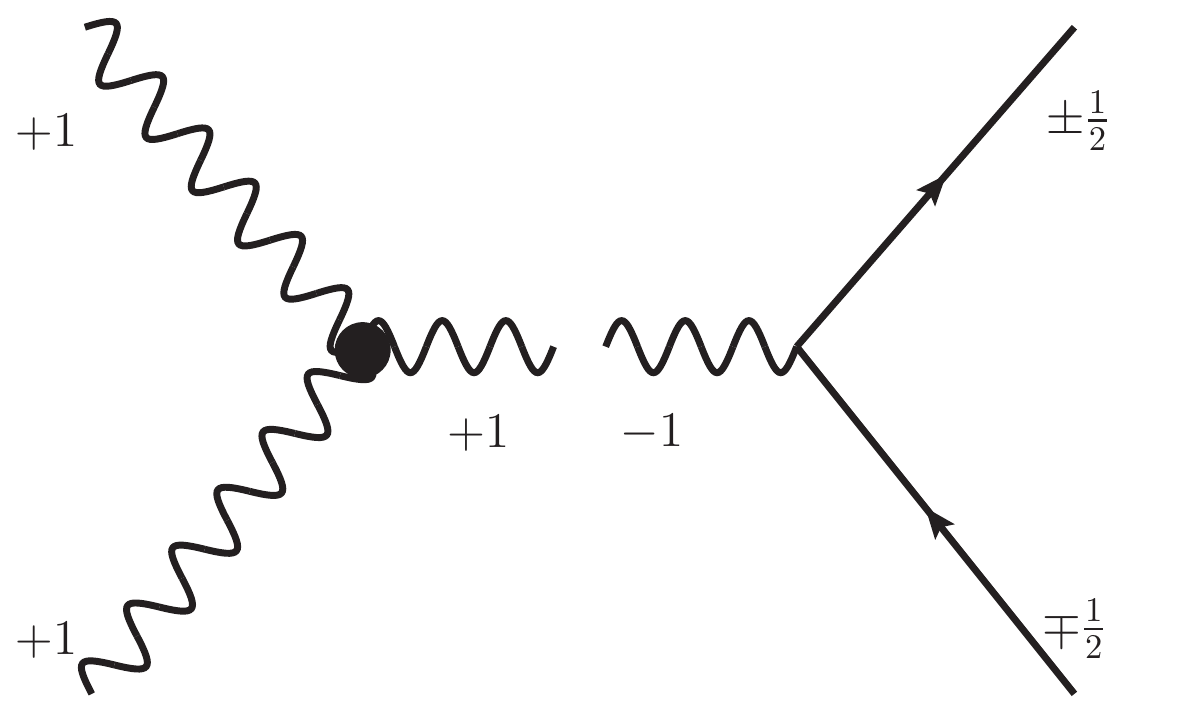}
\includegraphics[width=0.32\textwidth]{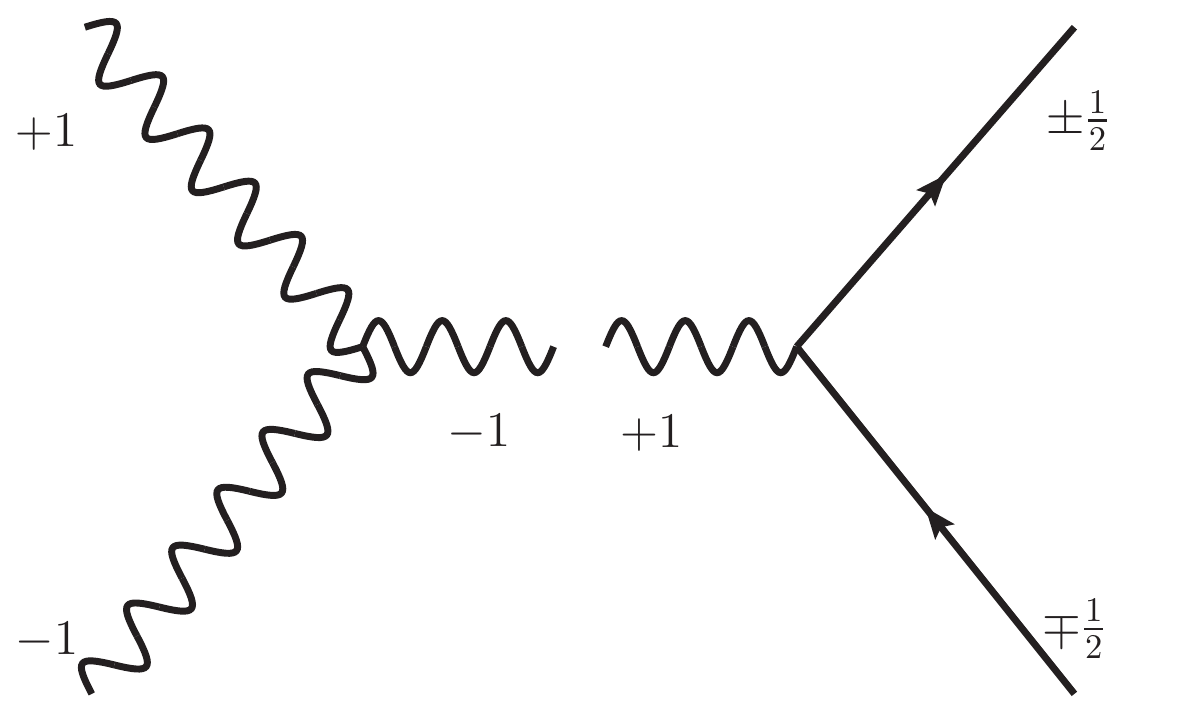}
\includegraphics[width=0.32\textwidth]{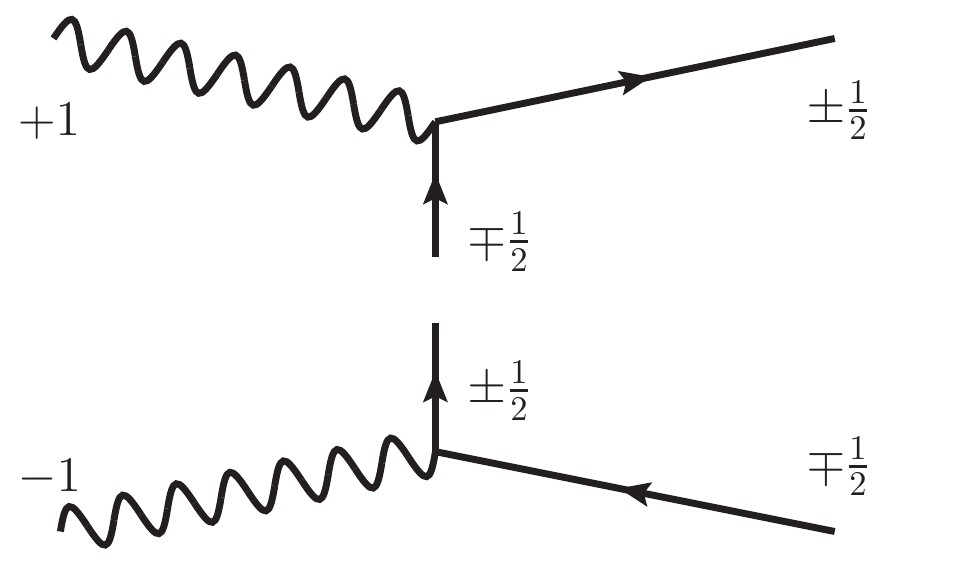}
\caption{\small Noninterference between BSM and SM amlitudes for $WW\rightarrow \ell\ell$. The blob denotes the insertion of the ${\rm tr}(W_{\mu\nu}^3)$ operator. The helicity assignment displayed is an example.}
\label{fig:aTGC:EWA}
\end{center}
\end{figure}
%
Apparently, the SM amplitudes of the $WW \rightarrow \ell\ell$ process in Fig.~\ref{fig:aTGC:EWA} do not interfere with the BSM amplitudes in the massless limit, known as noninterference~\cite{Azatov:2016sqh}. From the viewpoint of the aforementioned $2\rightarrow 4$ amplitudes, the amplitudes from the SM and BSM do interfere as total helicity allows. However, it can be shown rigorously that the interference vanishes upon the integration over the phase space if the relevant regime for the EWA is the dominant one.

The middle diagram in Fig.~\ref{fig:aTGC:EWA} can interfere with the BSM amplitude via helicity flips in the sub-leading order. It is consistent with that it can extend to the $2\rightarrow 4$ amplitude by attaching two quark currents upon helicity flips. 
On contrary, the extended amplitude with two attached quark currents of the third diagram in Fig.~\ref{fig:aTGC:EWA} can interfere with the corresponding BSM amplitude without any suppression as is evident in Fig.~\ref{fig:aTGC:2to4}.~\footnote{In the $2\rightarrow 4$ diboson process decaying into two pairs of fermions, the narrow width approximation allows to factorize the phase space of decaying on-shell gauge bosons from that of the hard process, and the process is subject to the noninterference. In this situation, the interference can appear, for instance, in the differential cross section of the azimuthal angle~\cite{Panico:2017frx,Azatov:2017kzw}.}  
In this work, we newly point out that the EW $\ell\ell$ process in VBF reveals a new sizable phase space which, otherwise, gets lost in the typical EWA limit, and thus interference can survive in the total cross section.

\section{Toy process for analytic study: Single lepton with an associated jet}
\label{sec:vbf:toy}
\begin{figure}[tp]
\begin{center}
\includegraphics[width=0.40\textwidth]{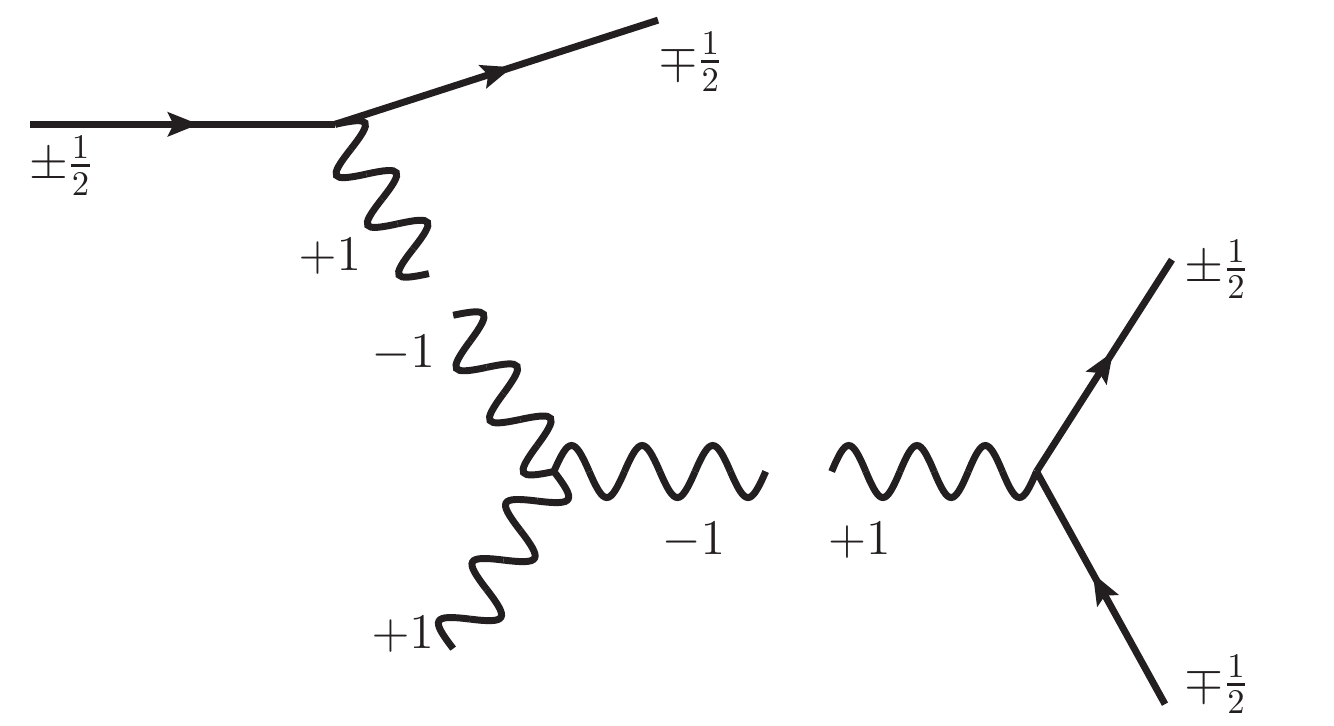} \hspace{1.3cm}
\includegraphics[width=0.40\textwidth]{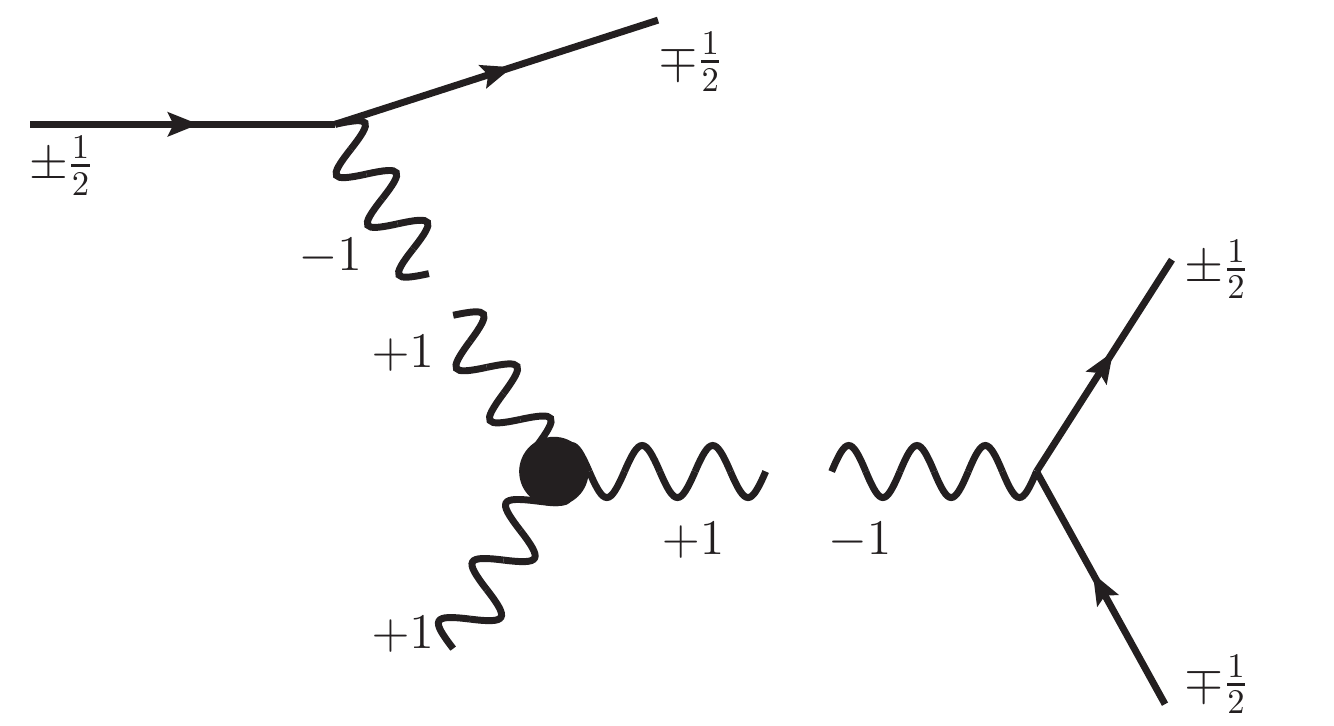}
\caption{\small The interference (noninterfernce) in the $2 \rightarrow 3$ process ($2 \rightarrow 2$ subprocess). The blob denotes the single insertion of the ${\rm tr}(W_{\mu\nu}^3)$ operator. The complete set of diagrams are shown in Fig.~\ref{fig:aTGC:2to3:diagrams}.}
\label{fig:aTGC:2to3:noninterference}
\end{center}
\end{figure}

\begin{figure}[tph]
\begin{center}
\includegraphics[width=0.34\textwidth]{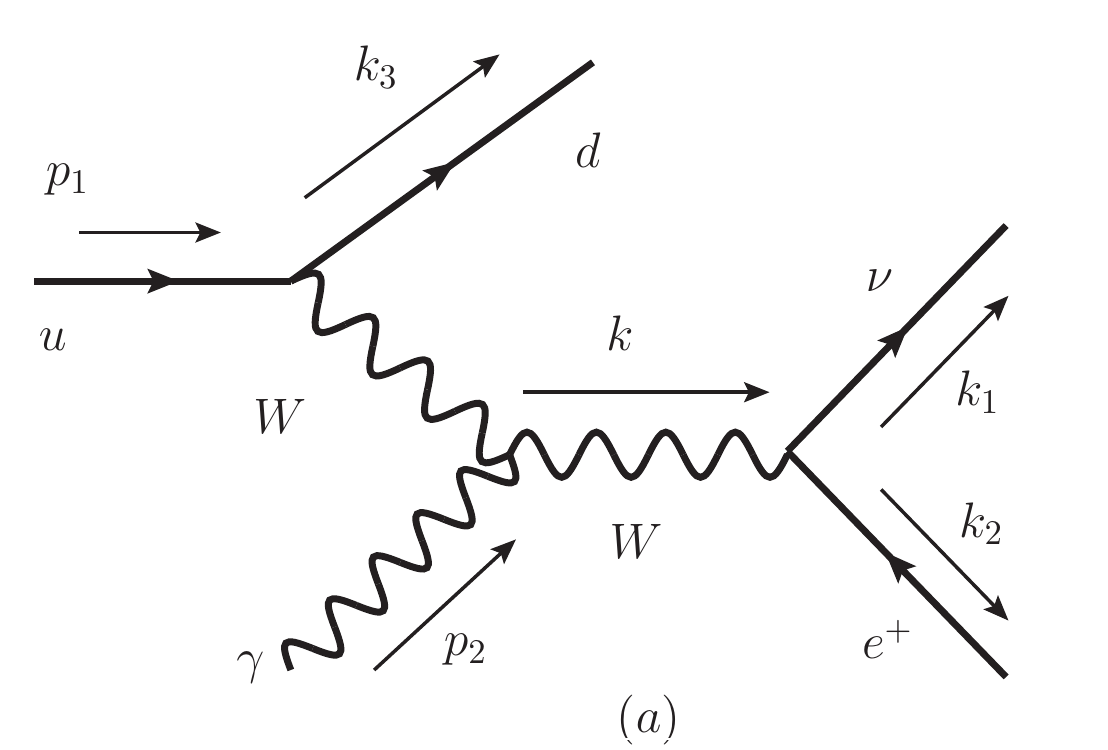} \hspace{1.3cm}
\includegraphics[width=0.34\textwidth]{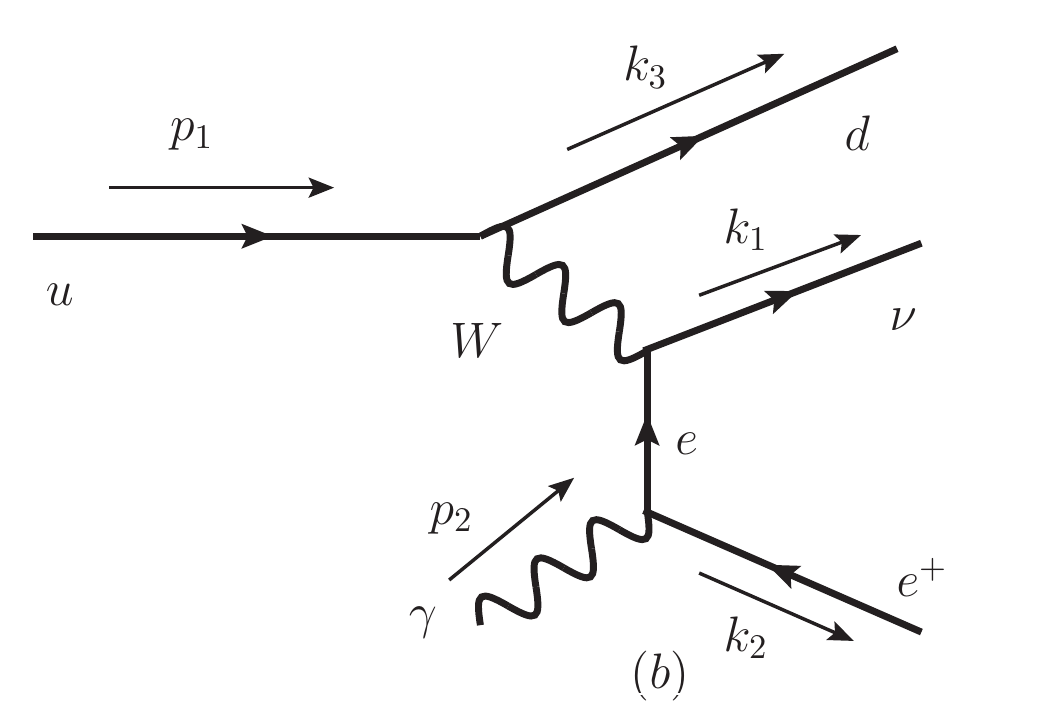} 
\\[10pt]
\includegraphics[width=0.34\textwidth]{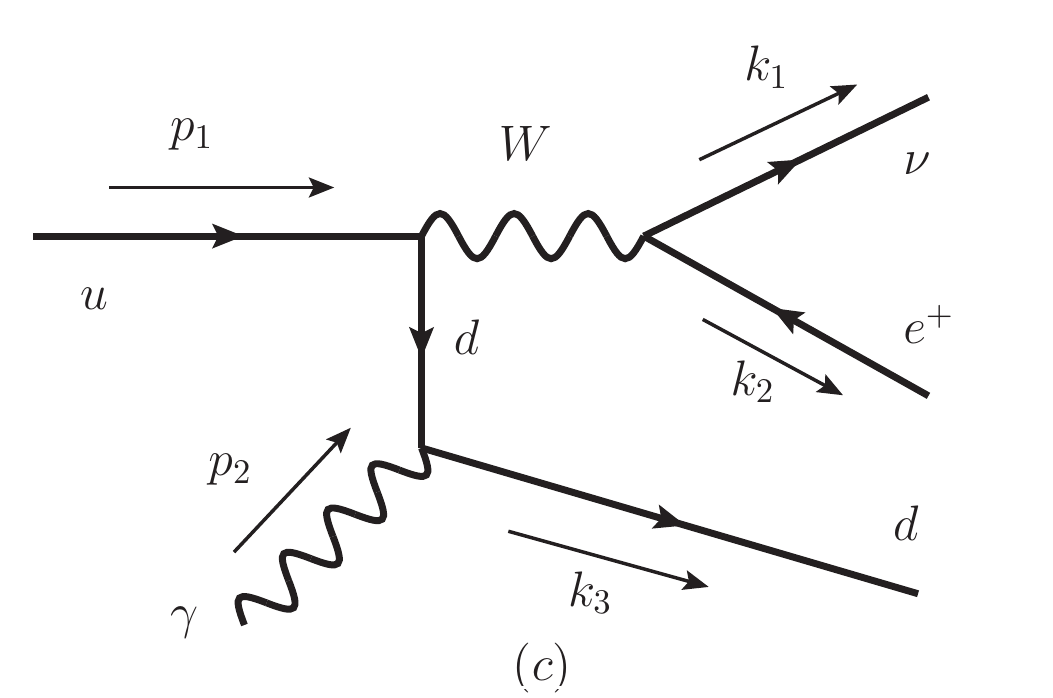} \hspace{1.3cm}
\includegraphics[width=0.31\textwidth]{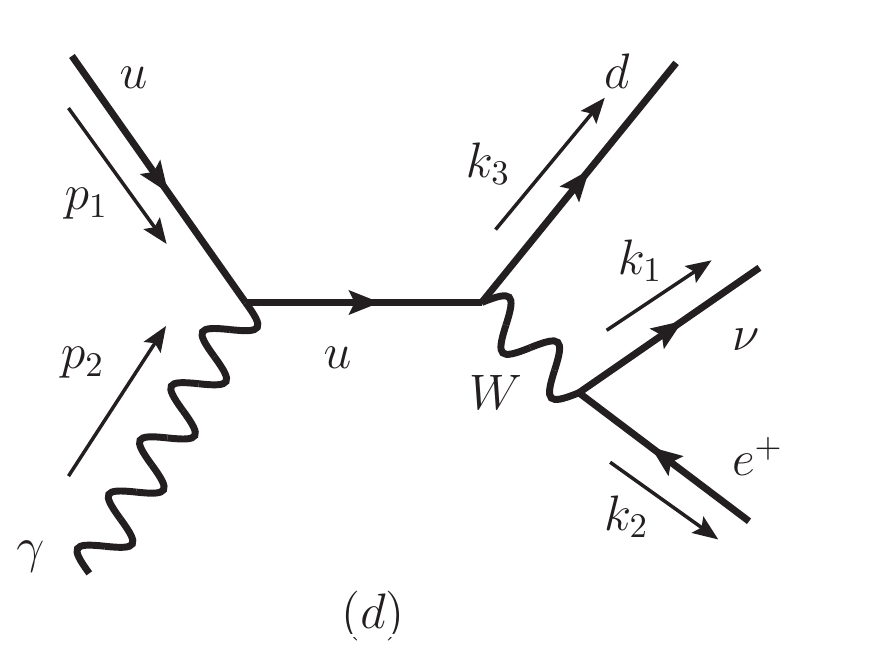}
\caption{\small The complete set of SM diagrams for the process $u\gamma \rightarrow d \nu e^+$. These four diagrams are required to guarantee the Ward identity and to get the correct high energy behavior. }
\label{fig:aTGC:2to3:diagrams}
\end{center}
\end{figure}

The purpose of this section is to analytically investigate (and numerically confirm) the helicity structure and related kinematics of simpler $2\rightarrow 3$ process $u\gamma \rightarrow d \nu e^+$ (see Fig.~\ref{fig:aTGC:2to3:noninterference} and Fig.~\ref{fig:aTGC:2to3:diagrams}) that captures the full effect of the quark current attached to a vector gauge boson. While the analytic calculation of the full $2\rightarrow 4$ process in Section~\ref{sec:vfb:twoleptons:inter} is beyond the scope of this work, our analytically calculable $2\rightarrow 3$ toy process
~\footnote{A similar explicit computation may be applicable to the $2\rightarrow 3$ process of $qq'\rightarrow \gamma W^* \rightarrow \gamma \ell \nu_\ell$ where the effect of the off-shell $W$ gauge boson on the interference, for instance, whether the interference between SM and BSM amplitudes can be resurrected in the inclusive cross section, can be explicitly understood.} 
provides the proof of concept for the resurrected interference and an intuition on the validity of the EWA in the SMEFT. We expect this simpler toy process to capture important missing properties when simply approximating with $2\rightarrow 2$ VBF process under the EWA assumption. 
We consider $u\gamma \rightarrow d \nu e^+$ since it involves with the exchange of only the $W$ gauge boson. A similar discussion is applied to $qV\rightarrow q'\ell\ell$ although the evaluation is more challenging.

The helicity assignments of two diagrams in Fig.~\ref{fig:aTGC:2to3:noninterference} indicate that the interference between the SM and BSM amplitudes in the $2\rightarrow 3$ process can be allowed. We separately consider the kinematic regions for the on-shell and off-shell intermediate $W$ gauge bosons decaying to $\ell \nu_\ell$ since they have different qualitative behaviors. For the resonant on-shell $W$ gauge boson, the $2\rightarrow 3$ process is factorized into the production of the on-shell $W$ gauge boson and its decay. It is expected that the inclusive cross section is subject to the noninterference and the interference at best can be resurrected only in the differential cross section of an angular observable (although it is difficult to be reconstructed in the experiment). On contrary, for the non-resonant $2\rightarrow 3$ process, the aforementioned factorization is not possible and the interference in principle can appear in the inclusive cross section.

The full set of SM diagrams of the EW $u\gamma \rightarrow d \nu e^+$ process are shown in Fig.~\ref{fig:aTGC:2to3:diagrams}. 
We classify the first two diagrams $a$ and $b$ as the process of interest that probe the hard $2 \rightarrow 2$ subprocess and the last two diagrams $c$ and $d$ as the radiation type. All four diagrams in Fig.~\ref{fig:aTGC:2to3:diagrams} are required to satisfy the Ward identity, namely $p_2\cdot (\mathcal{M}_a+\mathcal{M}_b+\mathcal{M}_c+\mathcal{M}_d)=0$. For the resonant intermediate $W$, the Ward identity can be shown to be satisfied among three diagrams, $p_2\cdot (\mathcal{M}_a+\mathcal{M}_c+\mathcal{M}_d)=0$, using the narrow width approximation.
We postpone all the details for the analytic calculation of $u\gamma \rightarrow d \nu e^+$ to Appendix~\ref{app:sec:2to3}. In what follows, we quote only the final result.

\subsection{Cross section for on-shell $W$ boson}
\label{sec:onshellW}
The diagrams $a$, $c$, and $d$ in Fig.~\ref{fig:aTGC:2to3:diagrams} mainly contribute to the resonant $2\rightarrow 3$ process where we can restrict the phase space to those in the $W$ mass window, or $k^2 = (2z -1)\hat{s} \approx m_W^2$ where $z = [1/2,\, 1]$ is the fraction of the total energy $\sqrt{\hat{s}}$ flowing into the $ \nu_e e^+$ system and $k$ is the four-momentum of it. The process can be factorized into the $2\rightarrow 2$ process of $u\gamma \rightarrow dW$ and the decay of $W$ to $\nu_e e^+$ using the narrow width approximation for the on-shell $W$ boson of the width $\Gamma_W$.

We evaluate the partonic differential cross section with respect to $\phi$ in the limit of $\hat{s} \gg m_W^2$ where $\phi$ is the angle between the planes made out of the forward quark current and the lepton current (see Fig.~\ref{fig:aTGC:coordinate1}).
The SM contribution is rather subtle to evaluate due to the forward singularity in the massless fermion limit. Its size is roughly given by
\begin{equation}
\begin{split}
\frac{d\hat{\sigma}_{\text{SM}}}{d\phi}  = \int_{\cos\theta_\text{min}}^{\cos\theta_\text{max}} d\cos\theta \frac{d^2\hat{\sigma}_{\text{SM}}}{d\phi d\cos\theta} \approx   
\frac{1}{2\cdot 2}\frac{1}{512 \pi^2}
\frac{8 \pi e^2 g^4}{3} \frac{m_W}{ \Gamma_W }\frac{1}{\hat{s}}\frac{1}{\delta}~,
\end{split}
\end{equation}
where $\delta = 2 p_{T\, \text{min}}^2/\hat{s}$ assuming $\delta \ll 1$ and it comes from the integration regularized by the $p_T$ cut of the forward quark, 
\begin{equation}
  \cos\theta_\text{max/min} = \pm \sqrt{1-\frac{p_{T\, \text{min}}^2}{\hat{s}(1-z)^2}} \approx \pm \left ( 1 - \frac{2\, p_{T\, \text{min}}^2}{\hat{s}} \right )\quad \text{for} \quad \hat{s} \gg m_W^2, \ p^2_{T\, \text{min}}~.
\end{equation}
On the other hand, the leading contribution to the partonic differential cross section for the interference in the high energy limit, $\hat{s} \gg m_W^2$, is estimated to be
\begin{equation}\label{eq:onW:int}
\begin{split}
\frac{d\hat\sigma_{\text{SM}\times\text{BSM}}}{d\phi} = 
 \frac{1}{2\cdot 2}\frac{\lambda_z}{512 \pi^4}\frac{\pi e^2 g^4}{3}\frac{2}{m_W \Gamma_W}
 \left [ \cos(2\phi) \left ( 2 - \log \frac{\hat{s}}{m_W^2} \right ) \right ]+ \mathcal{O}(\hat{s}^{-1/2})~.
\end{split}
\end{equation}
Upon the integration over the angle $\phi$, the interference term vanishes while it is recovered in the differential cross section with respect to $\phi$. The individual contributions to the total cross section for the interference are given by
\begin{equation}
\begin{split}
&\frac{d\hat\sigma_{\text{SM}\times\text{BSM}}(u_L\gamma_L \rightarrow d\nu e^+)}{d\phi} = 
 \frac{\lambda_z}{512 \pi^4}\frac{\pi e^2 g^4}{144}\frac{1}{m_W \Gamma_W}
 \\[5pt]
 & \hspace{3.5cm} \times \Big [ 9 \pi^2 \cos\phi + 16 \cos(2\phi) \left ( 5 - 3 \log \frac{\hat{s}}{m_W^2} \right ) \Big ]+ \mathcal{O}(\hat{s}^{-1/2})~,
 \\[5pt]
&\frac{d\hat\sigma_{\text{SM}\times\text{BSM}}(u_L\gamma_R \rightarrow d\nu e^+)}{d\phi} = 
 \frac{\lambda_z}{512 \pi^4}\frac{\pi e^2 g^4}{144}\frac{1}{m_W \Gamma_W}
 \\[5pt]
 & \hspace{3.5cm} \times \Big [- 9 \pi^2 \cos\phi + 16 \cos(2\phi) \left ( 7 - 3 \log \frac{\hat{s}}{m_W^2} \right ) \Big ]+ \mathcal{O}(\hat{s}^{-1/2})~, 
\end{split}
\end{equation}
where linear terms in $\cos\phi$ cancels upon the summation and there is no contribution from the right-handed quark in the massless limit.
In the same high energy limit, the quadratic term in the anomalous coupling $\lambda_z$ is approximately estimated to be
\begin{equation}
\frac{d\hat\sigma_{\text{BSM}^2}}{d\phi} = 
 \frac{1}{2\cdot 2}\frac{\lambda_z^2}{512 \pi^4} \frac{\pi e^2 g^4}{6} \frac{\hat{s}}{m_W^3 \Gamma_W}
\left ( 1 + \mathcal{O}(\hat{s}^{-1/2}) \right )~,
\end{equation}
where $\phi$ dependent terms are subdominant. For the quadratic terms in aTGC couplings, the leading contributions from both photon polarizations are the same. 

The energy flowing into the on-shell $W$ is $z \sqrt{\hat{s}} \sim \sqrt{\hat{s}}/2$ as usual in the high $\hat{s}$ limit because of $z \sim 1/2 + m_W^2/(2\hat{s}) \rightarrow 1/2$ for $\hat{s} \gg m_W^2$. The produced on-shell $W$ gets boosted with the transverse momentum of the order $\mathcal{O}(\sqrt{\hat{s}})$.
The boosted $W$ boson requires a large recoiling against a hard quark jet which likely invalidates the EWA as the jet can not be treated as a forward jet anymore. The process is subject to the noninterference, as is seen in Eq.~(\ref{eq:onW:int}), since it is basically $2\rightarrow 2$ process $u\gamma \rightarrow dW$ where the $W$ decay can be factorized. In this situation, the interference can be accessed through the differential distribution of $\phi$ of the forward jets. A large number of signal events in the $Z$ mass window may help.

When generalizing our toy process to the $2\rightarrow 4$ process with the intermediate resonant $W$ by attaching the fermion line to the photon, the situation becomes less obvious. Similarly the EWA  of either jet or both will not be valid if the boosted $W$ boson is considered. The interference can be  in principle possible as the helicity structure of the $2\rightarrow 3$ process for the on-shell $W$ production ($pp\rightarrow qq'W$) allows the amplitudes to interfere.

\subsection{Cross section for off-shell $W$ boson}
\label{sec:offshellW}
Alternatively, one can probe the high energy behavior of the anomalous coupling by directly accessing far off-shell region of $W$, or $k^2 = (2z -1)\hat{s} \gg m_W^2$. For this case, the full matrix element for the $2\rightarrow 3$ process needs to be considered. One new feature will be the resurrection of the interference in the inclusive cross section, and its size is expected to be proportional to the off-shellness of the $W$ boson.
While the analytic evaluation of the differential cross section in terms of $\phi$ is challenging due to the diagram $b$ in Fig.~\ref{fig:aTGC:2to3:diagrams}, we have managed to get the leading contribution only for the left-handed polarization of the photon in the high energy limit. Similarly to the previous Section~\ref{sec:onshellW}, we will use $\hat{s}$ to take a high energy limit. The cross section for the interference in the limit of $\hat{s} \gg m_W^2$ far away from the $W$ mass window, $k^2 \gg m^2_W$, is estimated to be
\begin{equation}\label{eq:off:inter}
\begin{split}
&\frac{d\hat\sigma_{\text{SM}\times \text{BSM}}(u_L\gamma_L \rightarrow d\nu e^+)}{d\phi} = 
 \frac{\lambda_z}{512 \pi^4} \frac{e^2 g^4}{m^2_W} \Big [ - \frac{2}{9} - \frac{\pi^2}{6} \cos\phi 
 \\[3pt]
 &\hspace{3.5cm}+ \frac{1}{18} \left ( \pi^2 - 26 + 22 \ln \frac{\hat{s}}{m_W^2} - 6 \ln^2 \frac{\hat{s}}{m_W^2} \right ) \cos(2\phi)
 \Big ] \left ( 1 + \mathcal{O}(\hat{s}^{-1})\right )~,
\end{split}
\end{equation}
where $\Gamma_W$ dependent terms are not shown as they contribute to the region of the $W$ mass window. Although the off-shell contribution is suppressed by the factor of $\mathcal{O} (\Gamma_W/m_W)$, or $\sim \mathcal{O}(1/m_W^{2})$, compared to the cross section from the $W$ mass window, the interference term can survive in the inclusive cross section even after the integration over all angular variables (see the first term in Eq.~(\ref{eq:off:inter})). 
The cross section for the quadratic term in $\lambda_z$ is given by
\begin{equation}\label{eq:off:quadr:L}
\begin{split}
\frac{d\hat\sigma_{\text{BSM}^2}(u_L\gamma_L \rightarrow d\nu e^+)}{d\phi} &= 
 \frac{\lambda_z^2\, e^2 g^4}{512 \pi^4}  \frac{\hat{s}}{m_W^4} \Big [ 
 \frac{1}{24} \left (-9 + 4 \ln \frac{\hat{s}}{m_W^2} \right ) \\
\\[3pt]
 & \hspace{1.5cm}
 - \frac{\pi^2}{48} \cos\phi 
 - \frac{1}{12} \cos(2\phi) \Big ]
\left ( 1 + \mathcal{O}(\hat{s}^{-1/2}) \right )~,
\end{split}
\end{equation} 
where $\phi$ dependence only appears in $\cos\phi$ and $\cos(2\phi)$ terms.
For the quadratic term, the analytic expression in the high energy limit outside the $W$ mass window can be obtained for both polarizations of the photon, and the summed and averaged cross section over helicities is given by
\begin{equation}\label{eq:off:quadr}
\begin{split}
\frac{d\hat\sigma_{\text{BSM}^2}}{d\phi} &= 
 \frac{1}{2\cdot 2}\frac{\lambda_z^2\, e^2 g^4}{512 \pi^4}  \frac{\hat{s}}{m_W^4} \Big [ 
 \frac{1}{216} \left (-143 + 60 \ln \frac{\hat{s}}{m_W^2} \right ) \\
\\[3pt]
 & \hspace{1.5cm}
 + \frac{\pi^2}{240} \cos\phi 
 - \frac{1}{12} \cos(2\phi) \Big ]
\left ( 1 + \mathcal{O}(\hat{s}^{-1/2}) \right )~.
\end{split}
\end{equation}

While we have shown the evidence of the resurrected interference in the inclusive cross section through the computation of the differential cross section in terms of $\phi$ only for the left-handed photon helicity, it may be more convenient to access directly to the inclusive cross section summed and averaged over helicities. For the direct analytic computation of the inclusive cross section, we have managed to get the final result for both helicities of the photon by performing the integration over $\phi$ first and the remaining variables later. The leading contribution of the summed and averaged cross section over helicities is given by
\begin{equation}\label{eq:off:inter:incl:L}
\begin{split}
\hat\sigma_{\text{SM}\times \text{BSM}} &= 
 \frac{1}{2\cdot 2} \frac{\lambda_z}{512 \pi^4} \frac{e^2 g^4}{m^2_W} \times  \frac{\pi}{3} \left (13 - 6 \ln \frac{\hat{s}}{m_W^2} \right ) + \cdots~,
\end{split}
\end{equation}
where $\cdots$ denotes the higher order in $\hat{s}$ and the logarithmic term is due to the contribution from the right-handed helicity of the photon. The summed and averaged cross section which is quadratic in $\lambda_z$ can be easily obtained by integrating Eq.~(\ref{eq:off:quadr}) over the angle $\phi$. 

For more off-shell $W$, more energy flows into the $e\nu$ system. 
We can isolate the behavior of the corresponding phase space by integrating over only the interval $z = [1-\varepsilon,\, 1]$ with $\varepsilon \ll 1$,
\begin{equation}\label{eq:int:ewa:Lgamma}
\begin{split}
 & \hat\sigma_{\text{SM}\times \text{BSM}}\ \frac{512 \pi^4}{\lambda_z} \frac{m^2_W} {2\pi e^2 g^4}
\\[3pt]
&\hspace{1.0cm}= 
- \frac{1}{3}  \varepsilon^2 + 
 \frac{1}{3} \frac{m_W^2} {\hat{s}} \left [ \left ( - 3 + 2 \ln \frac{2\varepsilon \hat{s}}{m_W^2} \right ) \varepsilon 
 +  \left ( -13 + 6 \ln \frac{2\varepsilon \hat{s}}{m_W^2} \right ) \varepsilon^2 + \cdots
 \right ] 
 + \cdots ~,
\end{split}
\end{equation}
where $\cdots$ denotes the higher order terms in $\varepsilon$ and $m_W^2/\hat{s}$.
In the high energy limit of $\hat{s} \rightarrow \infty$, the first constant term will eventually dominate, and it will appear as the energy growing interference in $\hat{\sigma}_{\text{SM}\times \text{BSM}}/\hat{\sigma}_\text{SM}$ assuming $\hat{\sigma}_\text{SM} \sim 1/\hat{s}$. The variable $m_{e\nu}$ may be considered to be more relevant one to take a high energy limit of the hard subprocess $W\gamma \rightarrow \nu e^+$ inside $u\gamma \rightarrow d \nu e^+$. 
Simply changing variable from $\sqrt{\hat{s}}$ to $m_{e\nu}$ in expressions obtained in the high $\hat{s}$ limit in Eqs.~\ref{eq:off:inter},~\ref{eq:off:quadr:L} and~\ref{eq:off:quadr} could be misleading or not well defined, for instance, a wide range of $\sqrt{\hat{s}}$ can be associated with a small value of $m_{e\nu}$ for $z \sim 1/2$ (see the right panel of Fig.~\ref{fig:int:toy:control:demo}).
The analytic computation of the interference in terms of $m_{e\nu}$, performed at this time starting from amplitudes, reveals a similar energy growing behavior to Eq.~(\ref{eq:int:ewa:Lgamma}).
%

\subsection{Numerical calculation of toy process and interference resurrection}
We numerically investigate the analytic behavior discussed in Sections~\ref{sec:onshellW} and~\ref{sec:offshellW}. To this end, we generate partonic level events for the EW $u\gamma \rightarrow d\nu e^+$ process using \textsc{\sc MadGraph}5\_aMC$@$NLO v2.6.7~\cite{Alwall:2014hca} only with the nominal $p_T$ cuts of $10$ GeV for the final quark, neutrino, and electron. As the noninterference is well established for the operator involving $\lambda_z$, the events for the interference are generated only for $\lambda_z$ coupling. While the separation of the off-shell region from the on-shell one in Section~\ref{sec:offshellW} was done just by dropping out all $\Gamma_W$ dependent terms by hand, we numerically control the separation using two variables $\Delta m_W = m_{e\nu} - m_W$ and $z$ for the purpose of the demonstration.

\begin{figure}[tph]
\begin{center}
\includegraphics[width=0.48\textwidth]{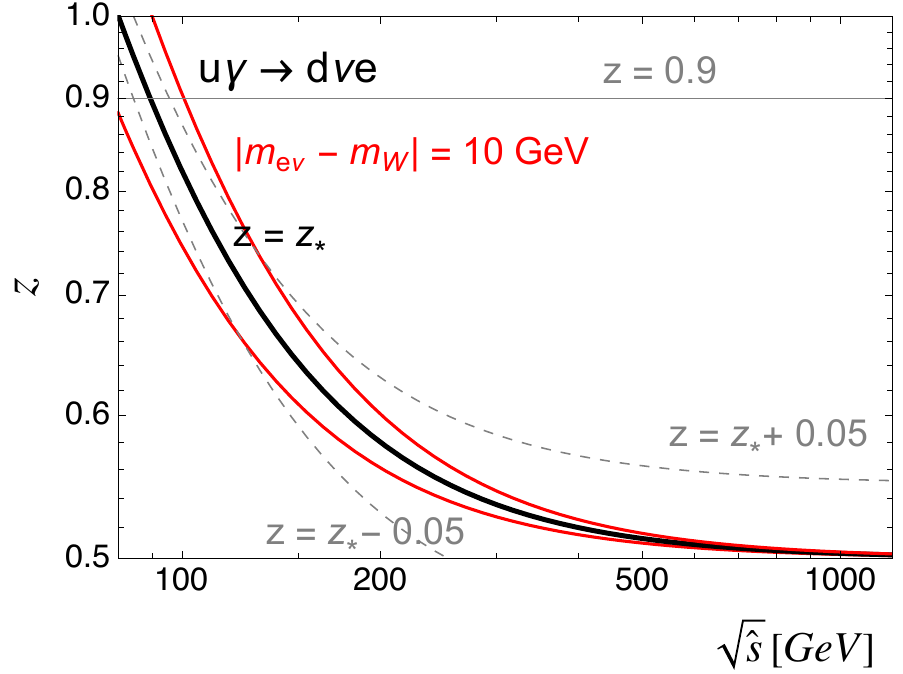} 
\includegraphics[width=0.495\textwidth]{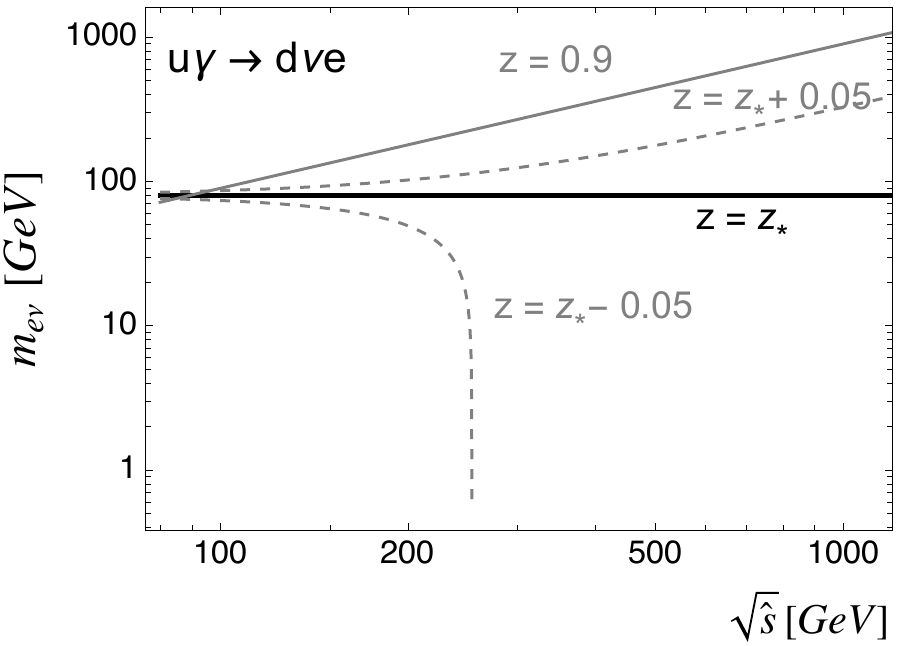}
\caption{\small Left: The fraction of the energy that flows into the $e\nu$ system, $z = E_{e\nu}/\sqrt{\hat{s}}$, as a function of $\sqrt{\hat{s}}$. The band bounded by red lines correspond to the $W$ mass window of 10 GeV. Black line denotes the $z$ value for the on-shell $W$ of the mass $m_W$. Right: the correlation between $m_{e\nu}$ and $\sqrt{\hat{s}}$ depending on the cut on $z$.}
\label{fig:int:toy:control:demo}
\end{center}
\end{figure}

The fraction of the energy, $z_* = 1/2 + m_W^2/(2\hat{s})$, carried by the on-shell $\nu_e e^+$ system is roughly order one for $\hat{s} \sim m_W^2$, and it rapidly drops to 1/2 with increasing $\sqrt{\hat{s}}$ as is seen in the left panel of Fig.~\ref{fig:int:toy:control:demo}. 
Due to the relation $m_{e\nu} = \sqrt{(2z-1) \hat{s}}$, two variables $m_{e\nu}$ and $\sqrt{\hat{s}}$ are comparable to each other only for a low $\sqrt{\hat{s}}$ where $2z-1$ is roughly order one and they can be very different for a large $\sqrt{\hat{s}}$ as $2z-1$ can be almost zero. The region bounded by red lines in the left panel of Fig.~\ref{fig:int:toy:control:demo} corresponds to the $z$ value for the $W$ mass window of 10 GeV, or $| m_{\ell\nu} (= \sqrt{(2z-1) \hat{s}} ) - m_W | < 10$ GeV. To access the off-shell region, we impose the cut on $z$ such as $|z - z_*| = |(m_{e\nu}^2  - m_W^2)/(2 \hat{s})| >$ 0.05 and $z > 0.9$ and they are shown by gray lines in Fig.~\ref{fig:int:toy:control:demo}. The selected region by $z  > 0.9$ isolates the phase space where most of the center of mass energy $\sqrt{\hat{s}}$ flows into the $\nu_e e^+$ system and $m_{e\nu}$ can be as large as $\sqrt{\hat{s}}$ as is seen in the right panel of Fig.~\ref{fig:int:toy:control:demo}. It is also the phase space where the condition for the typical EWA is expected to be satisfied.  While the cut of $z-z_* > 0.05$ makes it possible to access a deeper off-shell region in a large $\sqrt{\hat{s}}$ region than that specified by the $W$ mass window of 10 GeV (red lines in Fig.~\ref{fig:int:toy:control:demo}), most events are still populated near the lower $m_{e\nu}$ value than $\sqrt{\hat{s}}$. 

\begin{figure}[tph]
\begin{center}
\includegraphics[width=0.48\textwidth]{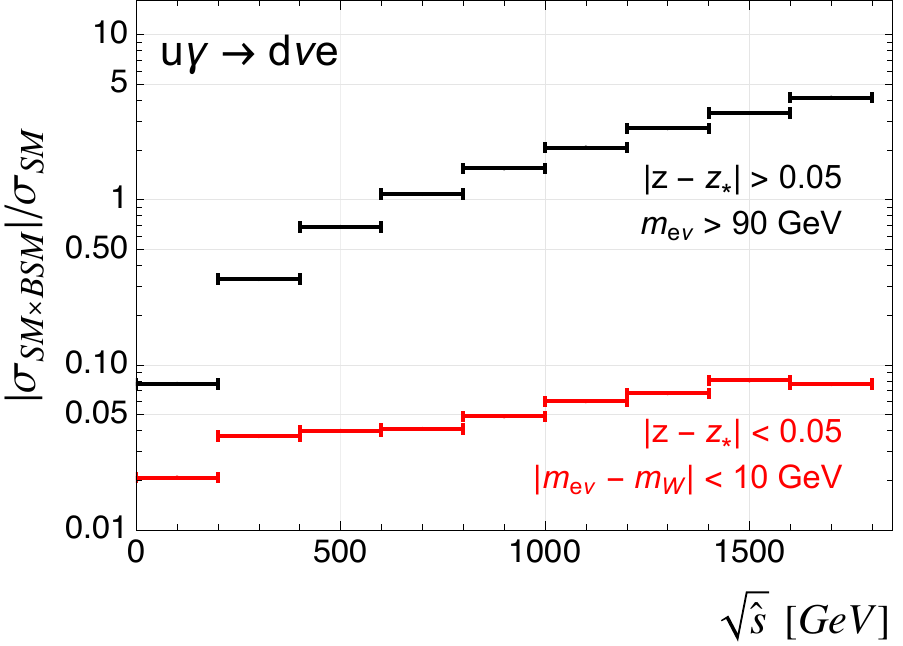}
\caption{\small The differential distribution of $|\sigma_{SM\times BSM}|/\sigma_{SM}$  in $\sqrt{\hat{s}}$ for the EW $u\gamma \rightarrow d\nu_e e^+$ at the parton level. Black lines demonstrate the interference at off-shell region specified as $|z-z_*| > 0.05$ and $m_{e \nu} > 90$ GeV. Red lines demonstrate the noninterference for the on-shell $W$ defined by  $|z-z_*| < 0.05$ and $|m_{e\nu} - m_W | < 10$ GeV. }
\label{fig:int:toy:shat:demo}
\end{center}
\end{figure}

Our numerical simulations of $|\sigma_{SM\times BSM}|/\sigma_{SM}$ binned in $\sqrt{\hat{s}}$ is illustrated in Fig.~\ref{fig:int:toy:shat:demo}~\footnote{In this work, we will not explore the sign of the interference and its sensitivity at the collider. The sign of the interference depends on the phase space (see Appendix~\ref{sec:app:int} for the related discussion).}.
As is evident by the red-colored almost flat distribution in Fig.~\ref{fig:int:toy:shat:demo}, the noninterference predicted for the on-shell $W$ in Section~\ref{sec:onshellW} is numerically confirmed. The black-colored distribution in Fig.~\ref{fig:int:toy:shat:demo} demonstrates the resurrected energy growing interference in the inclusive cross section for the off-shell $W$ and they agree with our expectation in Section~\ref{sec:offshellW}.

\begin{figure}[tph]
\begin{center}
\includegraphics[width=0.48\textwidth]{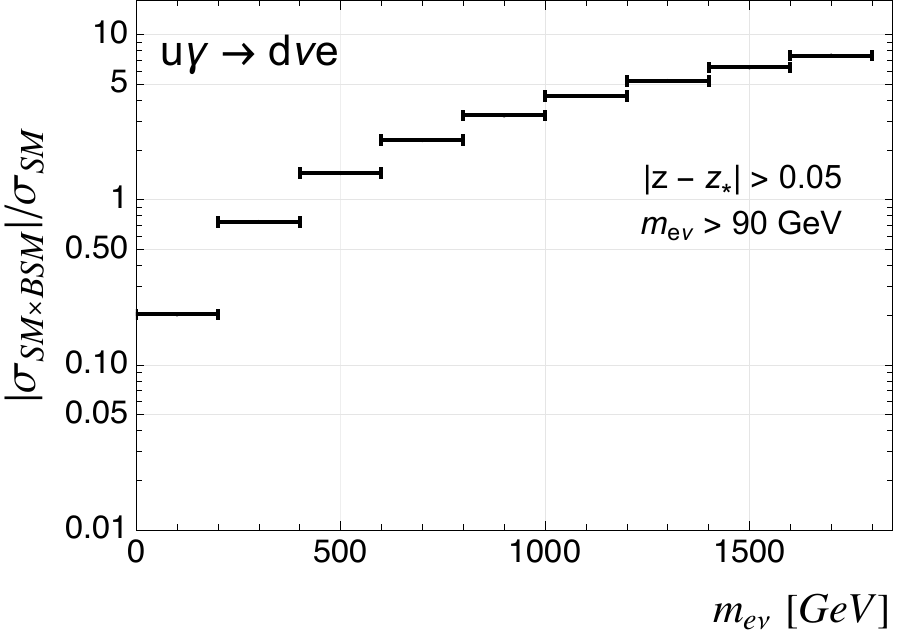}
\includegraphics[width=0.48\textwidth]{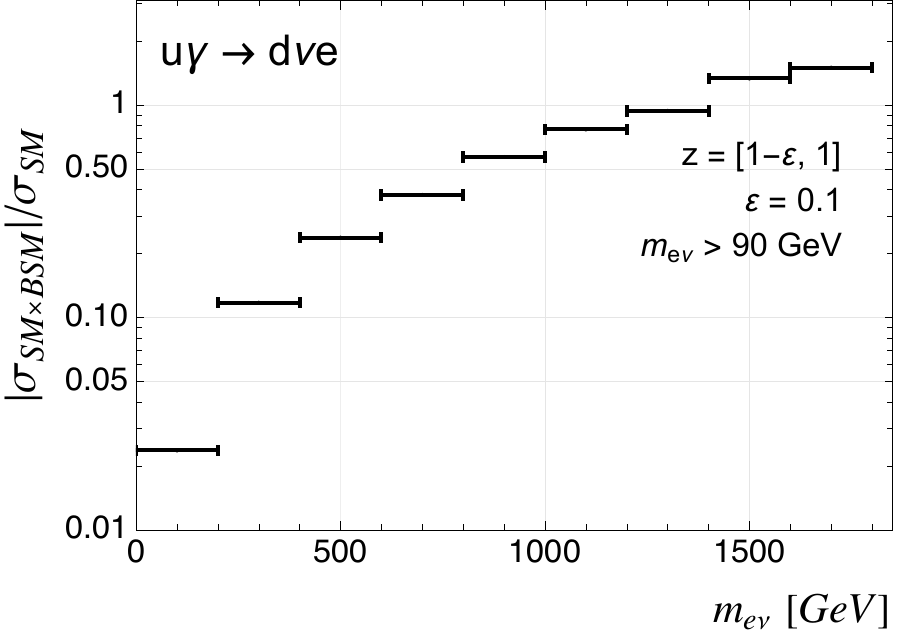}
\caption{\small The differential distribution of $|\sigma_{SM\times BSM}|/\sigma_{SM}$ in $m_{e\nu}$ for the EW $u\gamma \rightarrow d\nu_e e^+$ at the parton level. Events are restricted to satisfy $|z-z_*| > 0.05$ and $m_{e \nu} > 90$ GeV (left) or $z > 0.9$ and $m_{e \nu} > 90$ GeV (right).  }
\label{fig:int:toy:mll:demo}
\end{center}
\end{figure}

The left panel of Fig.~\ref{fig:int:toy:mll:demo}  illustrates the interference cross section with respect to the SM in terms of the $m_{e\nu}$ variable for the same phase space as those in Fig.~\ref{fig:int:toy:shat:demo}, and the energy-growing behavior is clearly seen.
In the right panel of Fig.~\ref{fig:int:toy:mll:demo}, we take a limit where almost all energy $\sqrt{\hat{s}}$ flows into the $e\nu$ system (see the solid gray line in the left panel of Fig.~\ref{fig:int:toy:control:demo}). 
As is clearly seen in the right panel of Fig.~\ref{fig:int:toy:mll:demo}, the energy growing interference term looks survive in this limit of the full $2\rightarrow 3$ process. 
However, this energy growing interference term allowed by the helicity selection rule of the full $2\rightarrow 3$ process will get lost if one simply assumes the EWA and works on the $2\rightarrow 2$ hard subprocess. Recall that the helicity selection rule of the $2\rightarrow 2$ subprocess does not allow the interference in the massless limit.

While we have exploited the variable $z$ to distinguish the phase spaces of the on-shell and off-shell regions, it can be traded for a combination of experimental variables. Using the transverse momentum of the forward quark, $p_T(q) = (1-z) \sqrt{\hat{s}}\sin\theta$ (with $\sin\theta = 1/\cosh\eta$), and $m_{e\nu}= \sqrt{2z-1} \hat{s}$, one can easily derive the relation,
\begin{equation}\label{eq:toy:pTcut}
  \frac{p_T(q)\cosh\eta}{m_{e\nu}} = \frac{1-z}{\sqrt{2z-1}} \leq  \frac{1-z_{\rm min}}{\sqrt{2z_{\rm min}-1}} \equiv \delta_{\rm min}~ 
  \rightarrow p_T(q) \leq \delta_{\rm min} \frac{m_{e\nu}}{\cosh\eta}~,
\end{equation}
where $z_{\rm min} = \{z_* + 0.05, \, 1-\varepsilon \}$ was used in the plots in Fig.~\ref{fig:int:toy:mll:demo} and $\eta$ is the pseudorapidity of the outgoing quark. Note that $z_*$ (thus $\delta_{\rm min}$ as well) is still a function of the experimentally inaccessible $\hat{s}$ although its dependence gets mild in the high $\hat{s}$ limit. For the hard cut on $z$, $\delta_{\rm min}$ becomes a constant. We have numerically checked that the cut $p_T(q) < 0.112\times (m_{e\nu}/\cosh\eta)$ is physically equivalent to $z > 1-\varepsilon$ (with $\varepsilon = 0.1$) and reproduces the same plot as the right panel of Fig.~\ref{fig:int:toy:mll:demo}. 

\subsection{Beyond the effective $W$ approximation}
\label{sec:ewa}

We provide a brief qualitative comparison with the derivation of the effective $W$ approximation in literature to understand better what really happens regarding the interference resurrection.  This reveals a nontrivial nature of the interference resurrected in the inclusive cross section that was demonstrated above. The full detail of the comparison is given in Appendix~\ref{app:sec:ewa}. Here, we quote only the final essence.
Motivated by the discussion in~\cite{Borel:2012by}, the total amplitude of $u\gamma \rightarrow d \nu e^+$ from the SM and BSM is decomposed into Fourier modes in $\phi$, or $\epsilon\cdot \mathcal{M} = \sum_n C_n\, e^{i n\phi}$ where the coefficient $C_n$ can be easily obtained by the residue theorem. 
In the forward quark limit, $\theta = \pi - \varepsilon$ with $\varepsilon \ll 1$ (with the abuse of the notation), the full amplitude can be expanded in powers of $\tilde{\varepsilon} = \varepsilon e^{-i\phi}$ and its complex conjugate $\tilde{\varepsilon}^*$ and so on,
\begin{equation}\label{eq:amp:varepsilon}
\begin{split}
   \epsilon\cdot \mathcal{M} &= \tilde{\varepsilon} \left ( \mathcal{M}^{(0,0)}_+ + \mathcal{M}^{(1,0)}_+ \tilde{\varepsilon} + \mathcal{M}^{(0,1)}_+ \tilde{\varepsilon}^* + \cdots~ \right )
   \\[4pt] 
   &\hspace{1.cm} + \tilde{\varepsilon}^* \left ( \mathcal{M}^{(0,0)}_- + \mathcal{M}^{(1,0)}_- \tilde{\varepsilon} + \mathcal{M}^{(0,1)}_- \tilde{\varepsilon}^* + \cdots~ \right ) +  \cdots,
\end{split}
\end{equation}
where $\epsilon$ is the polarization of the photon and all $\mathcal{M}^{(i,j)}$ have the same energy dependence. 
Two groups of terms, denoted by $\pm$, being proportional to the overall $\tilde{\varepsilon}$ and $\tilde{\varepsilon}^*$ in Eq.~(\ref{eq:amp:varepsilon}) are associated with the contributions from two transverse polarizations. The contributions associated with the longitudinal polarization and those suppressed by $\frac{m_W^2}{E^2}$ are denoted by $\cdots$ in Eq.~(\ref{eq:amp:varepsilon}).
Since $p_T(q) = p_\perp \sim (1-z)\sqrt{\hat{s}}\sin\varepsilon$ and $m^2_{e\nu} = (2z-1)\hat{s}$ in our toy process, taking $m_{e\nu} \sim E$ for $z\sim \mathcal{O}(1)$ as an energy of the hard subprocess, we have the relation $p_\perp \sim E\, \varepsilon$ from which we can relate $\tilde{\varepsilon}$ with $\tilde{p}_\perp \equiv p_\perp e^{-i\phi}$ introduced in~\cite{Borel:2012by}. Therefore, the expansion of our full amplitude in powers of $\tilde{\varepsilon}$ and its complex conjugate should be equivalent to the expansion in powers of $\frac{\tilde{p}_\perp}{E}$ and its complex conjugate as was done in~\cite{Borel:2012by},
\begin{equation}
\begin{split}
   \epsilon\cdot \mathcal{M} &= \frac{\tilde{p}_\perp}{E} \left ( \mathcal{M}^{(0,0)}_+ + \mathcal{M}^{(1,0)}_+ \frac{\tilde{p}_\perp}{E} + \mathcal{M}^{(0,1)}_+ \frac{\tilde{p}_\perp^*}{E} + \cdots~ \right )
   \\[4pt]
   &\hspace{1.cm} + \frac{\tilde{p}^*_\perp}{E} \left ( \mathcal{M}^{(0,0)}_- + \mathcal{M}^{(1,0)}_- \frac{\tilde{p}_\perp}{E} + \mathcal{M}^{(0,1)}_- \frac{\tilde{p}_\perp^*}{E} + \cdots~ \right ) + \cdots~.
\end{split}
\end{equation}
From our explicit expressions of the full amplitude with only the coupling $\lambda_z$ for the left-handed photon polarization as an illustration, given in Appendix~\ref{app:sec:ewa}, we find that
\begin{equation}
\begin{split}
   \mathcal{M}^{(0,0)}_{-,SM}  \neq 0 
   \quad \mathcal{M}^{(0,0)}_{-,BSM} = 0~,\quad
   \mathcal{M}^{(0,0)}_{+,SM} = 0
   \quad \mathcal{M}^{(0,0)}_{+,BSM} \neq 0~,
\end{split}
\end{equation}
which is consistent with the helicity selection rule, namely noninterference in the subprocess. The leading contribution to the interference that can survive in the total cross section comes from, when squaring the full amplitude,
\begin{equation}
  \propto (\tilde{\varepsilon} \tilde{\varepsilon}^*)^2\left ( \mathcal{M}^{(1,0)*}_{-,SM}  \mathcal{M}^{(0,1)}_{+,BSM} + h.c. \right )
  + \cdots~,
\end{equation}
whereas the leading SM and quadratic terms of the BSM are given by
\begin{equation}\label{app:eq:sm:bsmsq}
\begin{split}
  &\propto (\tilde{\varepsilon} \tilde{\varepsilon}^* ) \left | \mathcal{M}^{(0,0)}_{-,SM} \right |^2
+ (\tilde{\varepsilon} \tilde{\varepsilon}^*) \left | \mathcal{M}^{(0,0)}_{+,BSM} \right |^2 + \cdots~,
\end{split}
\end{equation}
where only leading terms that should survive in the inclusive cross section ($\phi$-independent terms) are shown.
Therefore, the leading contributions to the interference and quadratic terms in the inclusive cross section with respect to the SM cross section scale as, keeping only leading $\phi$-independent contributions,
\begin{equation}\label{eq:int:quad:wrt:sm}
 \frac{ \left |\epsilon\cdot{\mathcal{M}} \right |^2_{SM\times BSM}}{\left | \epsilon\cdot{\mathcal{M}} \right |^2_{SM}} \propto \lambda_z \varepsilon^2 \frac{E^2}{\Lambda^2}~,\quad
 \frac{ \left |\epsilon\cdot {\mathcal{M}} \right |^2_{BSM^2}}{ \left | \epsilon\cdot {\mathcal{M}} \right |_{SM}} \propto \lambda_z^2 \frac{E^4}{\Lambda^4}~,
\end{equation}
where only the interference with respect to the SM appears suppressed by $\varepsilon^2$ compared to the typical energy-growing behaviors. Note that there could be also terms suppressed by $\frac{m^2_W}{E^2}$ in the interference in Eq.~(\ref{eq:int:quad:wrt:sm}). Although our demonstration was done assuming the limit $\varepsilon \ll 1$, our exact result (for instance, Eq. (\ref{app:eq:amp23:exact})) extends to the situation with a sizeable $\varepsilon$ which can be thought of kind of the resummation, and this phase space is not caught in the EWA limit.
If only leading terms $\mathcal{M}^{(0,0)}_\pm$ were taken (as was in the derivation of the EWA in~\cite{Borel:2012by}), the interference between the SM and BSM would have been lost upon the integration over $\phi$ and the situation falls into the usual expectation from the typical EWA. In this work, however, we newly point out that the sizable interference terms to the total cross section can come from higher-order terms beyond the regime for the EWA. We suspect that our situation should belong to an exceptional case, the helicity selection rules, briefly mentioned in~\cite{Borel:2012by}, as a case where their derivation may become invalid.

\section{Numerical analysis of EW dilepton with two associated jets}
\label{sec:vbf:numeric}

In this section we numerically investigate the EW $\ell\ell$ + two jets process at the LHC. We take the CMS analysis in~\cite{CMS:2017dmo} as our baseline for both the validation of our analysis and the derivation of the sensitivity on aTGCs at the LHC~\footnote{Similar study by the CMS collaboration for the EW $\ell\nu_\ell$ + two jets process has been made in~\cite{CMS:2019nep}}. The detail of the event generation can be found in Appendix~\ref{sec:app:sim:detail}.

\subsection{Interference resurrection}
\label{sec:int:resurrection}

We can use the intuition from the EW $u\gamma \rightarrow d \nu e^+$ process in Section~\ref{sec:vbf:toy} to isolate the phase space that reveals the interference resurrection in the EW $\ell\ell +$ two jets process.
In the partonic EW $\ell\ell +qq'$ process, we can treat the $\ell\ell$ ($qq'$) system effectively as a single particle with the energy of $z\sqrt{\hat{s}}$ ($(1-z)\sqrt{\hat{s}}$) and the invariant mass of $m_{\ell\ell}$ ($m_{qq'}$). Similarly to our toy process in Section~\ref{sec:vbf:toy}, the variable $z$ represents the fraction of the total energy flowing into the dilepton system. Three momentum conservation, $\vec{p}_T(\ell\ell) = -\vec{p}_T(qq')$, in the center of mass frame of two initial quarks leads to $m^2_{\ell\ell} - m^2_{qq'} = (2z-1)\hat{s}$ where $z$ varies over the range $z = [\, m_{\ell\ell}/\sqrt{\hat{s}},\, 1-m_{qq'}/\sqrt{\hat{s}}\, ]$.
Similarly to the previous section, we start with the variable $z = 1/2 + (m^2_{\ell\ell} - m^2_{qq'})/(2\hat{s})$ to separate the off-shell phase space from the on-shell one where $z_* = 1/2 + (m^2_Z - m^2_{qq'})/(2\hat{s})$ at the $Z$ pole.  An appropriate cut on $z$ such as $|z-z_*| = |(m^2_{\ell\ell} - m^2_Z)/(2\hat{s})| > \Delta z$ or $z > z_{\rm min}$ will select the corresponding off-shell region, while ensuring a certain correlation between $m_{\ell\ell}$ and $\sqrt{\hat{s}}$.
Combining $m^2_{\ell\ell} - m^2_{qq'} = (2z-1)\hat{s}$ with the transverse momentum of the effective $qq'$ system $p_T(qq') = \sqrt{(1-z)^2 \hat{s} - m^2_{qq'}} \sin\theta_{qq'}$, the variable $z$ can be translated into the nontrivial combination of various kinematic variables via the relation,  
\begin{equation}\label{eq:dilep:pTcut}
 {\rm VBFhardness} \equiv \frac{m^2_{\ell\ell} - m^2_{qq'}}{p_T^2(qq') \cosh^2\eta_{qq'} + m^2_{qq'}} = \frac{2z-1}{(1-z)^2} \geq  \frac{2z_{\rm min}-1}{(1-z_{\rm min})^2} \quad \text{for}\quad z \geq z_{\rm min}~,
\end{equation} 
where the ratio is the monotonically increasing function, while it can have either sign, and $\sin\theta_{qq'} = 1/\cosh\eta_{qq'}$ was used to express in terms of the pseudorapidity of the $qq'$ system.  The positive value of the VBFhardness (or equivalently $z>1/2$) corresponds to the case where more than half the total energy flows into the dilepton system.
Just like the case of our toy process in Section~\ref{sec:vbf:toy}, $z_{\rm min}$ still has the $\hat{s}$ dependence if one intends to impose a cut on $|z-z_*|$ instead of a constant cut on $z$ itself.

\begin{figure}[tp]
\begin{center}
\includegraphics[width=0.48\textwidth]{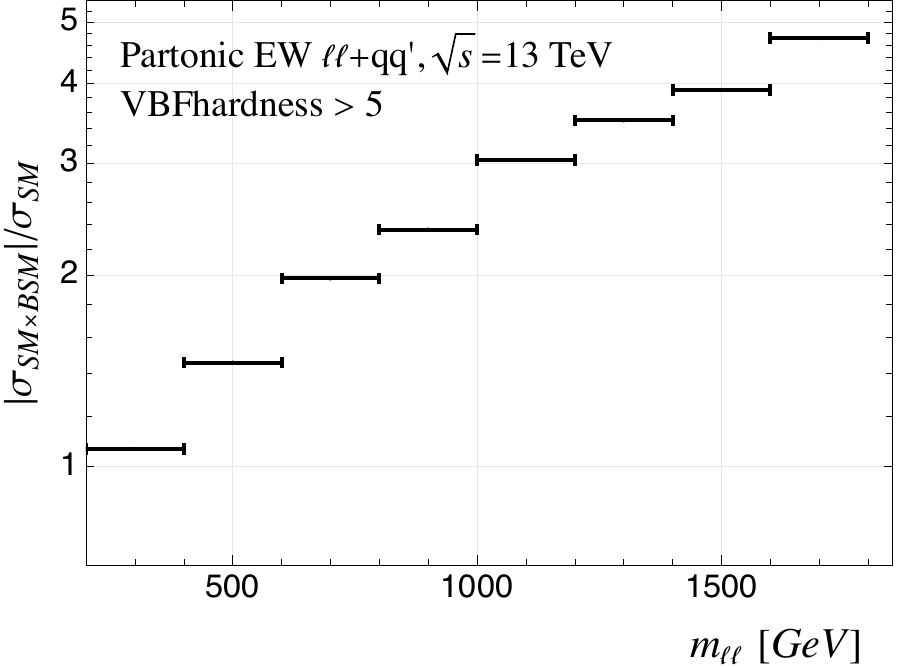}
\includegraphics[width=0.51\textwidth]{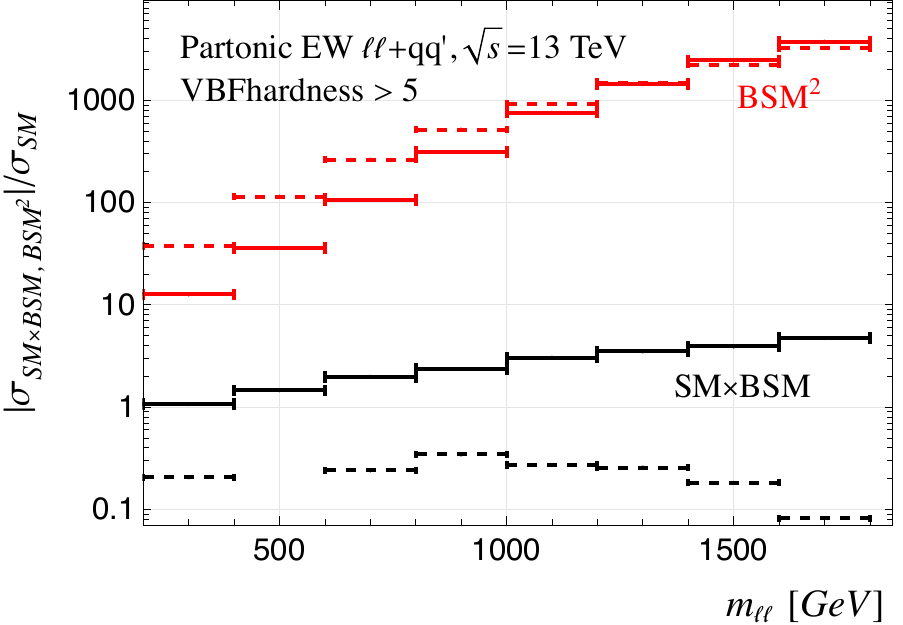}
\caption{\small The distributions of $|\sigma_{SM\times BSM}|/\sigma_{SM}$ in $m_{\ell\ell}$ for the partonic EW $\ell\ell + qq'$ (black lines in both panels) for the $\lambda_z$ coupling (other couplings are set to zero). Similarly for $|\sigma_{{BSM}^2}|/\sigma_{SM}$ (red lines).  Events for solid lines are restricted to those with VBFhardness $ > 5$ in Eq.~(\ref{eq:dilep:pTcut}) along with $p_T(q) >25$ GeV, $p_T(\ell) > 10$ GeV, and $m_{qq'} >120$ GeV. For dashed lines in the right panel, the VBFhardness cut is removed while others kept the same.}
\label{fig:int:dilep:mll:demo}
\end{center}
\end{figure}

\begin{figure}[tp]
\begin{center}
\includegraphics[width=0.48\textwidth]{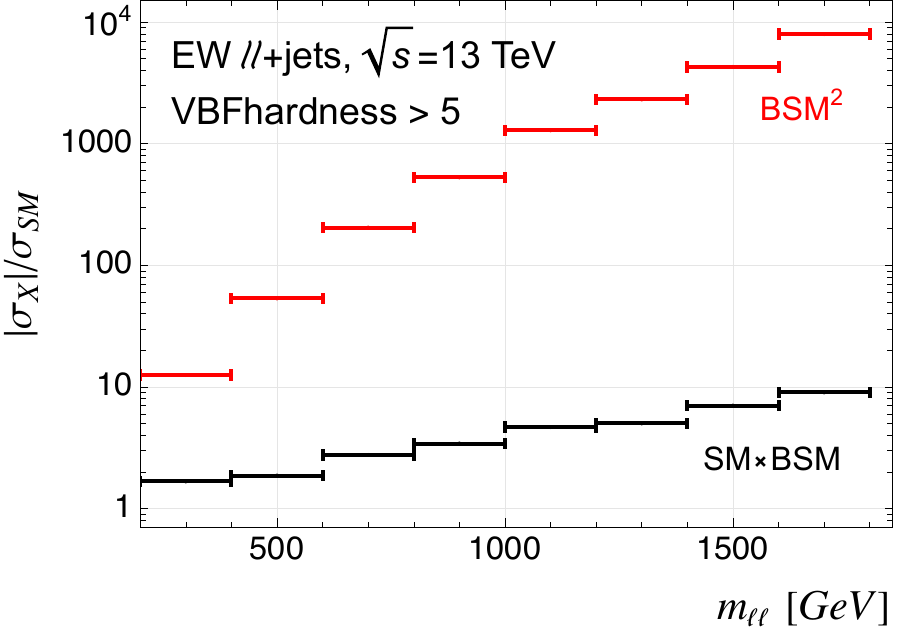}
\includegraphics[width=0.462\textwidth]{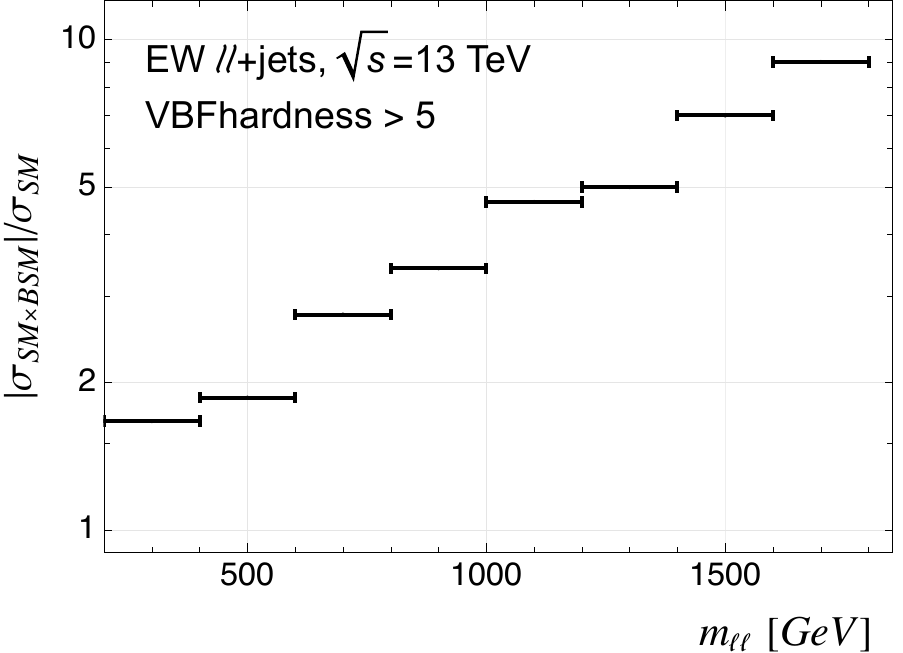}\\[10pt]
\includegraphics[width=0.48\textwidth]{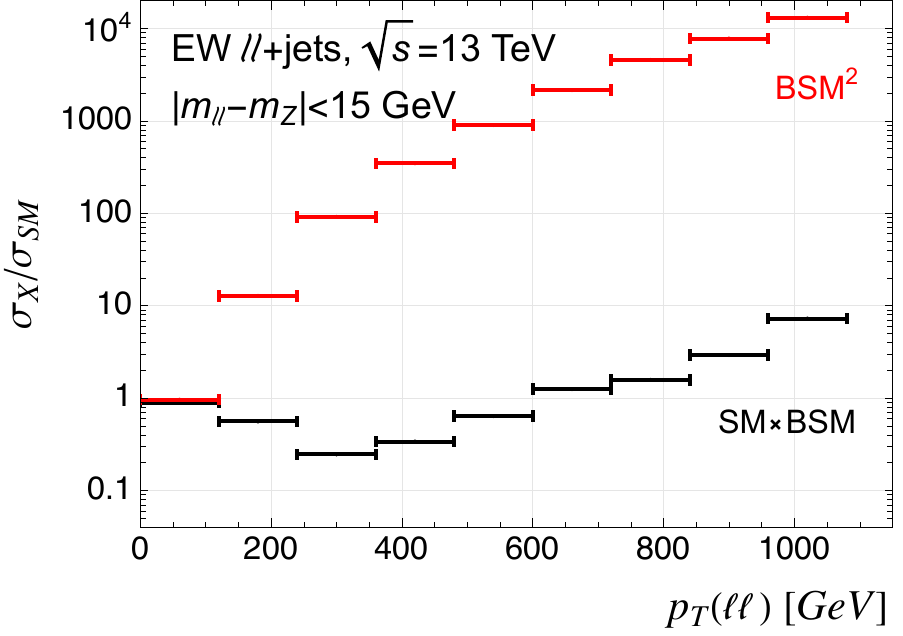}
\caption{\small $|\sigma_{X}|/\sigma_{SM}$ where $X =$ SM$\times$BSM (black) or ${\rm BSM}^2$ (red) for the EW $\ell\ell$ + two jets for the coupling $C_i = \lambda_z$ (other couplings are set to zero). Plots are made with events at the jet level after imposing the loosened cuts, compared to the CMS analysis~\cite{CMS:2017dmo}, $p_T(j) >30$ GeV, $p_T(\ell) >20$ GeV, $|\eta(j)| <4.5$ , $|\eta(\ell)| < 2.5$, and $m_{jj}>120$ GeV.}
\label{fig:int:dilep:mll:signal}
\end{center}
\end{figure}

As is evident in the right panel of Fig.~\ref{fig:int:dilep:mll:demo} (see black dashed lines), the interference does not reveal the energy growing behavior without a cut on the ratio in Eq.~(\ref{eq:dilep:pTcut}). As an illustration, the resurrected interference in the inclusive cross section for the $\lambda_z$ coupling is clearly shown in the left panel of Fig.~\ref{fig:int:dilep:mll:demo} for VBFhardness $> 5$ that corresponds to $z \geq z_{\rm min} = 0.71$. We checked that a similar energy growing interference appears in terms of $\sqrt{\hat{s}}$ as well. 
The same interference is displayed again with the quadratic cross section in the right panel of Fig.~\ref{fig:int:dilep:mll:demo}. The square of the interference term in this illustrative example in Fig.~\ref{fig:int:dilep:mll:demo} appears to have a milder energy growing behavior than the quadratic term itself. The interference would have been lost if one has not included the full effect of the forward quarks or not imposed a cut on a proper variable like the one in Eq.~(\ref{eq:dilep:pTcut}). 
In Fig.~\ref{fig:int:dilep:mll:signal}, we show the resurrected interference pattern continues to survive at the hadron level where the VBFhardness is constructed out of two forward jet candidates and lepton pairs. The CMS analysis in~\cite{CMS:2017dmo} derives the sensitivity on aTGC using the $p_T$ distribution of $Z$ only for the events inside the $Z$ mass window. In the bottom panel of Fig.~\ref{fig:int:dilep:mll:signal}, the interference and quadratic terms of the inclusive cross section are illustrated in $p_T(\ell\ell)$ only for the events in the $Z$ mass window $|m_{\ell \ell} - m_Z | < 15$ GeV.
%

\subsection{Validation against the CMS analysis and BDT analysis}
\label{sec:val:CMS}

We adopt the CMS analysis in~\cite{CMS:2017dmo} for the validation of our framework. Events with two isolated leptons (electrons or muons) and at least two jets are selected. A lepton is declared to be isolated if the ratio of the $p_T$-sum of all particles within the isolation cone $R_{iso} = 0.4$ around the lepton to the $p_T$ of the lepton is below 15\% and  25\% for electrons and muons, respectively. While two isolated leptons need to satisfy $p_T > 20$ GeV and $|\eta(\ell)| < 2.4$, and have the opposite electric charges, the harder lepton must pass the cut $p_T > 30$ GeV as well. 

The particles excluding the isolated leptons are clustered into jets by anti-$k_t$ algorithm~\cite{Cacciari:2008gp} with the distance parameter of $R_{jet} = 0.4$. Jets are required to satisfy $p_T(j) > 15$ GeV and $|\eta(j)| \leq 4.7$. Two hardest jets, called the tagging jets, are required to have $p_T(j) > 50$ GeV and $p_T(j) > 30$ GeV for the leading and subleading jets, respectively, and their invariant mass should satisfy $m_{jj} > 200$ GeV. The initial cuts in CMS analysis in~\cite{CMS:2017dmo} are defined as 
\begin{equation}\label{eq:cuts:ini}
\begin{split}
    &p_T(\ell_1) > 30\ \text{GeV}~, \quad  p_T(\ell_2) > 20\ \text{GeV}~,\quad |\eta(\mu)| < 2.4~,\quad |\eta(e)| < 2.1~,
    \\[3pt]
    &p_T(j_1) > 50\ \text{GeV}~, \quad p_T(j_2) > 30\ \text{GeV}~, \quad |\eta(j)| \leq 4.7~,
    \\[3pt]
     &| m_Z - m_{\ell\ell}| < 15\ \text{GeV}~,  \textrm{ and } \quad m_{jj} > 200\ \text{GeV}
\end{split}
\end{equation}
where the subscripts $1$ and $2$ mean leading and subleading objects, respectively. The event yields after imposing the initial cuts are given in Table~\ref{tab:CMS:Zjj:cuts:initial} where we included only two largest backgrounds.
\begin{table}[tbh]
\centering
  \renewcommand{\arraystretch}{1.1}
      \addtolength{\tabcolsep}{0.2pt} 
\scalebox{0.95}{
\begin{tabular}{|c|cc|}  
\hline
              & \multicolumn{2}{c|}{Initial}  \\
Sample  & \quad $ee$  & \quad $\mu\mu$ \\
\hline \hline
$t\bar{t}$  & 5454 (5363$\pm$48)        & 13962 (12938$\pm$81)    \\[5pt]
DY $Zjj$ (pythia8)  & 146147 (152750$\pm$510)  & 373731 (394640$\pm$880) \\[5pt]
EW $Zjj$ (pythia8)  & 2639 (2833$\pm$10)  & 6328 (6665$\pm$16)   \\[5pt]
\hline
\end{tabular}
}
\caption{\small Validation of our simulation at $\sqrt{s}$ =13 TeV assuming 35.9 fb$^{-1}$ of the integrated luminosity. The numbers in parenthesis are CMS values for comparison. The $k$-factor of 1.7 was applied for the $t\bar{t}$ process.}
\label{tab:CMS:Zjj:cuts:initial}
\end{table}
The smaller yield of the $ee$ channel is due to the lower selection efficiency of electrons. We adopted the $p_T$-dependent electron selection efficiency~\cite{CMS-DP-2017-004} in our analysis, while setting the selection efficiency for muons to unity. The electron selection efficiency is roughly $0.7 - 0.8$ for the $p_T$ of interest.

Having our analysis validated with the initial cuts, we move onto the BDT analysis.
The CMS analysis introduces two additional variables. Event balance variable, $R(p_T^{\text{hard}})$, is defined as
\begin{equation}
  R(p_T^{\text{hard}}) = \frac{\left | \vec{p}_{Tj_1} + \vec{p}_{Tj_2} + \vec{p}_{TZ} \right |}{\left | \vec{p}_{Tj_1} \right | + \left | \vec{p}_{Tj_2} \right | + \left | \vec{p}_{TZ} \right |}~
\end{equation}
The $z^*$ Zeppenfeld variable is defined as
\begin{equation}
   z^* = \frac{y^*}{\Delta y_{jj}}~,
\end{equation}
where $y^* = y_Z - \frac{1}{2} \left ( y_{j_1} + y_{j2} \right )$. Additionally, the quark-gluon discrimination is applied to two tagging jets. Instead of constructing a likelihood function for the $q$/$g$ discrimination and use it in the BDT analysis afterwards as done in the CMS analysis~\cite{CMS:2013kfa}, we directly use the three input variables to the likelihood in our BDT. They are multiplicity, jet shapes, and the fragmentation function. The jet shape variable is defined as  
\begin{equation}
 \sigma = \sqrt{\sigma_1^2 + \sigma_2^2}~\quad \text{with}\quad 
 \sigma_1 = (\lambda_1/\sum_i p^2_{T,i})^{1/2}~,\quad
 \sigma_2 = (\lambda_2/\sum_i p^2_{T,i})^{1/2}~,
\end{equation}
where the sum runs over the jet constituents. $\lambda_1$ and $\lambda_2$ are the two eigenvalues of the matrix with the elements,
$M_{11} = \sum_i p^2_{T,i} \Delta \eta_i^2$, $M_{22} = \sum_i p^2_{T,i} \Delta \phi_i^2$, and $M_{12} = M_{21} = - \sum_i p^2_{T,i}\Delta \eta_i \Delta \phi_i$ where $\Delta \eta_i$ and $\Delta\phi_i$ are the pseudorapidity and azimuthal distances between a constituent and the average direction which is defined as the $p^2_{T,i}$-weighted direction of jet constituents in $\eta-\phi$ space.
The fragmentation function is captured by the variable,
\begin{equation}
    p_T D = \frac{\sqrt{\sum_i p^2_{T,i}}}{\sum_i p_{T,i}}~,
\end{equation}
where the sum runs over the jet constituents. For the multiplicity we count all charged and neutral constituents of a jet whose energy is above 1 GeV, and it is denoted as $n_{\rm tracks}(j)$.

Similarly to the CMS analysis in~\cite{CMS:2017dmo}, we use the following set of the BDT variables to train and test our signal and background samples with the initial cuts in Eq.~(\ref{eq:cuts:ini}):
\begin{equation}\label{eq:bdt:set}
\begin{split}
 &m_{jj}~, \quad |\Delta \eta_{jj}|~,\quad p_T(jj)~, \quad R(p^{\rm hard}_T)~,\quad z^*(Z)~,
  \\[3pt]
   &n_{\rm tracks}(j_{1,2})~,\quad p_TD(j_{1,2})~,\quad \sigma_1 (j_{1,2})~,
\end{split}
\end{equation} 
where $m_{jj}$, $\eta_{jj}$, and $p_T(jj)$ are the invariant mass, pseudorapidity, and transverse momentum of two leading jets system, respectively. 
To simplify our analysis and at the same time to take full advantage of kinematic distribution to efficiently suppress the largest QCD Drell-Yan background, we first train and test over the EW $\ell\ell$ + jets in the SM as a signal and the remaining samples as the background using the gradient boosting algorithm (BDTG) provided by the {\tt TMVA} package~\cite{Hocker:2007ht}. Since the signal and the dominant background have the largest population in the $Z$ mass window with the small transverse momentum, the BSM effect is expected to be small. This rejects the QCD Drell-Yan and top pair backgrounds as much as possible. We impose an appropriate cut on the BDT variable, that was computed in the previous training, for all the samples of EW $\ell\ell$ + jets in the SM and BSM, and background processes. While it is nontrivial to exactly reproduce the outcome of the CMS BDT analysis, the outcome of our BDT training, illustrated in Fig.~\ref{fig:BDTG:output} in Appendix~\ref{app:sec:BDT}, shows the clear separation between the signal and background.

\begin{figure}[tp]
\begin{center}
\includegraphics[width=0.48\textwidth]{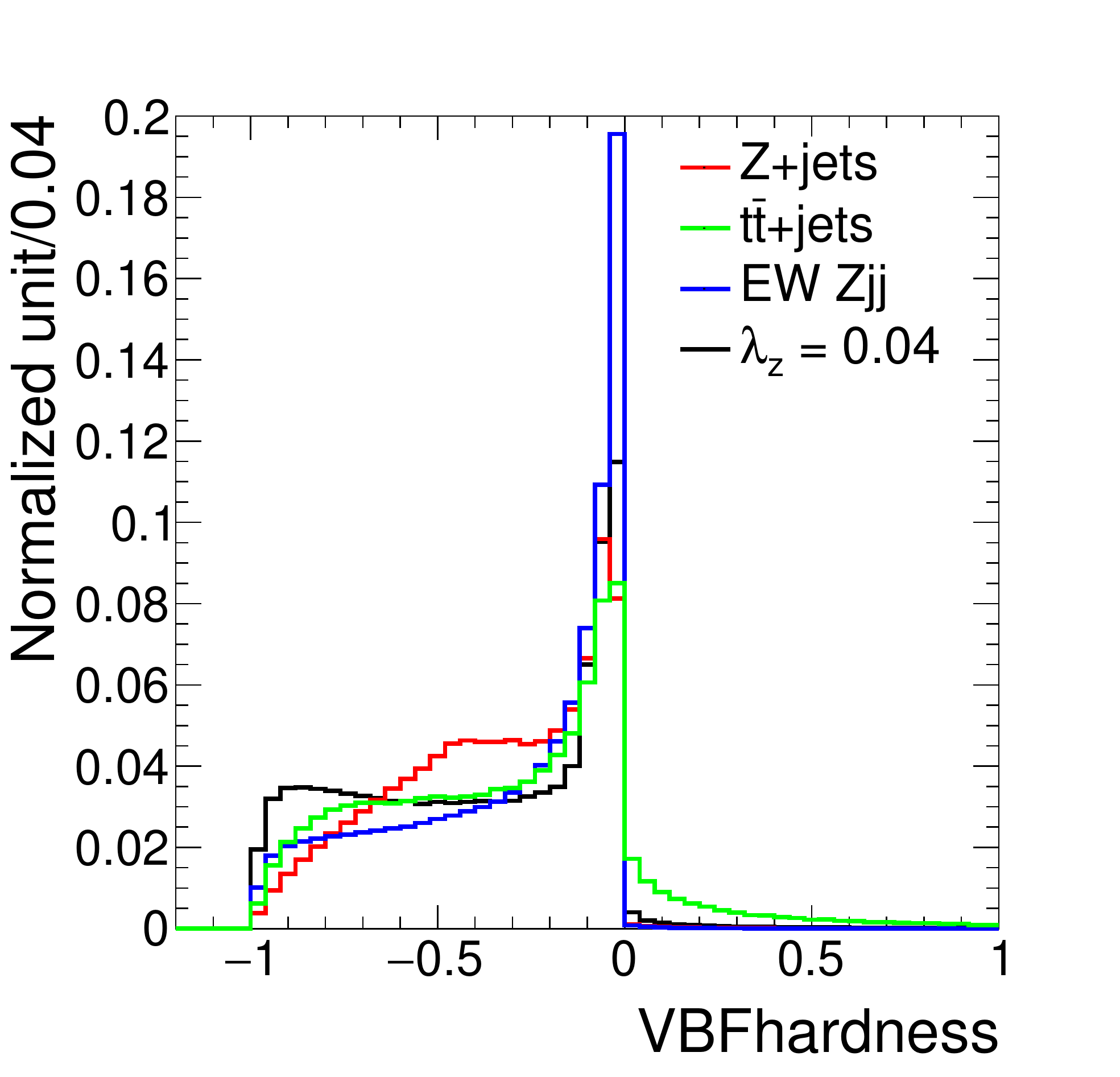}
\includegraphics[width=0.48\textwidth]{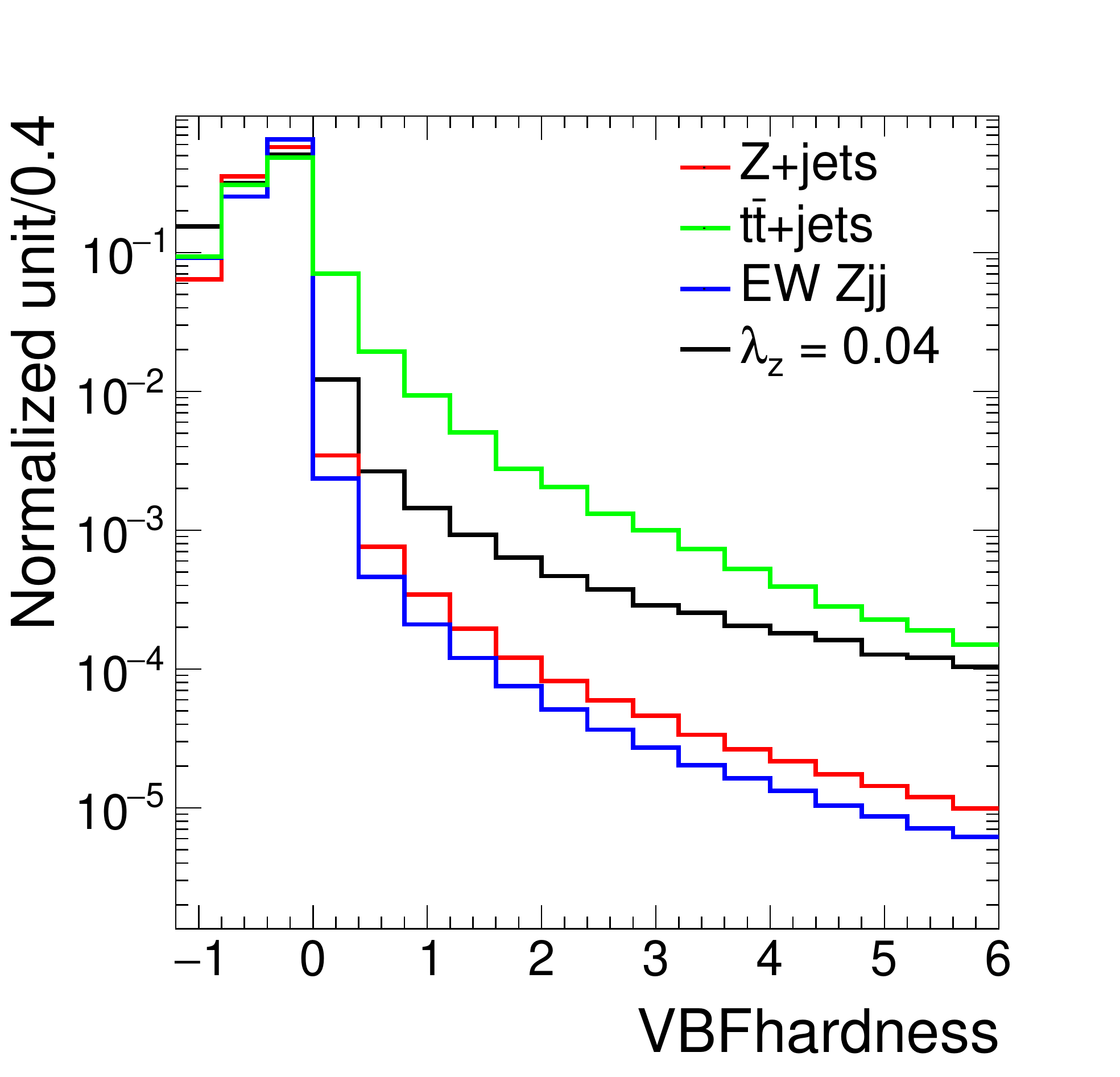}
\caption{\small The normalized distribution of VBFhardness for the EFT signal for $\lambda_z = 0.04$, EW dilepton (denoted by EW $Zjj$), $t\bar{t}$+jets, and QCD Drell-Yan backgrounds (denoted by $Z$+jets) after imposing $p_T(j) >30$ GeV, $p_T(\ell) >20$ GeV, $|\eta(j)| <4.5$ , $|\eta(\ell)| < 2.5$, and $m_{jj}>120$ GeV. Right panel is logarithmic plot of the left panel in a large VBFhardness range.}
\label{fig:vbfhardness}
\end{center}
\end{figure}

We do not add our newly introduced VBFhardness in Eq.~(\ref{eq:dilep:pTcut}) to the BDT variable set although it has a correlation with $m_{jj}$, $\eta_{jj}$, and $p_T(jj)$. Since we take the EW $\ell\ell$ + jets in the SM as a signal in the training, we expect its effect on the signal/background discrimination to be mild as is indicated in Fig.~\ref{fig:vbfhardness}. While the VBFhardness variable helps in resurrecting the interference, its effect should be small as well in the situation where the sensitivity of aTGCs is mainly driven by the quadratic terms. It will be relevant in case where the sensitivity is derived by the interference cross section. As is seen in Fig.~\ref{fig:vbfhardness}, although a proper cut may reduce the signal rate, VBFhardness seems to be a good discriminator for the EFT signal as it controls the amount of energy going into the dilepton subsystem. It will be important at the HL-LHC or future collider and we leave more dedicated analysis for the future study.

\begin{figure}[tp]
\begin{center}
\includegraphics[width=0.48\textwidth]{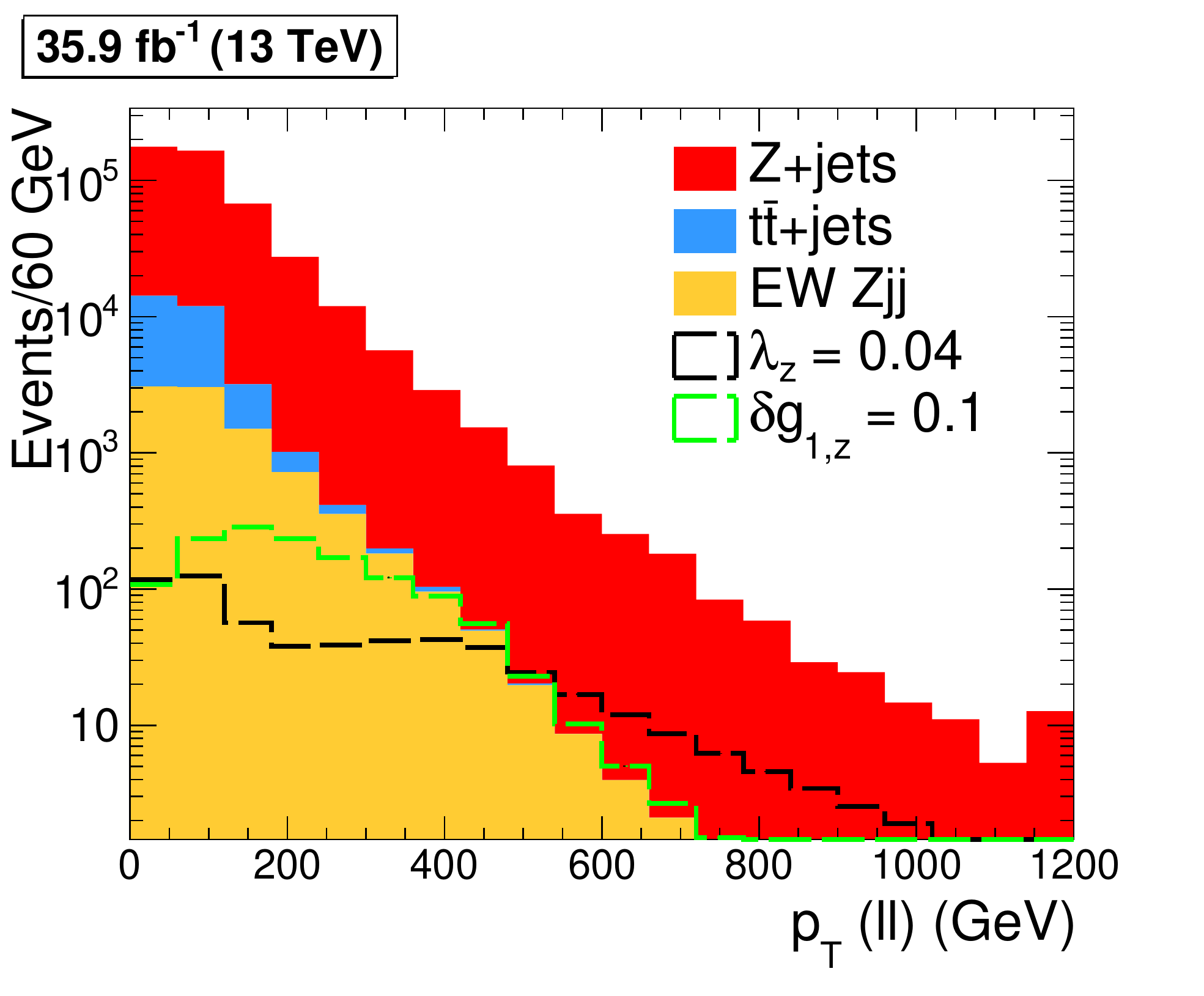}
\includegraphics[width=0.48\textwidth]{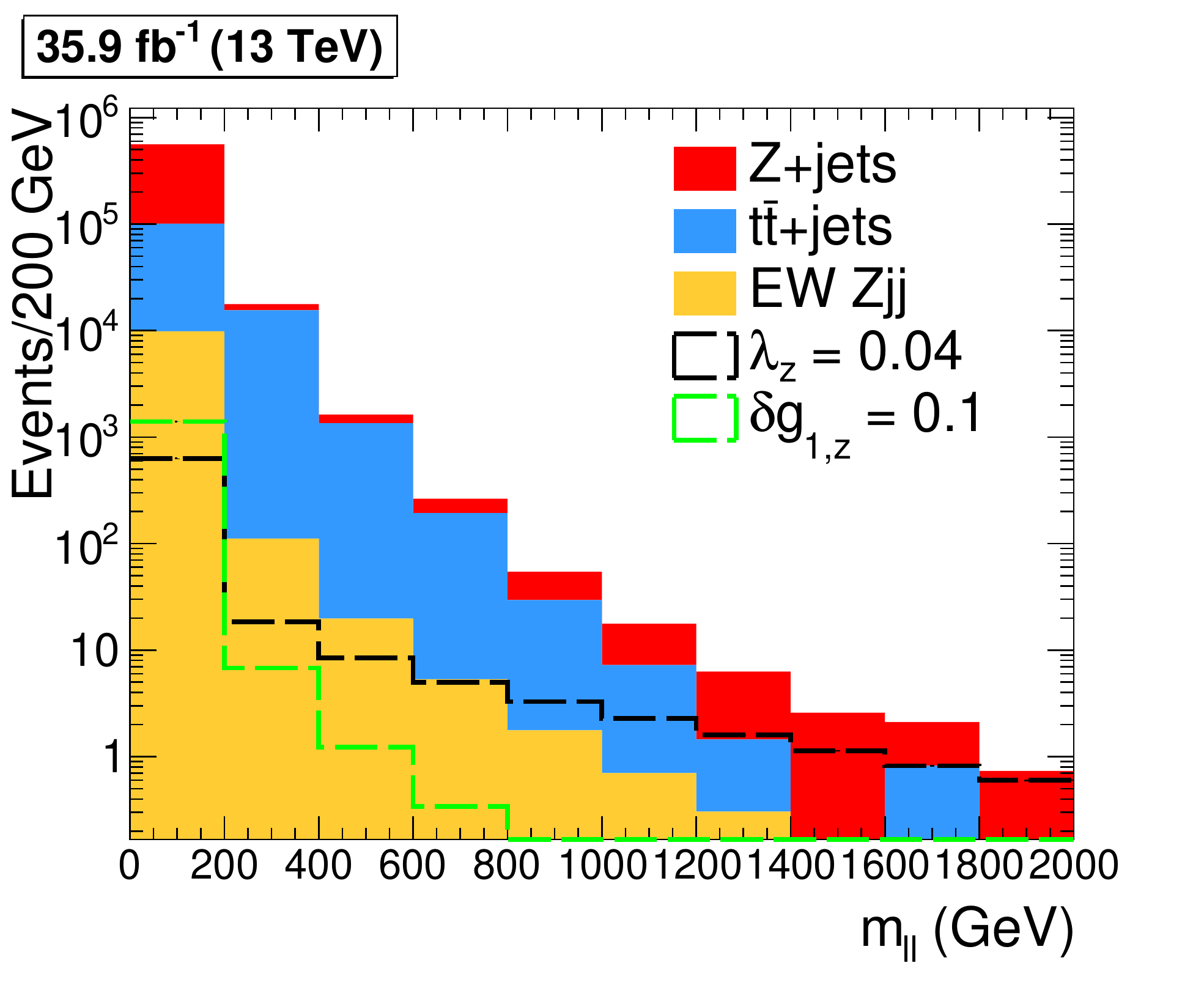}
\caption{\small The distributions of $p_T(\ell\ell)$ (left) and $m_{\ell\ell}$ (right) at 13 TeV, using the integrated luminosity of 35.9$^{-1}$, for backgrounds and two selected EFT benchmark signals with the SM contribution subtracted. Events are restricted to those satisfying CMS initial cuts in Eq.~(\ref{eq:cuts:ini}).}
\label{fig:diff:ptandmll:benchmark}
\end{center}
\end{figure}

\subsection{Sensitivity to aTGC at the LHC}
\label{sec:aTGC:limit}

To evaluate sentivity to aTGC, we construct 1D templates binned either in $p_T(\ell\ell)$ and $m_{\ell\ell}$. Events are distributed over 20 equal-spaced bins of $p_T(\ell\ell)$ between 0 and 1200 GeV where the last bin contains events beyond 1200 GeV. $\ell$ includes both electrons and muons~\footnote{On the contrary, the CMS analysis in~\cite{CMS:2017dmo} separately distribute events in 15 bins in $p_T(\ell\ell) = [0,\, 900]$ GeV and 20 bins in $[0,\, 1200]$ GeV for electrons and muons, respectively.}.
We also newly construct templates of $m_{\ell\ell}$ with 10 equal-spaced bins between 0 and 2000 GeV where the last bin contains events beyond 2000 GeV. The distributions of backgrounds and two selected EFT benchmark points (with the SM contribution subtracted) are illustrated in Fig.~\ref{fig:diff:ptandmll:benchmark}.
\begin{table}[tbh]
\centering
  \renewcommand{\arraystretch}{1.1}
      \addtolength{\tabcolsep}{0.2pt} 
\scalebox{0.86}{
\begin{tabular}{c|cc|c|cc|c}  
   \multicolumn{7}{l}{Using the template analysis of $p_T(\ell\ell)$ in the $Z$ mass window at 13 TeV, $\mathcal{L} =$ 35.9 fb$^{-1}$} \\
\hline
              &\multicolumn{3}{c|}{No BDT cut}								& \multicolumn{3}{c}{BDT $>$ 0.6} \\
aTGC  & \quad $68\%$ CL  \quad  & \quad $95\%$ CL  \quad & \quad $95\%$ CL (Linear) \quad & \quad $68\%$ CL \quad  & \quad $95\%$ CL  \quad  & \quad $95\%$ CL (Linear) \quad \\
\hline \hline
$\lambda_z$			& $[-0.026,\, 0.025]$		& $[-0.036,\, 0.036]$	& $[-0.20,\, 0.20]$	& $[-0.015,\, 0.016]$		& $[-0.025,\, 0.026]$  & $[-0.099,\, 0.1]$ \\[3pt]
$\delta g_{1,z}$			& $[-0.069,\, 0.040]$		& $[-0.130,\, 0.068]$	& $[-0.096,\, 0.097]$	& $[-0.029,\, 0.024]$		& $[-0.066,\, 0.043]$ 	 & $[-0.051,\, 0.051]$\\[3pt]
$\delta \kappa_z$		& $[-0.18,\, 0.19]$		& $[-0.29,\, 0.32]$	& $[-0.41,\, 0.41]$	& $[-0.089,\, 0.095]$		& $[-0.16,\, 0.18]$ 	 & $[-0.18,\, 0.18]$\\[3pt]
\hline
\end{tabular}
}
\caption{\small One-dimensional limits on aTGCs at 68\% and 95\% CL. Linear denotes the limits obtained using only the interference cross section between the SM and BSM amplitudes.}
\label{tab:template:ptll}
\end{table}
%
\begin{table}[tbh]
\centering
  \renewcommand{\arraystretch}{1.1}
      \addtolength{\tabcolsep}{0.2pt} 
\scalebox{0.86}{
\begin{tabular}{c|cc|c|cc|c}  
   \multicolumn{7}{l}{Using the template analysis of $m_{\ell\ell}$ at 13 TeV, $\mathcal{L} =$ 35.9 fb$^{-1}$} \\
\hline
              &\multicolumn{3}{c|}{No BDT cut} & \multicolumn{3}{c}{BDT $>$ 0.6} \\
aTGC  & \quad $68\%$ CL \quad  & \quad $95\%$ CL \quad & \quad $95\%$ CL (Linear)\quad & \quad $68\%$ CL \quad  & \quad $95\%$ CL \quad & \quad $95\%$ CL (Linear)\quad \\
\hline \hline
$\lambda_z$			& $[-0.031,\, 0.029]$		& $[-0.045,\, 0.043]$	&	$[-0.22,\, 0.22]$		& $[-0.025,\, 0.023]$		& $[-0.039,\, 0.035]$ & $[-0.13,\, 0.13]$ \\[3pt]
$\delta g_{1,z}$			& $[-0.074,\, 0.056]$		& $[-0.13,\, 0.094]$	&	$[-0.13,\, 0.13]$		& $[-0.033,\, 0.029]$		& $[-0.067,\, 0.052]$ & $[-0.062,\, 0.063]$\\[3pt]
$\delta \kappa_z$		& $[-0.099,\, 0.099]$		& $[-0.14,\, 0.15]$	&	$[-0.56,\, 0.56]$		& $[-0.062,\, 0.062]$		& $[-0.097,\, 0.098]$ & $[-0.26,\, 0.26]$\\[3pt]
\hline
\end{tabular}
}
\caption{\small Similar caption to Table~\ref{tab:template:ptll}.}
\label{tab:template:mll}
\end{table}
We construct a log likelihood in terms of aTGCs, assuming the Poisson distribution,
\begin{equation}
  -2 \Delta \log \mathcal{L} (\lambda_z, \delta g_{1,z}, \delta \kappa_z) ~,
\end{equation}
where $\Delta$ indicates that the minimum is subtracted. We include only the statistical uncertainty since the systematic uncertainty in each bin is not reported in~\cite{CMS:2017dmo} and the overall size of it in Table~\ref{tab:CMS:Zjj:cuts:initial} looks subdominant to the statistical one.

The 68\% and 95\% CL intervals of an individual aTGC, where two others are set to zero without the marginalization, are presented in Table~\ref{tab:template:ptll} and~\ref{tab:template:mll}.
For the result with the BDT cut, we estimated the sensitivity with the incremental BDT cut starting with a mild value, and did not find visible improvement with a stronger BDT cut than 0.6. For $\lambda_z$, the 95\% CL interval from BDT $> 0.6$ is worse than the expected value of the CMS one, or $\lambda_z^{\rm CMS} = [\, -0.014,\, 0.014\, ]$~\cite{CMS:2017dmo}~\footnote{\label{fn:compare:cms}Comparing two distributions of $p_T(Z)$ in Fig. 8 of~\cite{CMS:2017dmo} (separately displayed for electrons and muons) and Fig. 11 (summed over both leptons), our signal to background ratio looks rather smaller than the CMS one in a high $p_T$ region where a large statistical power is expected. We suspect that this discrepancy could be partly due to the different configuration for simulation of the aTGC signal and lepton selection efficiency and so on. As our estimation is conservative, we leave it as-is.}. 
For the $\delta g_{1,z}$ coupling, our analysis gives roughly comparable with the CMS one, $\delta g^{\rm CMS}_{1,z} = [-0.053,\, 0.061\, ]$~\cite{CMS:2017dmo}.
\begin{figure}[tp]
\begin{center}
\includegraphics[width=0.40\textwidth]{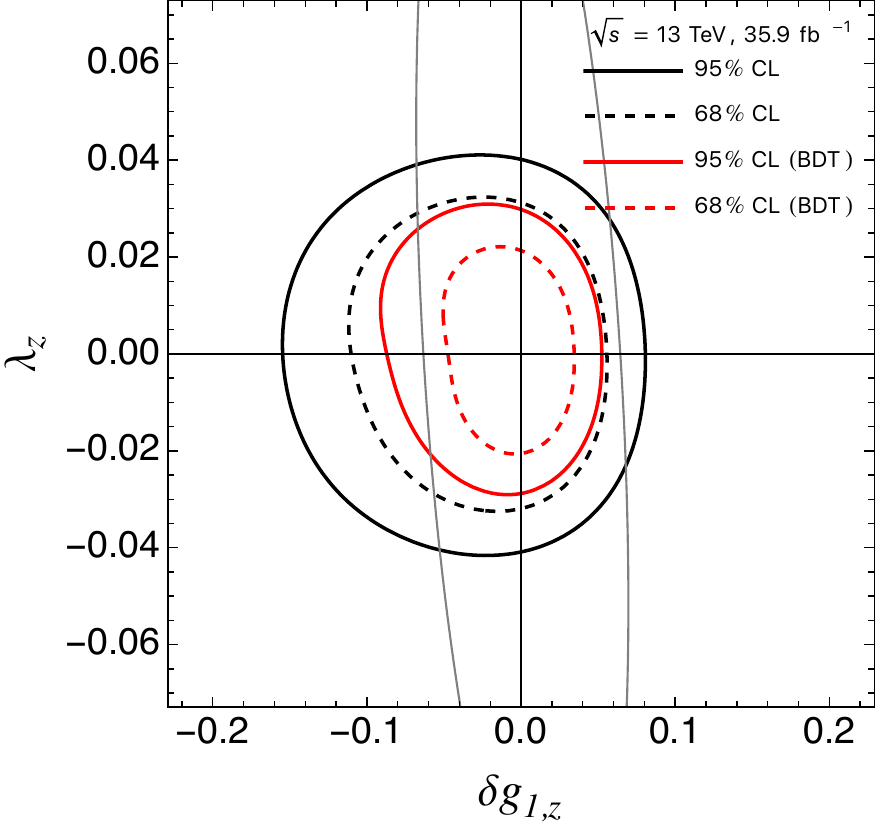}\quad
\includegraphics[width=0.40\textwidth]{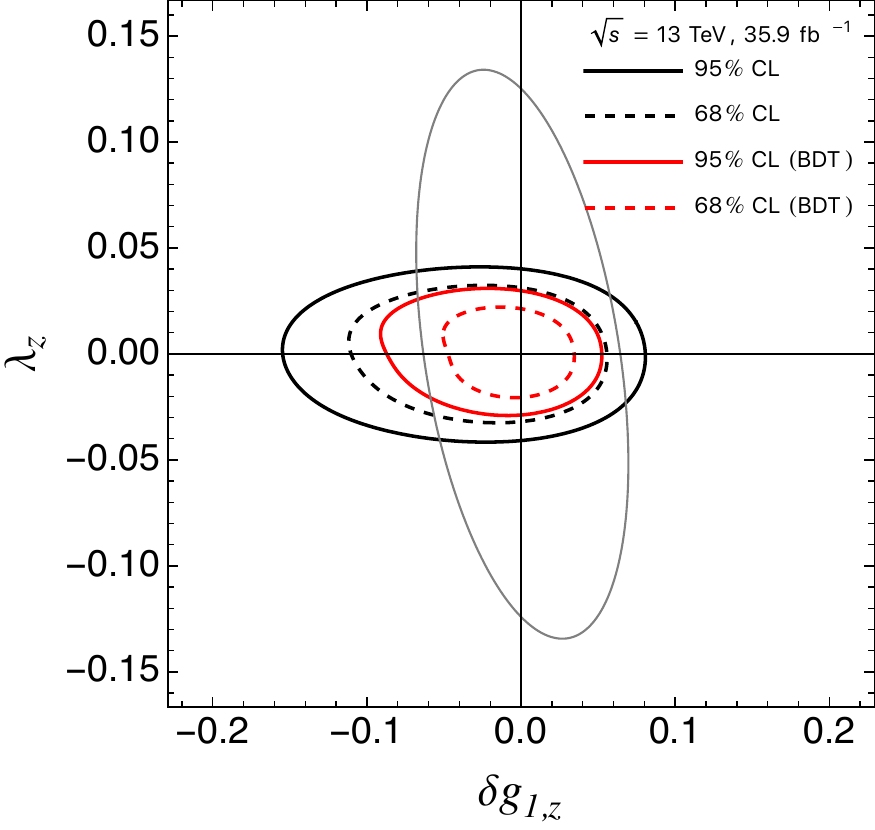}
\\[10pt]
\includegraphics[width=0.39\textwidth]{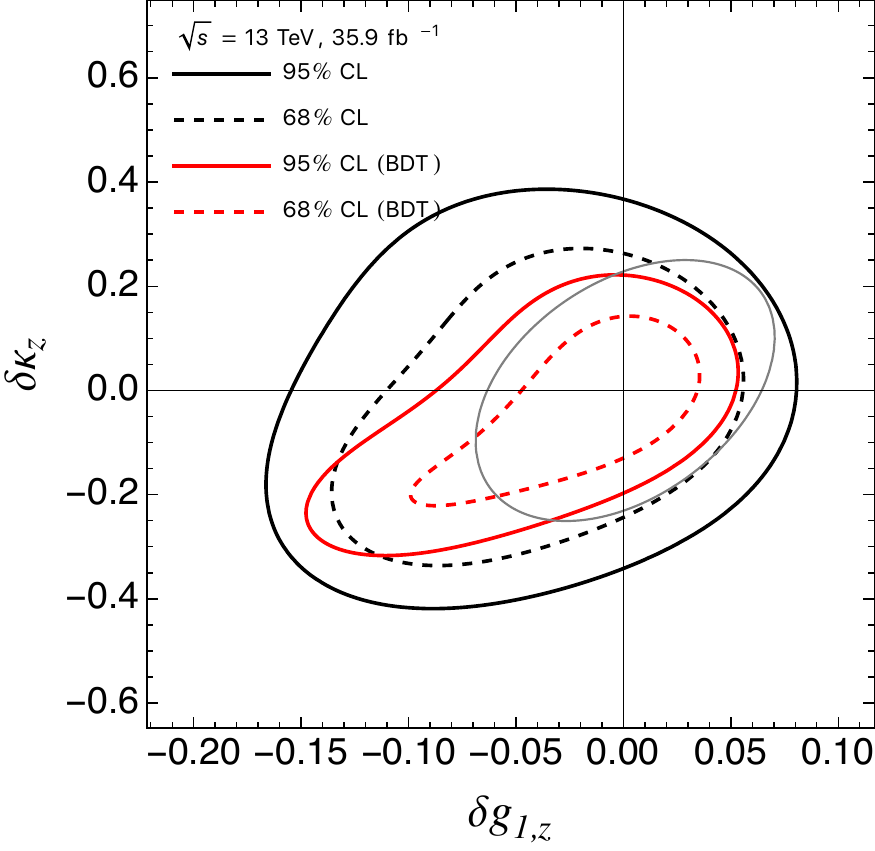}\quad
\includegraphics[width=0.395\textwidth]{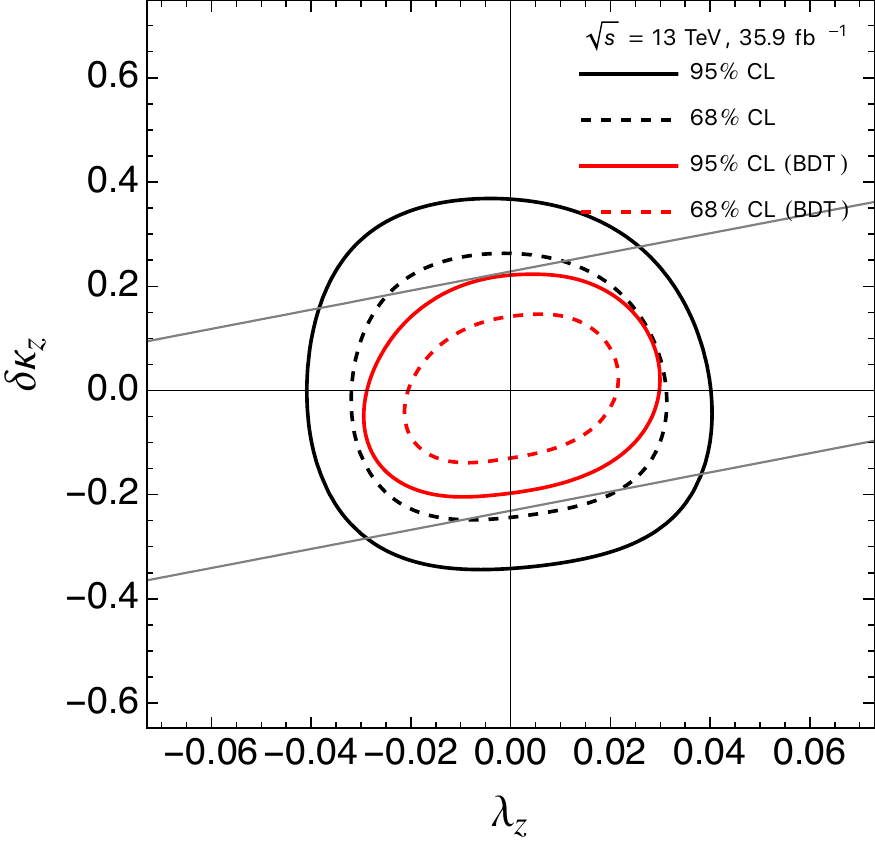}
\caption{\small Two-dimensional limits on aTGCs at 68\% (dashed) and 95\% CL (solid) regions obtained using the binned analysis of $p_T(\ell\ell)$ in the $Z$ mass window, assuming the integrated luminosity of 35.9 fb$^{-1}$ at $\sqrt{s} = 13$ TeV. Compared to the red solid lines, thin gray lines were obtained only with the interference term which is linear in the aTGC coupling for the BDT $> 0.6$.}
\label{fig:template2D:ptll:cms}
\end{center}
\end{figure}
The two-dimensional exclusion regions from the binned analysis of $p_T(\ell\ell)$ in the $Z$ mass window are illustrated in Fig.~\ref{fig:template2D:ptll:cms} where the remaining coupling is set to zero without the marginalization. The gray lines in Fig.~\ref{fig:template2D:ptll:cms} illustrate the exclusion region at 95\% CL using only linear terms in aTGCs in our parametrization of the cross section (see Eq.~(\ref{eq:EFT:xsec})). It indicates that the sensitivity of $\lambda_z$ is dominantly driven by the quadratic term whereas the effect of the quadratic term is less pronounced for two other aTGC couplings.

%
\begin{figure}[tp]
\begin{center}
\includegraphics[width=0.40\textwidth]{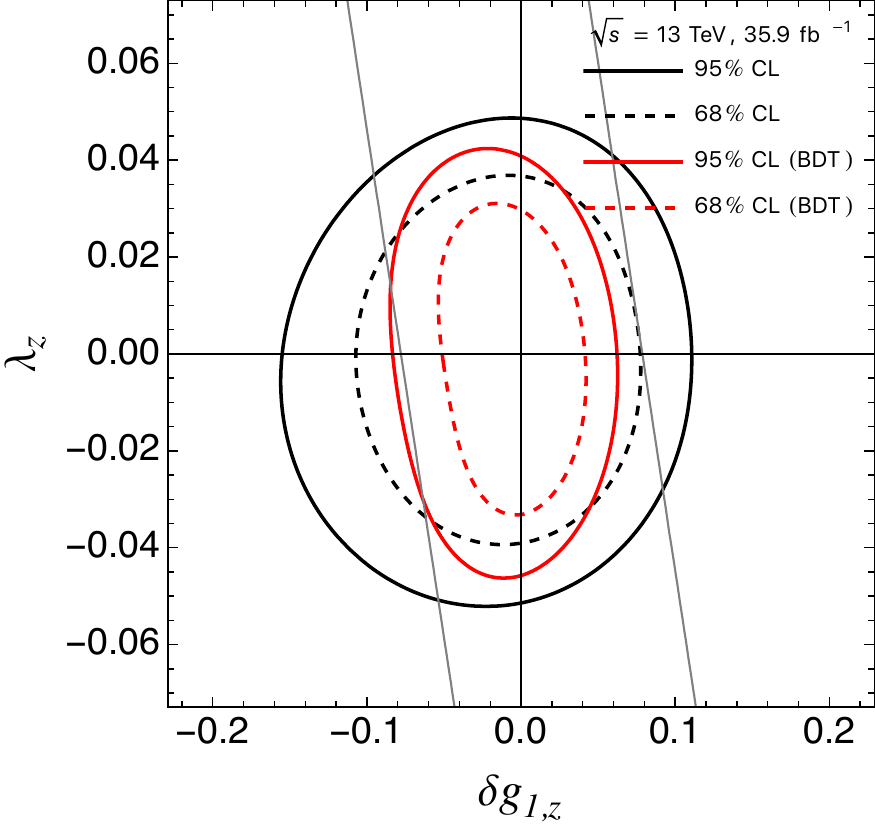}\quad
\includegraphics[width=0.376\textwidth]{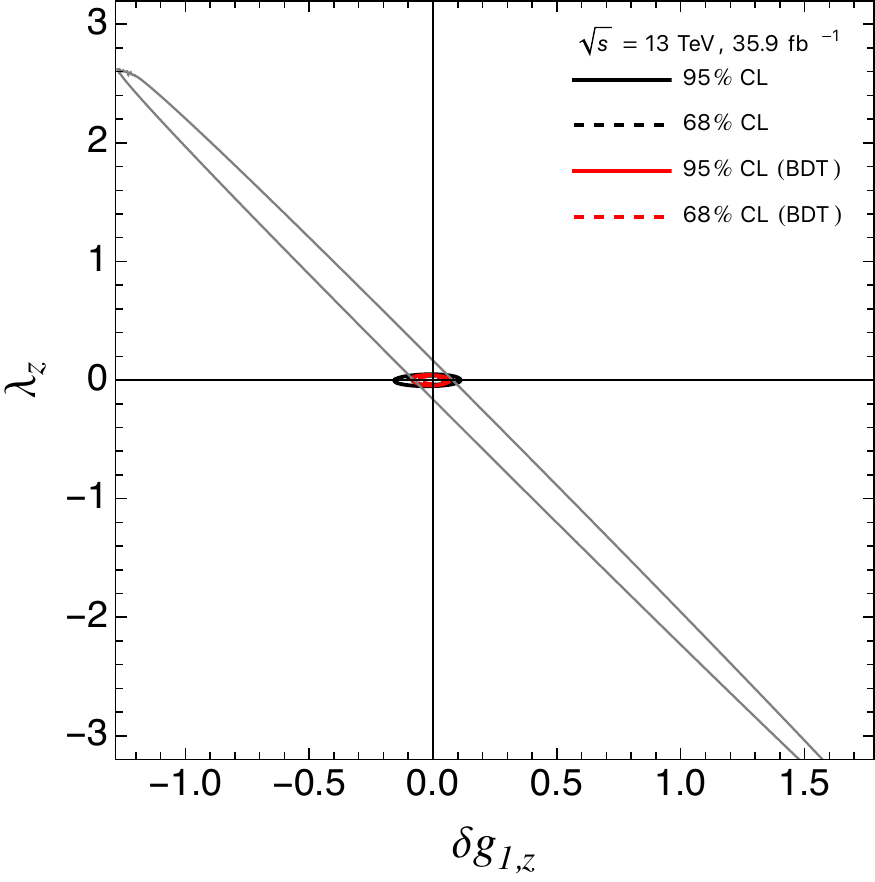}
\\[10pt]
\includegraphics[width=0.39\textwidth]{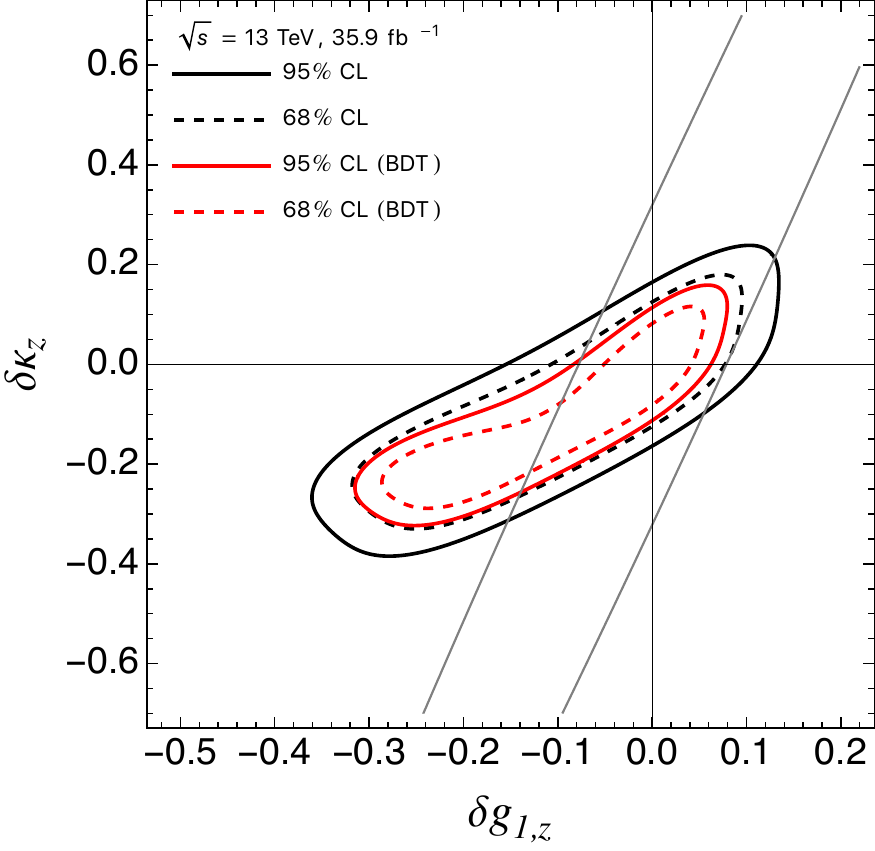}\quad
\includegraphics[width=0.395\textwidth]{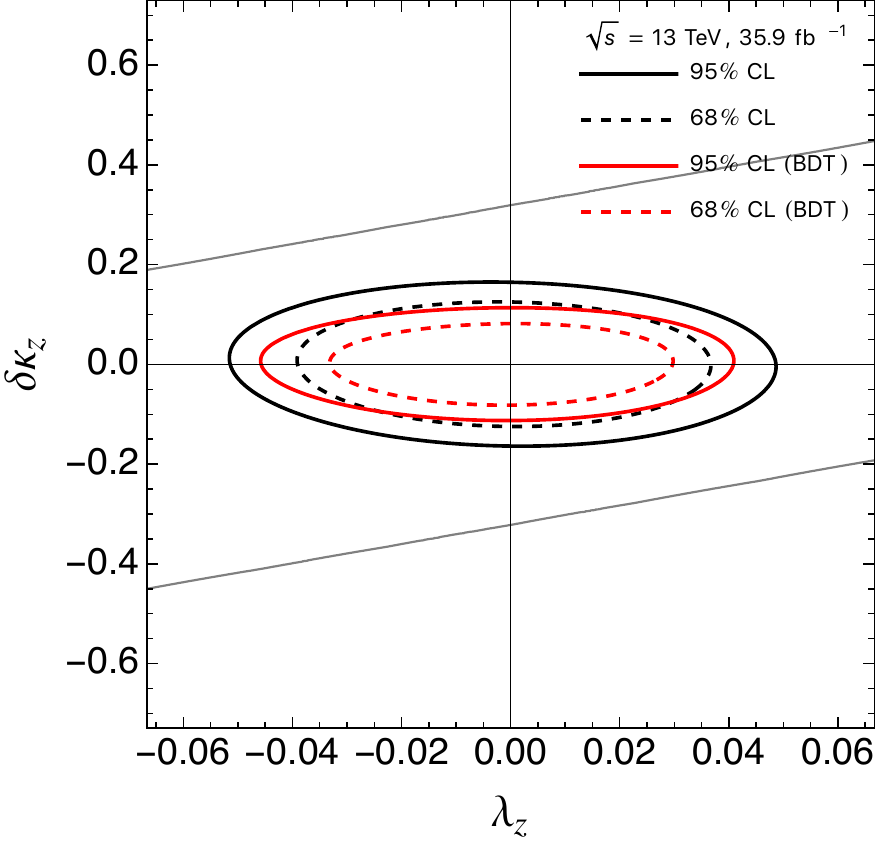}
\caption{\small Two-dimensional limits on aTGCs at 68\% (dashed) and 95\% CL (solid) regions obtained using the binned analysis of $m_{\ell\ell}$, assuming the integrated luminosity of 35.9 fb$^{-1}$ at $\sqrt{s} = 13$ TeV. Compared to the red solid lines, thin gray lines were obtained only with the interference term which is linear in the aTGC coupling for the BDT $> 0.6$. No cuts on VBFhardness was imposed.}
\label{fig:template:mll:noVBFhardness}
\end{center}
\end{figure}
%

We newly derive the sensitivity using the binned analysis of $m_{\ell\ell}$. As discussed in Section~\ref{sec:int:resurrection}, the invariant mass of the dilepton system has the relation $m^2_{\ell\ell} - m^2_{jj}= (2z -1) \hat{s}$, where $m_{jj}$ is the invariant mass of two forward jets, $z$ is the fraction of the total energy of the partonic system carried by the $\ell\ell$ system, and $m_{\ell\ell}$ alone does not guarantee the hardness of the $\ell\ell$ subsystem. However, while a nominal cut on the VBFhardness (see Eq.~(\ref{eq:dilep:pTcut}) for the definition) ensures that at least some amount of the total energy goes into the $\ell\ell$ subsystem and greatly helps recovering the interference, as is clearly seen in Fig.~\ref{fig:int:dilep:mll:demo}, it may not improve the situation for the case where the sensitivity is dominantly driven by the quadratic terms.
For this reason, we have not exploited VBFhardness. The 68\% and 95\% CL intervals of an individual aTGC are presented in Table~\ref{tab:template:mll}. 
From the comparison between Tables~\ref{tab:template:ptll} and~\ref{tab:template:mll}, we observe that $\delta \kappa_z$ is better constrained by the binned analysis of $m_{\ell\ell}$ whereas $\lambda_z$ and $\delta g_{1,z}$ are better constrained by the analysis using the distribution of $p_T(\ell\ell)$.

\begin{figure}[tp]
\begin{center}
\includegraphics[width=0.40\textwidth]{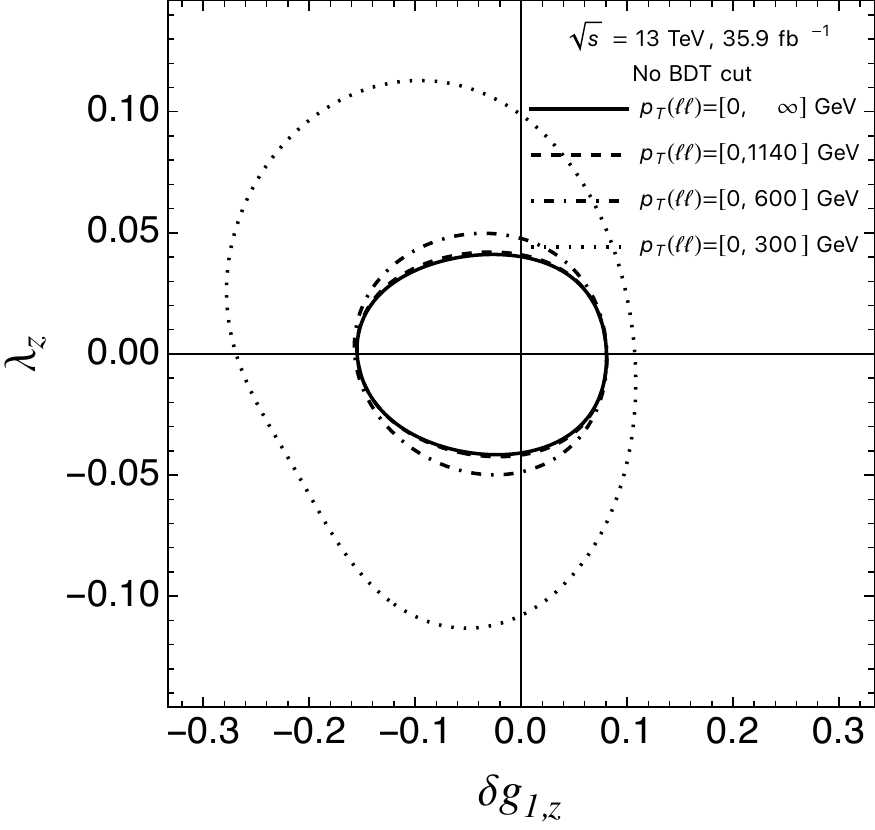}\quad
\includegraphics[width=0.40\textwidth]{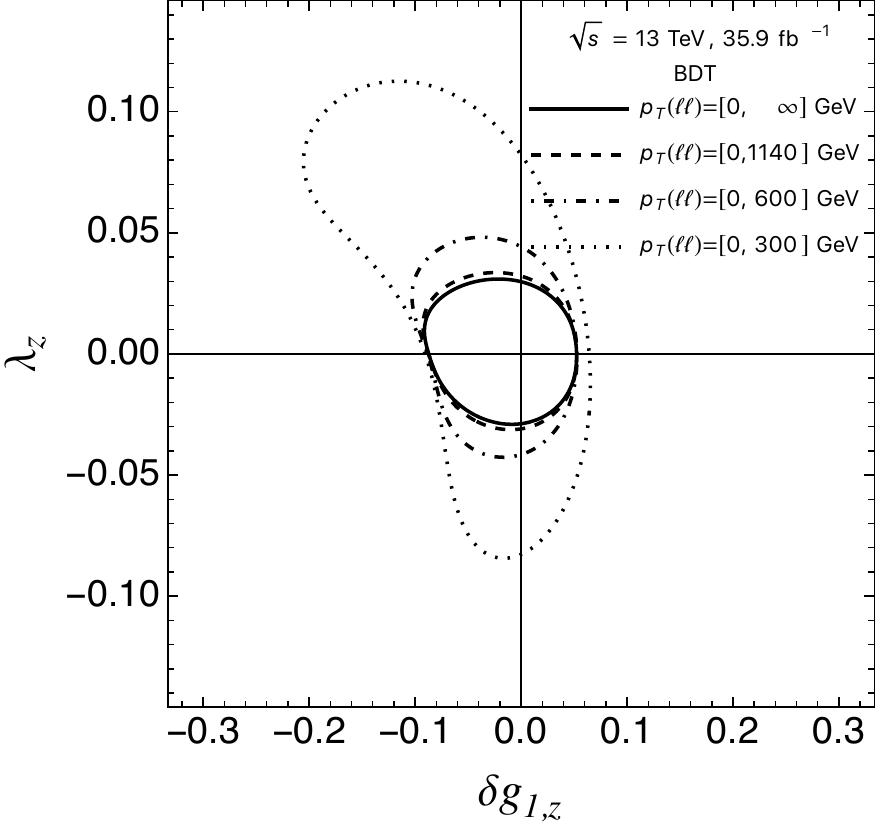}
\\[5pt]
\includegraphics[width=0.40\textwidth]{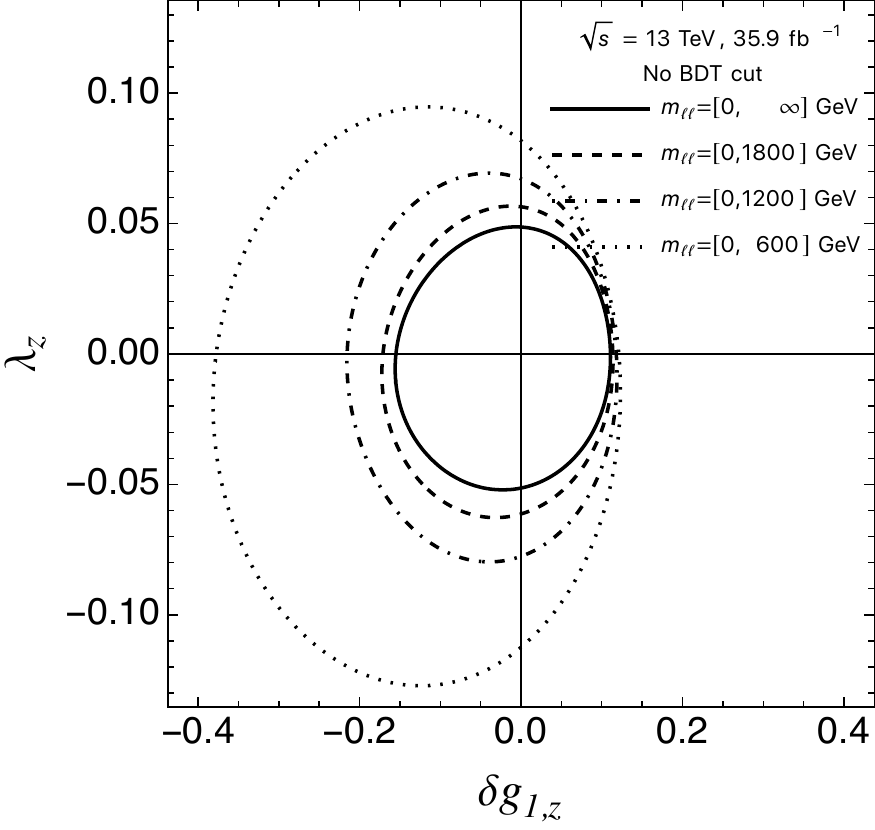}\quad
\includegraphics[width=0.40\textwidth]{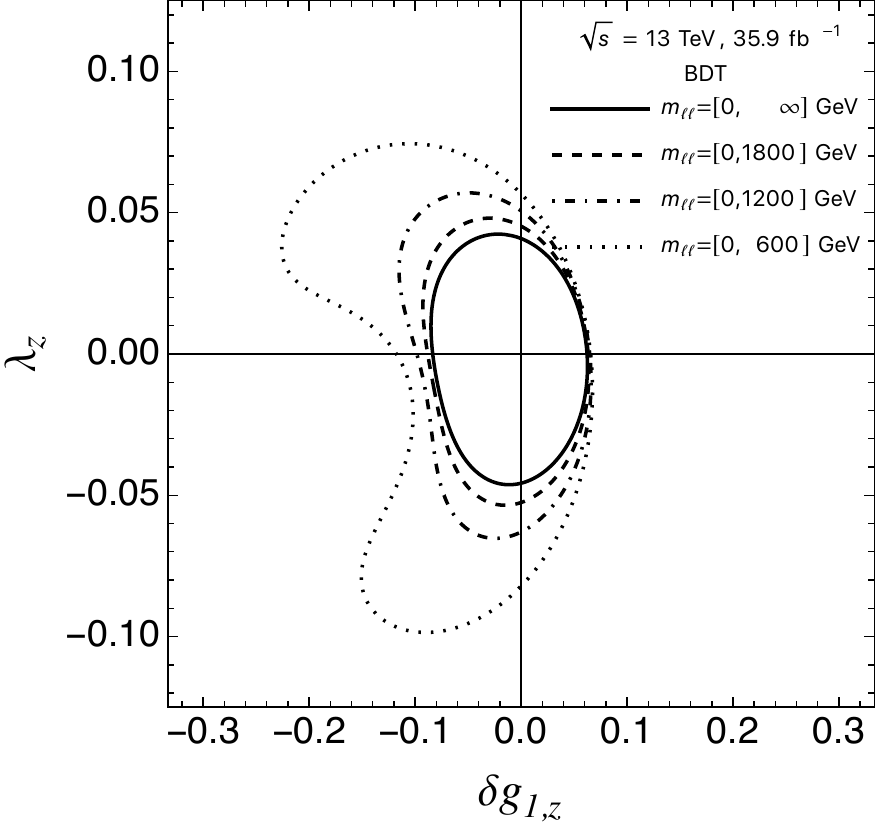}
\caption{\small Breakdown of $p_T(\ell\ell)$ (top) and $m_{\ell\ell}$ (bottom) categories in the plane $(\lambda_z, \, \delta g_{1,z})$, assuming the integrated luminosity of 35.9 fb$^{-1}$ at $\sqrt{s} = 13$ TeV. Curves of various styles indicate the 95\% CL contours.}
\label{fig:ptll:2d:c1c2:nBins:BDT06}
\end{center}
\end{figure}

The two-dimensional exclusion regions from the binned analysis of $m_{\ell\ell}$ are illustrated in Fig.~\ref{fig:template:mll:noVBFhardness} where similarly the remaining coupling was set to zero without the marginalization. Unlike the case using $p_T(\ell\ell)$ in Fig.~\ref{fig:template2D:ptll:cms}, the sensitivity, for instance, of $\lambda_z$ is significantly weakened (see upper right panel of Fig.~\ref{fig:template:mll:noVBFhardness}) when the quadratic term is removed. This is due to the interference suppression as illustrated by the black dashed line in the right panel of Fig.~\ref{fig:int:dilep:mll:demo}.
The situation is contrasted to those obtained using the binned analysis with $p_T(\ell\ell)$. As observed in the bottom panel of Fig.~\ref{fig:int:dilep:mll:signal}, the discrepancy between the interference and quadratic terms in the $p_T(\ell\ell)$ distribution is less pronounced, compared to the current case, and the interference term itself also shows the $p_T$-growing behavior.

Fig.~\ref{fig:ptll:2d:c1c2:nBins:BDT06} illustrates how the sensitivity in the plane  $(\lambda_z, \, \delta g_{1,z})$ changes as some of the higher bins are removed in the binned analysis of $p_T(\ell\ell)$ and $m_{\ell\ell}$, respectively, for two cases without (left panels of Fig.~\ref{fig:ptll:2d:c1c2:nBins:BDT06}) and with the BDT cut (right panels of Fig.~\ref{fig:ptll:2d:c1c2:nBins:BDT06}). This practice is meaningful especially for $m_{\ell\ell}$ as the EFT cutoff can be directly imposed on the $m_{\ell\ell}$ variable.
For the case with the BDT cut, sensitivity to $\delta g_{1,z}$ mostly comes from the first small number of bins, corresponding to the well below sub-TeV in both $p_T(\ell\ell)$ and $m_{\ell\ell}$ whereas a wider range of the energy contributes to the sensitivity to $\lambda_z$. On the contrary, for the case without the BDT cut, $\delta g_{1,z}$ becomes sensitive to the wide range of the energy.  


\begin{table}[tb]
\centering
  \renewcommand{\arraystretch}{1.1}
      \addtolength{\tabcolsep}{0.2pt} 
\scalebox{0.81}{
\begin{tabular}{c|cc|c|cc|c} 
   \multicolumn{7}{l}{13 TeV, $\mathcal{L} =$ 300 fb$^{-1}$} \\
   \multicolumn{7}{l}{Using the template analysis of $p_T(\ell\ell)$ in the $Z$ mass} \\
\hline
              &\multicolumn{3}{c|}{No BDT cut}								& \multicolumn{3}{c}{BDT $>$ 0.6} \\
aTGC  & \quad $68\%$ CL  \quad  & \quad $95\%$ CL  \quad & \quad $95\%$ CL (Linear) \quad & \quad $68\%$ CL \quad  & \quad $95\%$ CL  \quad  & \quad $95\%$ CL (Linear) \quad \\
\hline \hline
$\lambda_z$			& $[-0.017,\, 0.017]$	& $[-0.025,\, 0.024]$	&	$[-0.070,\, 0.070]$	& $[-0.0076,\, 0.0081]$		& $[-0.012,\, 0.012]$	&$[-0.035,\, 0.035]$	\\[3pt]
$\delta g_{1,z}$			& $[-0.019,\, 0.016]$	& $[-0.042,\, 0.029]$	&	$[-0.033,\, 0.033]$	& $[-0.0093,\, 0.0087]$	& $[-0.019,\, 0.017]$	&$[-0.018,\, 0.018]$	\\[3pt]
$\delta \kappa_z$		& $[-0.069,\, 0.072]$	& $[-0.13,\, 0.14]$	&	$[-0.14,\, 0.14]$		& $[-0.032,\, 0.033]$		& $[-0.062,\, 0.065]$	&$[-0.064,\, 0.064]$	\\[3pt]
\hline
   \multicolumn{7}{l}{Using the template analysis of $m_{\ell\ell}$} \\
\hline
$\lambda_z$			& $[-0.017,\, 0.016]$	& $[-0.025,\, 0.023]$	&$[-0.075,\, 0.075]$	& $[-0.013,\, 0.012]$	& $[-0.022,\, 0.018]$	&$[-0.045,\, 0.046]$	\\[3pt]
$\delta g_{1,z}$			& $[-0.025,\, 0.022]$	& $[-0.051,\, 0.040]$	&$[-0.047,\, 0.047]$	& $[-0.011,\, 0.011]$	& $[-0.023,\, 0.020]$	&$[-0.022,\, 0.022]$	\\[3pt]
$\delta \kappa_z$		& $[-0.054,\, 0.054]$	& $[-0.080,\, 0.080]$	&$[-0.19,\, 0.19]$	& $[-0.031,\, 0.030]$	& $[-0.049,\, 0.048]$	&$[-0.089,\, 0.089]$	\\[3pt]
\hline
%
%
   \multicolumn{7}{l}{} \\
   \multicolumn{7}{l}{13 TeV, $\mathcal{L} =$ 3000 fb$^{-1}$} \\
   \multicolumn{7}{l}{Using the template analysis of $p_T(\ell\ell)$ in the $Z$ mass} \\
\hline
$\lambda_z$			& $[-0.0077,\, 0.0072]$	& $[-0.011,\, 0.011]$	&$[-0.022,\, 0.022]$	& $[-0.0036,\, 0.0039]$	& $[-0.0056,\, 0.0060]$	&$[-0.011,\, 0.011]$	\\[3pt]
$\delta g_{1,z}$			& $[-0.0055,\, 0.0052]$	& $[-0.011,\, 0.010]$	&$[-0.011,\, 0.011]$	& $[-0.0029,\, 0.0028]$	& $[-0.0057,\, 0.0055]$	&$[-0.0057,\, 0.0057]$	\\[3pt]
$\delta \kappa_z$		& $[-0.023,\, 0.023]$		& $[-0.044,\, 0.045]$	&$[-0.045,\, 0.045]$	& $[-0.010,\, 0.010]$		& $[-0.020,\, 0.020]$		&$[-0.020,\, 0.020]$	\\[3pt]
\hline
   \multicolumn{7}{l}{Using the template analysis of $m_{\ell\ell}$} \\
\hline
$\lambda_z$			& $[-0.0090,\, 0.0077]$	& $[-0.013,\, 0.012]$	&$[-0.024,\, 0.024]$	& $[-0.0060,\, 0.0053]$	& $[-0.0096,\, 0.0085]$	&$[-0.014,\, 0.014]$	\\[3pt]
$\delta g_{1,z}$			& $[-0.0076,\, 0.0077]$	& $[-0.015,\, 0.014]$	&$[-0.015,\, 0.015]$	& $[-0.0035,\, 0.0034]$	& $[-0.0070,\, 0.0067]$	&$[-0.0069,\, 0.0069]$	\\[3pt]
$\delta \kappa_z$		& $[-0.025,\, 0.025]$		& $[-0.040,\, 0.040]$	&$[-0.062,\, 0.062]$	& $[-0.013,\, 0.013]$		& $[-0.022,\, 0.022]$		&$[-0.028,\, 0.028]$	\\[3pt]
\hline
\end{tabular}
}
\caption{\small One-dimensional limits on aTGCs at 68\% and 95\% CL at 13 TeV using the integrated luminosity of $\mathcal{L} =$ 300 fb$^{-1}$ and $\mathcal{L} =$ 3000 fb$^{-1}$. No cut on VBFhardness was imposed.}
\label{tab:projection:LHC}
\end{table}

\begin{figure}[tph]
\begin{center}
\includegraphics[width=0.328\textwidth]{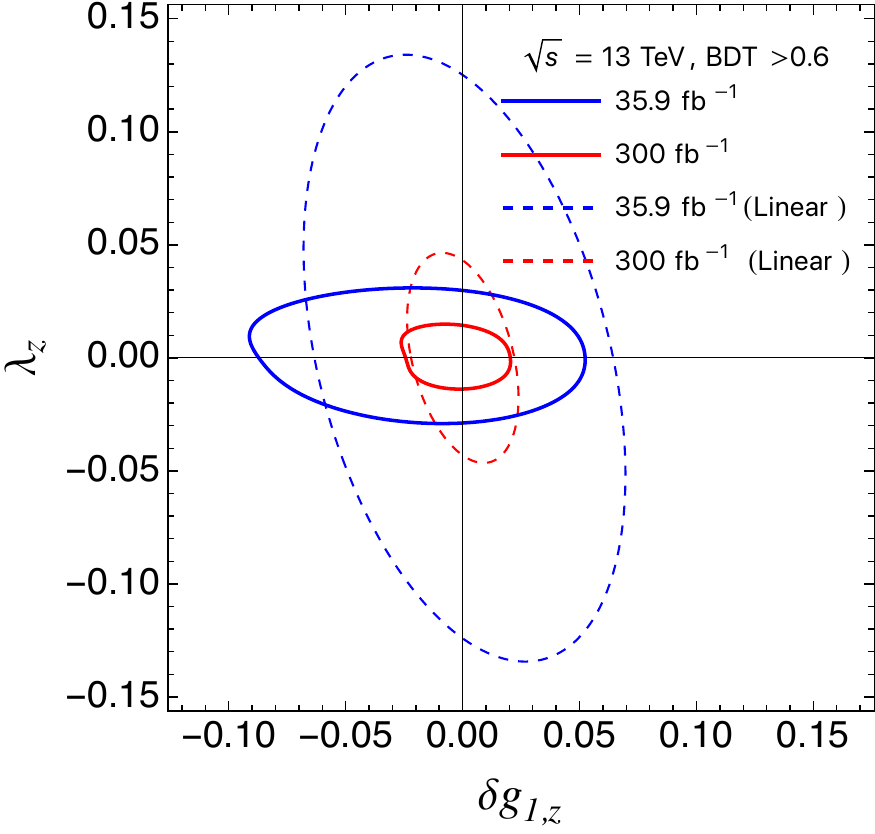}
\includegraphics[width=0.32\textwidth]{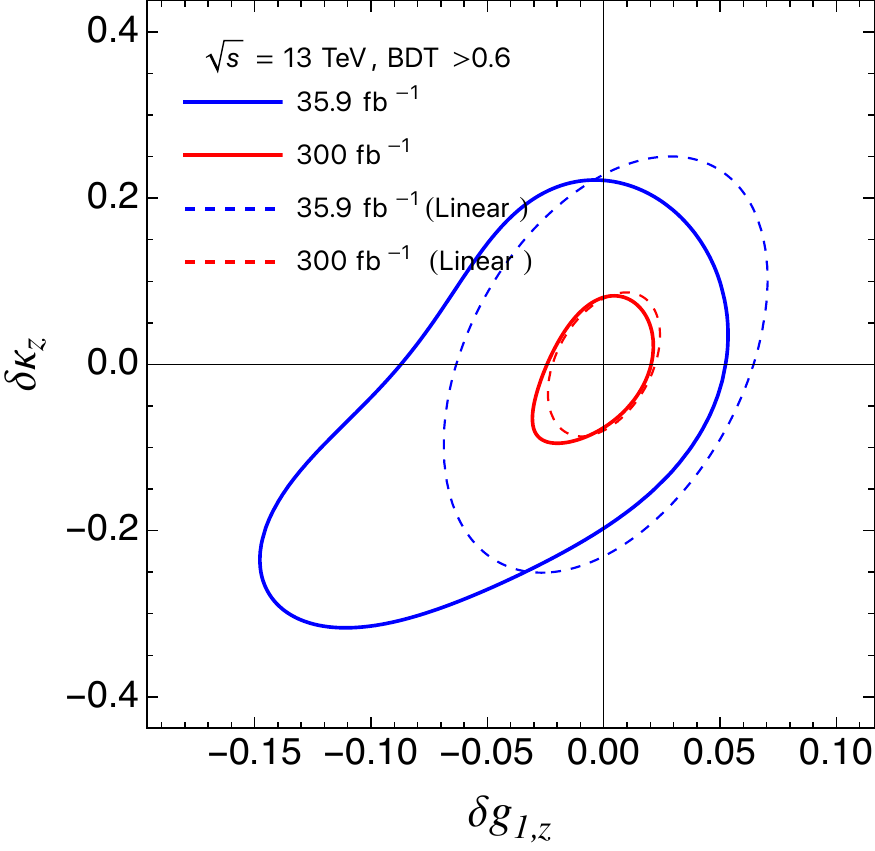}
\includegraphics[width=0.324\textwidth]{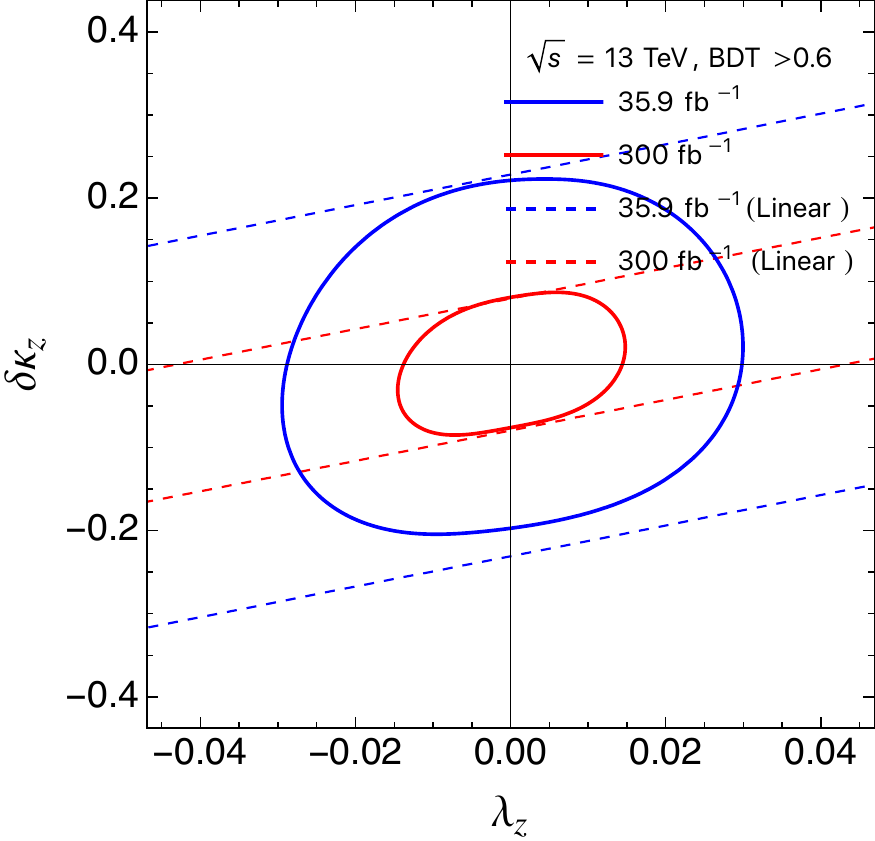}
\\[5pt]
\includegraphics[width=0.328\textwidth]{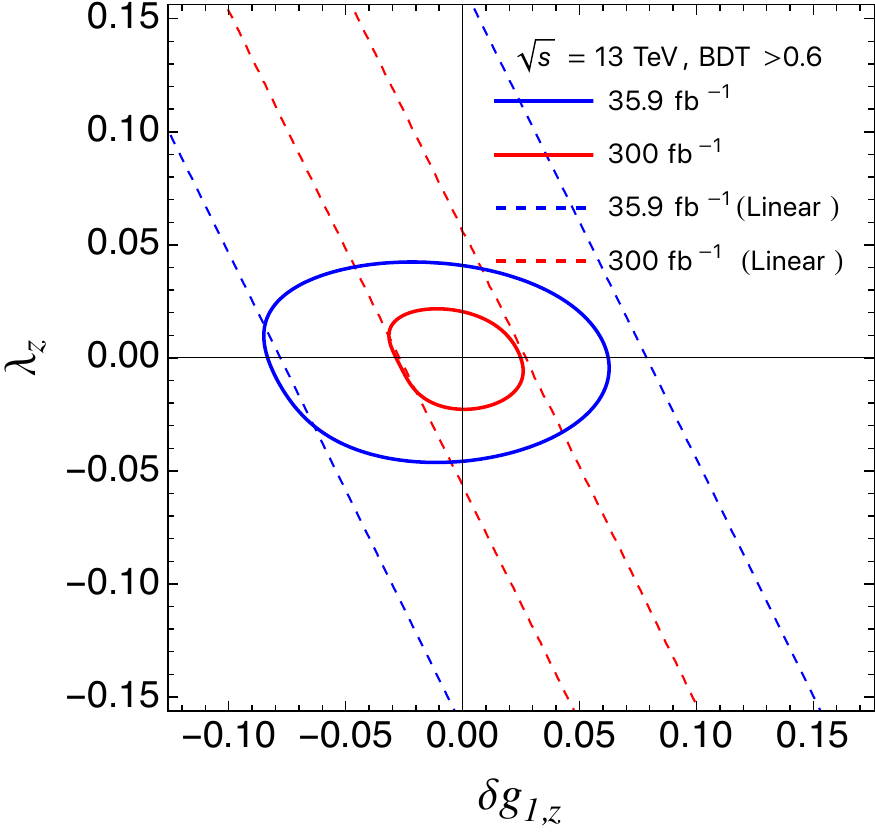}
\includegraphics[width=0.32\textwidth]{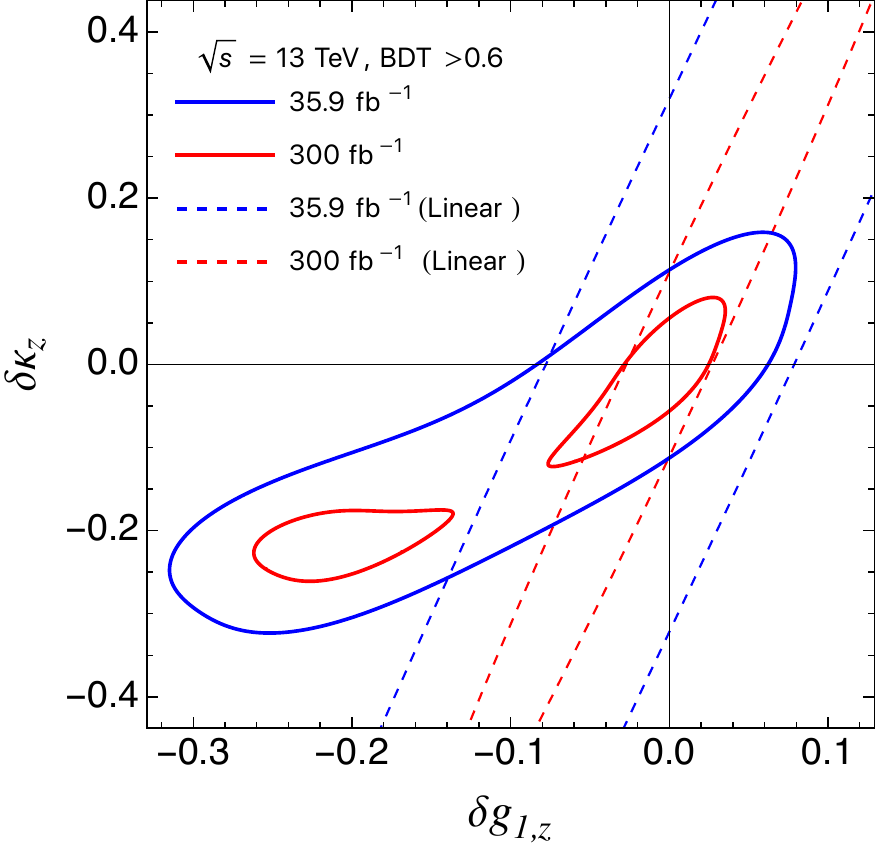}
\includegraphics[width=0.324\textwidth]{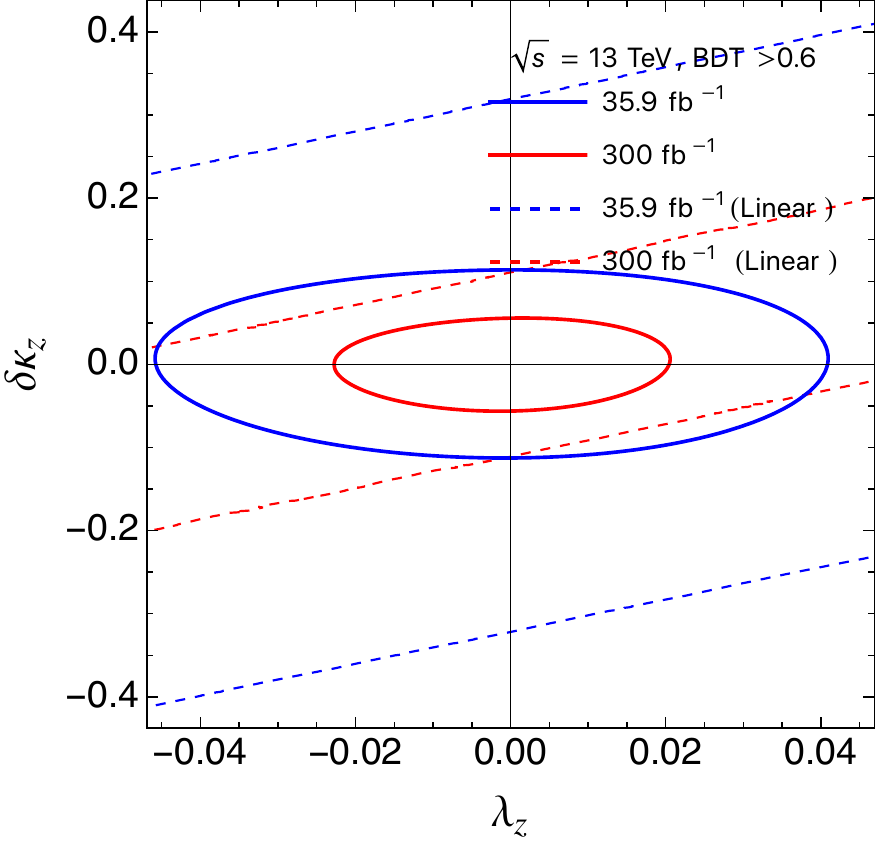}
\caption{\small The two-dimensional contours at 95\% CL, obtained using the binned analysis of $p_T(\ell\ell)$ (upper) and $m_{\ell\ell}$ (bottom), assuming the integrated luminosities of 35.9 fb$^{-1}$ and 300 fb$^{-1}$ at $\sqrt{s} = 13$ TeV. The dashed lines were obtained only with the interference term which is linear in the aTGC coupling.}
\label{fig:projection:LHC}
\end{center}
\end{figure}
%

We derive the sensitivity at the LHC and HL-LHC, assuming an integrated luminosity of 300 fb$^{-1}$ and 3 ab$^{-1}$, respectively. We assume that the systematic errors remain to be negligible, and we include only the statistical uncertainty. Our projection for the LHC and the HL-LHC is illustrated in Table~\ref{tab:projection:LHC}. The 95\% CL contours in the two-dimensional plane are shown in Fig.~\ref{fig:projection:LHC} where upper two plots were obtained by the template analysis of $p_T(\ell\ell)$ and the bottom ones using $m_{\ell\ell}$. The comparison between two analyses for $\delta g_{1,z}$ and $\delta\kappa_z$, namely, one by total cross section up to the quadratic order in aTGC and the other only with the interference cross section, indicates that the sensitivity is mainly driven by the linear term for the case of $p_T(\ell\ell)$. While, for the case of $m_{\ell\ell}$, the role of the interference hardly becomes important except for $\delta g_{1,z}$ where the other two couplings were set to zero, the VBFhardness may help making the interference more important. Although, as is evident in Fig.~\ref{fig:vbfhardness} a cut on VBFhardness may reduce the signal rate, loosening other cuts may compensate it and it can be an important variable at the HL-LHC regarding the interference.

\subsection{Sensitivity to EFT operators}

In this section, we also derive the sensitivity to dimension-6 operators from the EW $\ell\ell$+jets process. 
For a clear comparison, we adopt the same basis as in~\cite{CMS:2017dmo}, namely HISZ basis~\cite{Hagiwara:1993ck}.  Three operators that contributes to aTGCs are  given by
\begin{equation}
 \frac{C_{WWW}}{\Lambda^2} {\rm tr} ( \hat{W}_{\mu\nu} \hat{W}_{\nu\rho} \hat{W}_{\rho\mu} )~,\quad
 \frac{C_W}{\Lambda^2} (D_\mu H)^\dagger \hat{W}^{\mu\nu}(D_\nu H)~,\quad
 \frac{C_B}{\Lambda^2} (D_\mu H)^\dagger \hat{B}_{\mu\nu}(D_\nu H)~,
\end{equation}
where $\hat{W}_{\mu\nu} = W_{\mu\nu}^a \frac{\sigma^a}{2} g$ and $\hat{B}_{\mu\nu} = B_{\mu\nu} \frac{1}{2} g'$ were defined in~\cite{Hagiwara:1993ck}.
They contribute to aTGCs via the relations~\cite{Degrande:2012wf},
\begin{equation}
\begin{split}
  \lambda_z = \lambda_\gamma = C_{WWW}\frac{3g^2 m_W^2}{2 \Lambda^2}~,\quad
  \delta g_{1,z} = \left ( C_W + C_B \right ) \frac{m_W^2}{2 \Lambda^2}~,\quad
  \delta \kappa_\gamma = C_W \frac{m_W^2}{2 \Lambda^2} ~,
\end{split}
\end{equation}
from which we derive the sensitivity to the EFT operators. They are summarized in Table~\ref{tab:36ifb:eftop} for the integrated luminosity of 35.9 fb$^{-1}$.
The sensitivity on the ${\rm tr} W_{\mu\nu}^3$ operator has been measured by the recent CMS $W\gamma$ analysis~\cite{CMS:2021cxr}, using the differential distributions of the azimuthal angle and transverse momentum of the photon. The corresponding operator in~\cite{CMS:2021cxr}  in the Warsaw basis~\cite{Grzadkowski:2010es} was defined as 
\begin{equation}
C_{3W} \epsilon_{ijk}W_{\mu\nu}^i W_{\nu\rho}^j W_{\rho\mu}^k~,
\end{equation}
which connects to $C_{WWW}$ via 
$C_{3W} = \frac{g^3}{4} \frac{C_{WWW}}{\Lambda^2} \sim 0.07 \times \frac{C_{WWW}}{\Lambda^2}$.
We can translate the observed sensitivity on $C_{WWW}/\Lambda^2 = [-2.6,\, 2.6]$ (TeV$^{-2}$) at 95\% CL from the CMS $\ell\ell +$ jets analysis~\cite{CMS:2017dmo}, using the integrated luminosity of 35.9 fb$^{-1}$, in terms of $C_{3W}$, importantly taking into account roughly four times more data of 139 fb$^{-1}$:~\footnote{$1/\sqrt{1.97} \sim 1/\sqrt{2} $ is multiplied since roughly four times more luminosity is equivalent to increasing the signal by the factor of 2 which translates to the improvement of $C_{3W}$ by the factor of $\sqrt{2}$, assuming that the cross section is dominated purely by the quadratic term.}
\begin{equation}\label{eq:CMS:ll:recast}
   C_{3W,~\text{assuming}\, 139 \text{fb}^{-1}}^{\text{translated from EW}\, \ell\ell+jets} \sim  0.07\times [-2.6,\, 2.6]\times \frac{1}{\sqrt{1.97}}  = [-0.13,\, 0.13]~.
\end{equation}
That is, the CMS result from EW $\ell\ell +$ jets, assuming 139 fb$^{-1}$ of data, looks roughly two times worse than the sensitivity from the CMS $W\gamma$, namely $C_{3W} = [-0.062,\, 0.052]$ (TeV$^{-2}$) at 95 \% CL~\cite{CMS:2021cxr}. 
Our analysis of the EW $\ell\ell +$ jets gives the three times worse result compared to the CMS $\ell\ell +$ jets analysis, or $C_{3W,~ 139 \text{fb}^{-1}}^{\text{translated from our EW}\, \ell\ell + jets} = [-0.303,\, 0.314]$ (TeV$^{-2}$), which is roughly six times worse than the CMS $W\gamma$ result. We think that, given the discrepancy between the CMS analysis~\cite{CMS:2017dmo} and our re-analysis (see footnote~\ref{fn:compare:cms} and~\ref{app:fn:compare:simul}), whether the EW $\ell\ell +$ jets process is practically relevant or not for the $\text{tr}(W_{\mu\nu}^3)$ operator, compared to the diboson process, remains inconclusive. The sensitivity on the same operator from the $WW,\, WZ$ processes, using the integrated luminosity of 137 fb$^{-1}$, in the ATLAS analysis~\cite{ATLAS:2021ohb} reports $C_{3W} = [-0.14,\, 0.15]$ (TeV$^{-2}$) at 95 \% CL which is similar to the CMS EW $\ell\ell +$ jets analysis~\cite{CMS:2017dmo} as is seen in Eq.~(\ref{eq:CMS:ll:recast}). The recent CMS $WZ$ analysis \cite{CMS:2021icx}, using the integrated luminosity of 137 fb$^{-1}$, reports the observed limits $C_{WWW}/\Lambda^2 = [-1.0,\, 1.2]$ (TeV$^{-2}$) at 95\% CL which translates to $C_{3W,~ 137 \text{fb}^{-1}}^{\text{translated from}\, WZ} \sim 0.07 \times [-1.0,\, 1.2] =  [-0.07,\, 0.084]$ (TeV$^{-2}$) which is consistent with the CMS $W\gamma$ analysis~\cite{CMS:2021cxr}.

\begin{table}[tb]
\centering
  \renewcommand{\arraystretch}{1.1}
      \addtolength{\tabcolsep}{0.2pt} 
\scalebox{0.81}{
\begin{tabular}{c|cc|c|cc|c} 
   \multicolumn{7}{l}{13 TeV, $\mathcal{L} =$ 35.9 fb$^{-1}$} \\
   \multicolumn{7}{l}{Using the template analysis of $p_T(\ell\ell)$ in the $Z$ mass} \\
\hline
aTGC  &\multicolumn{3}{c|}{No BDT cut}								& \multicolumn{3}{c}{BDT $>$ 0.6} \\
(TeV$^{-2}$)  & \quad $68\%$ CL  \quad  & \quad $95\%$ CL  \quad & \quad $95\%$ CL (Linear) \quad & \quad $68\%$ CL \quad  & \quad $95\%$ CL  \quad  & \quad $95\%$ CL (Linear) \quad \\
\hline \hline
$C_{WWW}/\Lambda^2$	& $[-6.35,\, 6.18]$	& $[-9.0,\, 8.8]$	& $[-49.7,\, 49.8]$	& $[-3.79,\, 3.95]$	& $[-6.08,\, 6.30]$	&$[-24.5,\, 24.7]$	\\[3pt]
$C_W/\Lambda^2$		& $[-27.3,\, 12.5]$	& $[-46.1,\, 21.2]$	& $[-31.4,\, 31.5]$	& $[-10.2,\, 7.68]$	& $[-25.8,\, 13.9]$	&$[-16.9,\, 17.1]$	\\[3pt]
$C_B/\Lambda^2$		& $[-30.8,\, 12.4]$	& $[-47.1,\, 21.0]$	& $[-31.7,\, 31.8]$	& $[-10.7,\, 7.71]$	& $[-30.1,\, 13.8]$	&$[-17.1,\, 17.4]$	\\[3pt]
\hline
   \multicolumn{7}{l}{Using the template analysis of $m_{\ell\ell}$} \\
\hline
$C_{WWW}/\Lambda^2$	& $[-7.65,\, 7.13]$	& $[-11.1,\, 10.5]$	& $[-53.4,\, 53.5]$	& $[-6.21,\, 5.58]$	& $[-9.71,\, 8.68]$	&$[-32.2,\, 32.5]$	\\[3pt]
$C_W/\Lambda^2$		& $[-40.2,\, 20.2]$	& $[-40.2,\, 34.7]$	& $[-50.4,\, 50.5]$	& $[-14.5,\, 10.6]$	& $[-90.8,\, 19.2]$	&$[-23.5,\, 23.7]$	\\[3pt]
$C_B/\Lambda^2$		& $[-32.1,\, 21.1]$	& $[-32.1,\, 35.5]$	& $[-55.1,\, 55.2]$	& $[-17.5,\, 11.3]$	& $[-88.8,\, 20.1]$	&$[-25.7,\, 25.9]$	\\[3pt]
\hline
\end{tabular}
}
\caption{\small One-dimensional limits on dimension-6 operators at 68\% and 95\% CL at 13 TeV using the integrated luminosity of $\mathcal{L} =$ 35.9 fb$^{-1}$. $C_X/\Lambda^2$ ($X=WWW,\, W,\, B$) in the TeV$^{-2}$. No cut on VBFhardness was imposed.}
\label{tab:36ifb:eftop}
\end{table}

The coupling $\lambda_z$ is probed only by the transverse modes in the diboson process and thus it is subject to the noninterference issue. Whereas the couplings $\delta g_{1,\, z}$ and $\delta \kappa_z$ are also probed by the longitudinal polarization of the gauge bosons (see~\cite{Falkowski:2016cxu}, for instance), and they are not necessarily subject to the same issue. 
The CMS $WZ$ analysis \cite{CMS:2021icx}, using 137 fb$^{-1}$ of data, obtained the observed limits, $C_W/\Lambda^2 = [-2.5,\, 0.3]$ (TeV$^{-2}$) and $C_B/\Lambda^2 = [-43,\, 113]$ (TeV$^{-2}$) at 95\% CL. The CMS EW $\ell \ell +$ jets analysis~\cite{CMS:2017dmo}, using 35.9 fb$^{-1}$ of data, obtained the observed limit $C_W/\Lambda^2 = [-8.4,\, 10.1]$ (TeV$^{-2}$) with no limit on $C_B/\Lambda^2$. Our analysis of EW $\ell\ell +$ jets, assuming 139 fb$^{-1}$ of data, leads to $C_W/\Lambda^2 = [-10.2,\, 7.69]$ (TeV$^{-2}$) and $C_B/\Lambda^2 = [-10.7,\, 7.72]$ (TeV$^{-2}$).

\section{Conclusion}

\begin{figure}[tp]
\begin{center}
\includegraphics[width=0.48\textwidth]{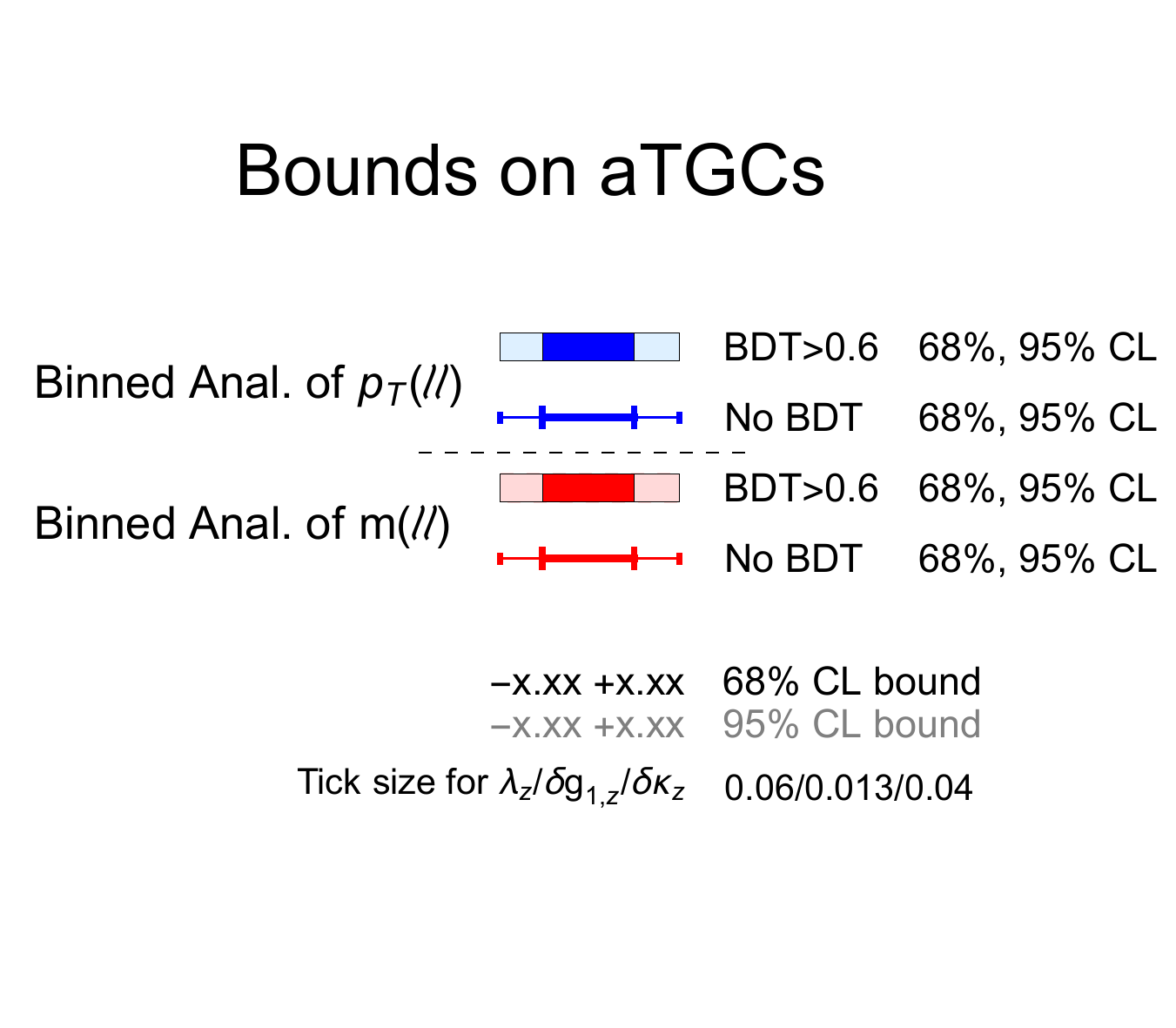}
\includegraphics[width=0.51\textwidth]{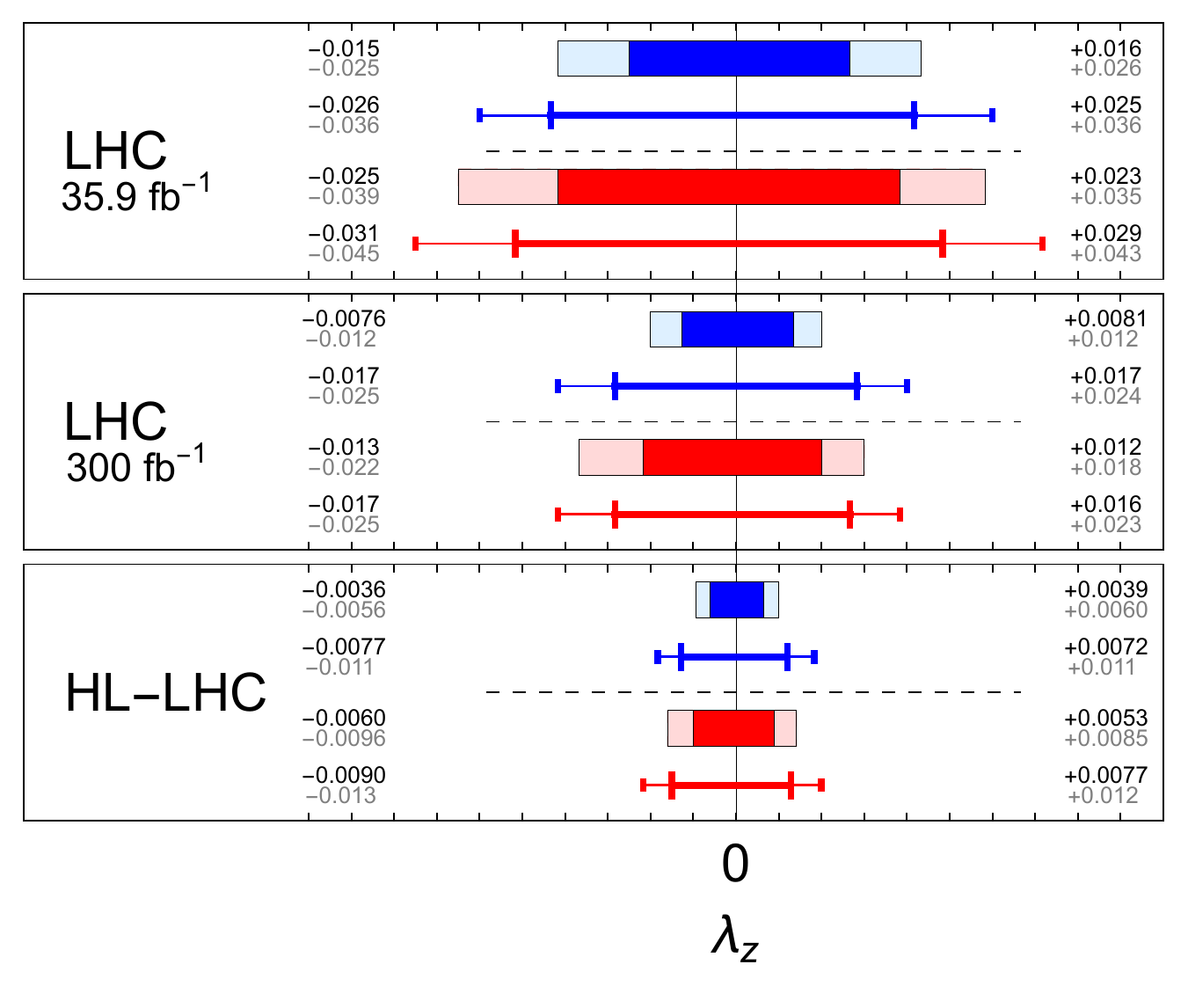}
\\
\includegraphics[width=0.48\textwidth]{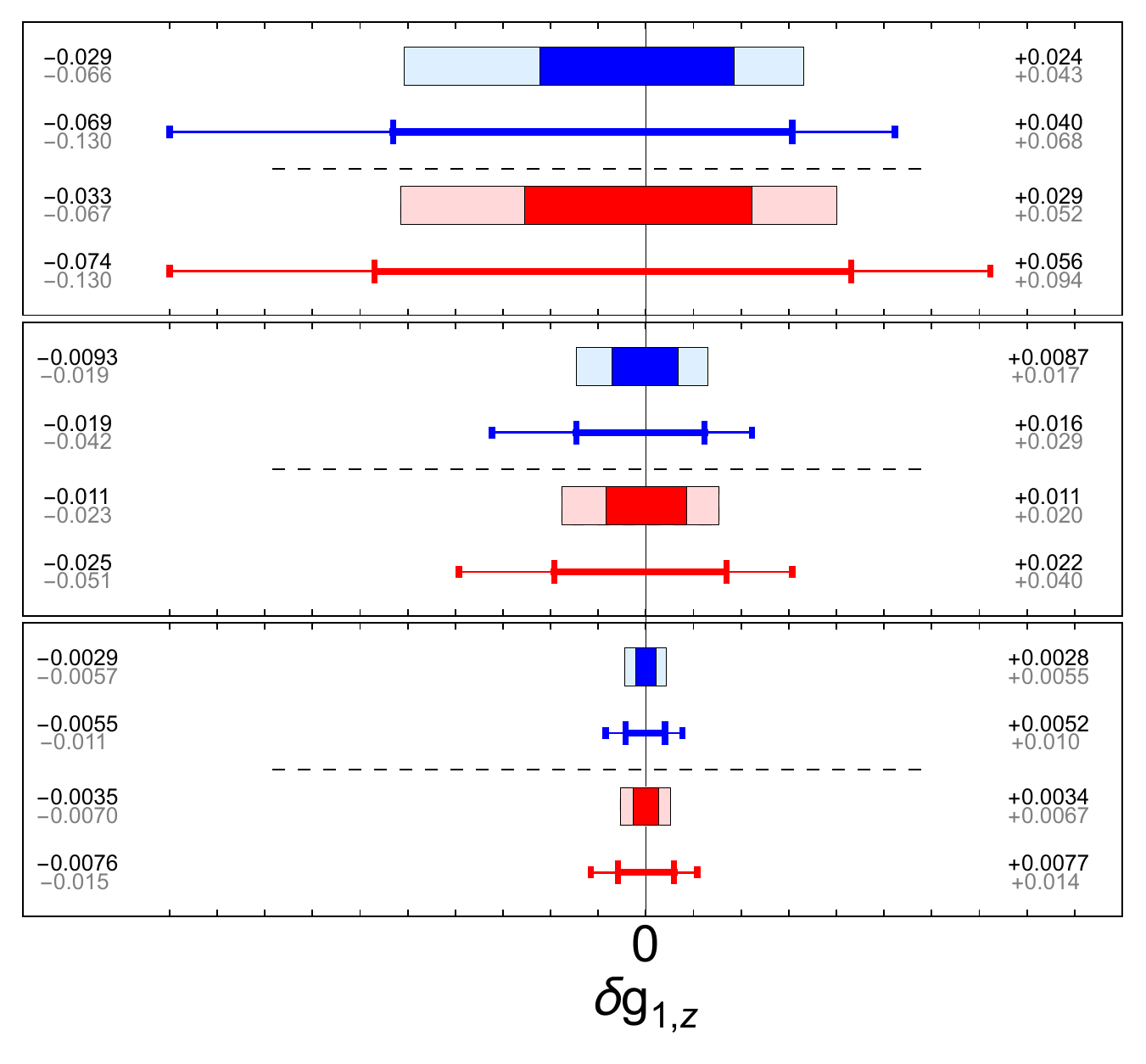}
\includegraphics[width=0.5108\textwidth]{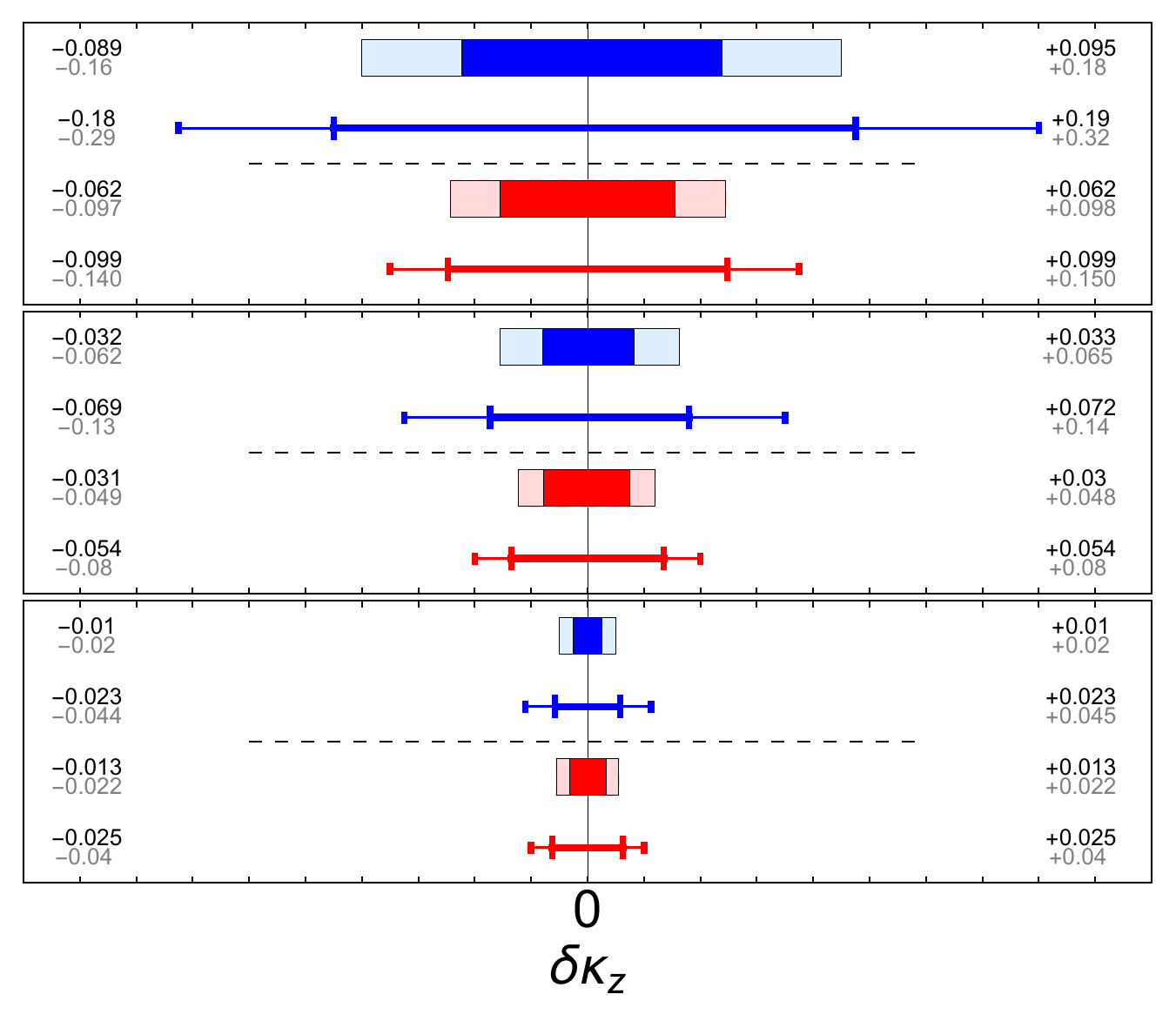}
\caption{\small The visual presentation of the sensitivity of aTGCs at 13 TeV, assuming three different luminosities, given in Tables~\ref{tab:template:ptll},~\ref{tab:template:mll}, and~\ref{tab:projection:LHC}.}
\label{fig:atgc1:chart}
\end{center}
\end{figure}

We have explored the EW dilepton production with two associated jets for the precision measurement of aTGC couplings. 
As was explicitly shown (both analytically and numerically) in this work, the full amplitude, including the forward quarks that radiate off the vector gauge bosons, exhibits the interference in the inclusive cross section. It reveals an intriguing feature regarding the interference between the SM and BSM amplitudes (that is subject to the helicity selection rule), or the sizable interference in the total cross section can arise from beyond the relevant regime for the EWA.
For the purpose of the interference resurrection in our dilepton production in vector boson fusion, we have introduced a new variable, VBFhardness, that can control the amount of energy flowing into the dilepton system. Using this variable, we have demonstrated that the interference clearly appears when an appropriate cut is applied. 
As a proof-of-concept example for the interference resurrection in the inclusive cross section, we have performed the analytic study using the simpler toy process, or $u\gamma \rightarrow d\nu e^+$, which was numerically confirmed as well. In the same toy process, we have newly identified that the sizable interference term in the total cross section arises beyond the relevant regime for the EWA which apparently looks negligible in the EWA limit.

We have derived the sensitivity to aTGCs for three scenarios of the LHC and HL-LHC, assuming the integrated luminosity of $35.9$ fb$^{-1}$, $300$ fb$^{-1}$, and $3000$ fb$^{-1}$. In addition to the template analysis using the transverse momentum of the dilepton, we also carried out the template analysis using the invariant mass of the dilepton in this work. While the bounds on $\lambda_z$ and $\delta g_{1,z}$ from the dilepton invariant mass are rather weaker than those from the transverse momentum of the dilepton system, the situation is opposite for $\delta \kappa_z$. The final result of the one-dimensional bounds at 68\% and 95\% CL is summarized in Fig.~\ref{fig:atgc1:chart}.
Our analysis using the dilepton invariant mass may further be optimized. Vetoing $b$-jets could help suppress top-enriched backgrounds. Exploiting VBFhardness may help in enhancing the role of the interference with respect to the quadratic terms in aTGCs.
Our results were compared with the existing limits from the CMS and ATLAS diboson processes in terms the EFT operators. While the sensitivity from the diboson process seems apparently stronger than the one from EW $\ell\ell +$ jets for the measurement of the $\text{tr}(W_{\mu\nu}^3)$ operator involving only the transverse polarizations, some other directions seem to be better constrained in our process.

\section*{Acknowledgments}
MS thanks A. Azatov and D. Marzocca for valuable discussions. Especially, MS thanks A. Azatov for the explanation of his previous work regarding the interference resurection. JP, MS, and MU were supported by National Research Foundation of Korea under Grant Number NRF-2021R1A2C1095430. JY and HH were supported by the National Research Foundation of Korea (NRF) grant funded by the Korea government (MSIT) (No. 2020R1C1C1005916).

\appendix

\section{Details on simulation}
\label{sec:app:sim:detail}

\subsection{Signal and background generation}
The aTGC interaction in Eq.~(\ref{eq:atgc}) is implemented in \textsc{\sc FeynRules}~\cite{Alloul:2013bka} from which we generate the UFO output for the \textsc{\sc MadGraph}. Electroweak $\ell^+\ell^- jj$ samples were simulated at leading order (LO) by \textsc{\sc MadGraph}5\_aMC$@$NLO v2.6.7~\cite{Alwall:2014hca} ({\tt QED=4, QCD=0}) with the default factorization and renormalization scales, interfaced with the \textsc{\sc Pythia8} v8.306 for the parton shower and hadronization.  For the parton distribution function, the NNPDF30 (lo\_as0130)~\cite{NNPDF:2014otw} is used. The linear (or interference) and quadratic terms in aTGC in our parametrization of the cross section in Eq.~(\ref{eq:EFT:xsec}) were separately simulated by using flags {\tt TGC$^2$ = 1} and {\tt TGC$^2$ = 2}, respectively~\footnote{\label{app:fn:compare:simul}On the other hand, the CMS analysis~\cite{CMS:2017dmo} generated aTGC signal samples (differently from ours) effectively over $5\times 5\times 5$ grid of $c_{WWW}/\Lambda^2 \times c_{W}/\Lambda^2 \times c_{B}/\Lambda^2$ which were equivalent to our aTGCs. We suspect that this could be partly responsible for the discrepancy between our sensitivity of aTGCs and that in~\cite{CMS:2017dmo}.}, where {\tt TGC} denotes the order of aTGC interaction. The phase space was restricted to those satisfying $m_{\ell\ell} >50$ GeV, $p_T(j)>25$ GeV, and $m_{jj}>120$ GeV at the generation level~\footnote{To guarantee enough statistics and the smoothness of the differential distribution in the high invariant mass tail, events were generated separately for multiple intervals of $m_{\ell\ell}$ and combined. Similarly for the EW $\ell\ell j j$ samples in the SM.}. 

All background samples were similarly simulated at leading order (LO) by \textsc{\sc MadGraph}5\_aMC$@$NLO v2.6.7~\cite{Alwall:2014hca} with the default factorization and renormalization scales, interfaced with the \textsc{\sc Pythia8}. The NNPDF30 (lo\_as0130) was used. The QCD Drell-Yan process $\gamma^*/Z(\ell^+ \ell^-)+$jets samples where jets arise from QCD interaction were matched using $k_T$-jet MLM matching at LO up to three extra jets in 5-flavor.  $k$-factor of 1.23 was applied~\cite{CMS:2017dmo}. 
The $t\bar{t}$ samples were matched using $k_T$-jet MLM matching ({\tt QCUT = 45} GeV) at LO up to two extra jets in 5-flavor and the total cross section was rescaled to match the NLO value from Powheg~\cite{Nason:2004rx} by applying the $k$-factor of 1.7.

\section{Computation detail of $qV \rightarrow q' \nu \ell$}
\label{app:sec:2to3}

\subsection{Choice of four momenta and amplitudes}

The polarization vectors of the photon are obtained by rotating $\epsilon_{L/R} = \frac{1}{\sqrt{2}} (0,\, 1,\, \pm i,\, 0 )$ (for the massless momenta moving to $-z$ axis) with angle $\theta$ about $y$-axis (similarly angle $\phi$ about $z$-axis). 
\begin{equation}
 \epsilon_{L/R}^\mu (p_2) = \frac{1}{\sqrt{2}} \left ( 0,\, \cos\theta \cos\phi \mp i \sin\phi, \, \cos\theta \sin\phi \pm i \cos\phi, \, -\sin\theta \right )~.
\end{equation}
The spinor solutions in our coordinate system are
\begin{equation}
\begin{split}
  \bar{u}_L(k_1) &= \hat{s}^{1/4} \left ( 0,\, 0,\,  - \sqrt{2z-1}\sin\frac{\psi}{2},\, \cos\frac{\psi}{2} \right )~,
  \\[3pt]
  v_L(k_2) &= \hat{s}^{1/4} \left ( \sqrt{2z-1}\cos\frac{\psi}{2},\, \sin\frac{\psi}{2},\, 0,\, 0 \right )^T ~,
  \\[3pt]
  u_L(p_1) &= \hat{s}^{1/4} \left ( -\sin\frac{\theta}{2},\, e^{i\phi} \cos\frac{\theta}{2},\, 0,\, 0 \right )^T~,
  \\[3pt]
  \bar{u}_L(k_3) &=  \hat{s}^{1/4} \sqrt{2(1-z)}  \left ( 0,\, 0,\, -1,\, 0 \right )~,
\end{split}
\end{equation}
where $T$ denotes the transpose.
%
%
\begin{figure}[tph]
\begin{center}
\includegraphics[width=0.45\textwidth]{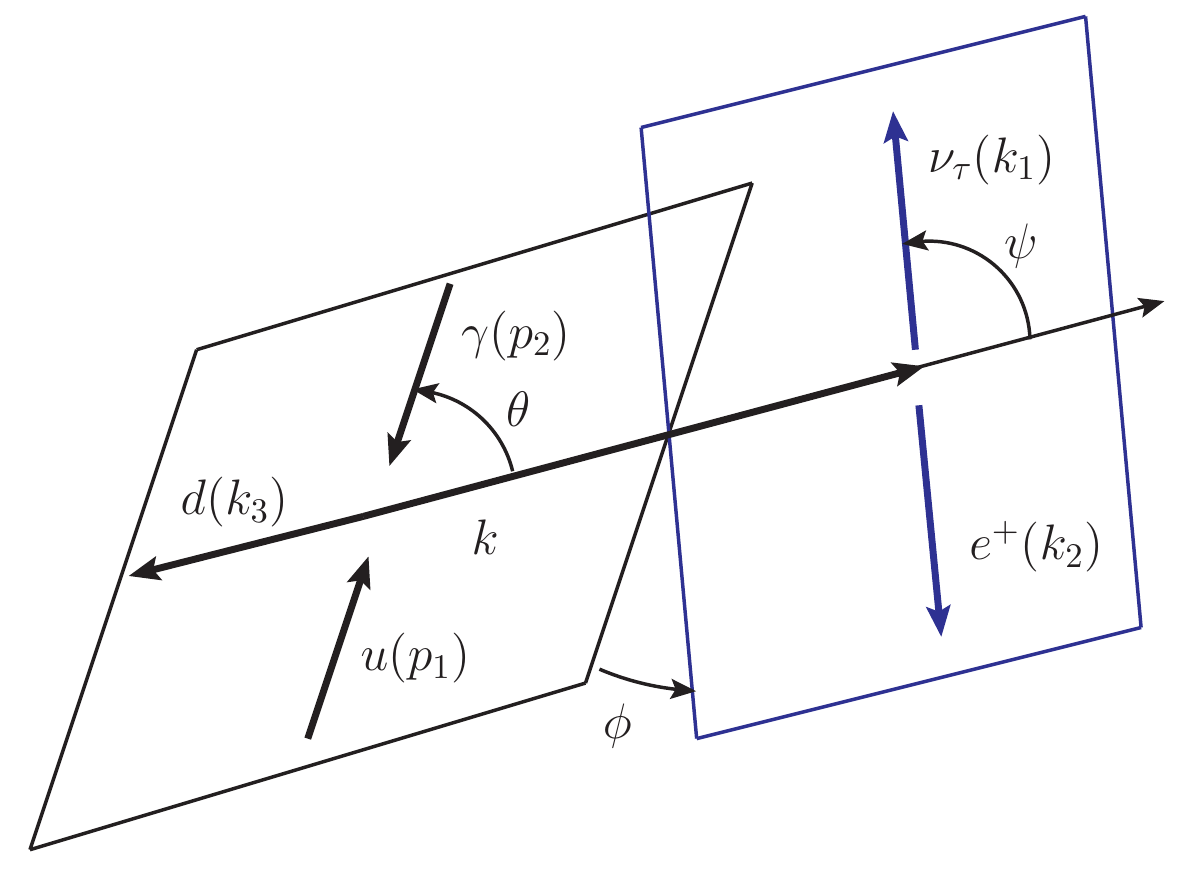}
\caption{\small The angular configuration of the illustrative toy process, $u\gamma \rightarrow d \nu e^+$.}
\label{fig:aTGC:coordinate1}
\end{center}
\end{figure}
We choose the following four momenta of the particles in our $2\rightarrow 3$ process, $u\gamma \rightarrow d \nu e^+$ and they are illustrated in Fig.~\ref{fig:aTGC:coordinate1}.
\begin{equation}\label{app:eq:momenta:2to3:Lab}
\begin{split}
  p_1^\mu =&\ \frac{\sqrt{\hat{s}}}{2} \left ( 1,\, \sin\theta\cos\phi,\, \sin\theta\sin\phi,\, \cos\theta \right )~,
  \\[5pt] 
  p_2^\mu =&\ \frac{\sqrt{\hat{s}}}{2} \left ( 1,\, -\sin\theta\cos\phi,\, -\sin\theta\sin\phi,\, -\cos\theta \right )~,
  \\[5pt]
  k_1^\mu =&\ \frac{\sqrt{\hat{s}}}{2} \left ( z+ (1-z) \cos\psi,\,  \sqrt{(2z-1)} \sin\psi,\, 0,\, (1-z) + z \cos\psi \right )~,
  \\[5pt]
  k_2^\mu =&\ \frac{\sqrt{\hat{s}}}{2} \left ( z - (1-z) \cos\psi,\, - \sqrt{(2z-1)} \sin\psi,\, 0,\, (1-z) - z \cos\psi \right )~,
  \\[5pt]
  k_3^\mu =&\ \sqrt{\hat{s}} \left ( 1-z,\, 0,\, 0,\, -(1-z) \right )~,
  \\[5pt]
  k^\mu =&\ \sqrt{\hat{s}} \left ( z,\, 0,\, 0,\, (1-z) \right )~,
\end{split}
\end{equation}
where the momentum $k$ has the invariant mass of $m^2_k = (2z-1)\hat{s}$. Note that the $2\rightarrow 3$ process can be effectively factorized into $2\rightarrow 2$ and $1\rightarrow 2$ via an intermediate momentum $k$. The momenta $k_1$ and $k_2$ in Eq.~(\ref{app:eq:momenta:2to3:Lab}) are obtained by boosting those in the $\nu e$ rest frame,
\begin{equation}
\begin{split}
  k_1^\mu =&\ \frac{m_k}{2} \left ( 1,\, \sin\psi,\, 0,\, \cos\psi\ \right )~,
  \\[5pt]
  k_2^\mu =&\ \frac{m_k}{2} \left ( 1,\, -\sin\psi,\, 0,\, -\cos\psi \right )~,
\end{split}
\end{equation}
along the $z$-axis with the boosting factor,
\begin{equation}\label{app:eq:boost:k1k2}
   k_z = \gamma_z\, m_k \beta_z \rightarrow \gamma_z = \frac{k^0}{m_k} = \frac{z}{\sqrt{2z-1}}~.
\end{equation}
When the intermediate $W$ emitted from the quark line is produced nearly on shell, $z$ is nearly fixed to be
\begin{equation}
  z \sim \frac{1}{2} \left ( 1 + \frac{m^2_W}{\hat{s}} \right )~.
\end{equation}
%
The helicity amplitudes for four diagrams in Fig.~\ref{fig:aTGC:2to3:diagrams} are given by
\begin{equation}
\begin{split}
 i \epsilon\cdot\mathcal{M}_a =&\ \bar{u}_L(k_3) \Big (  i \frac{g}{\sqrt{2}} \gamma^\rho  \Big ) u_L (p_1 ) 
  \frac{- i \eta^{\rho\nu}}{q^2 - m^2_W} 
  \\[3pt]
   & \times \epsilon_\lambda (p_2)\, i\, e \Big \{  \Big [ \eta^{\mu\nu} (q - k)^\lambda - (2+\delta \kappa_\gamma ) ( p_2^\mu \eta^{\nu\lambda} - p_2^\nu \eta^{\mu\lambda} )
 + \eta^{\nu\lambda} k^\mu - \eta^{\mu\lambda} q^\nu \Big ] 
 \\[3pt]
 &+  \frac{\lambda_z}{m^2_W} 
 \Big [ ( p_2^\mu \eta^{\nu\lambda} - p_2^\nu \eta^{\mu\lambda} ) (k \cdot q) 
   + (q^\lambda \eta^{\mu\nu} - q^\mu \eta^{\nu\lambda} )(k\cdot p_2) 
 \\[3pt]
 & \hspace{2cm} + (k^\nu \eta^{\mu\lambda} - k^\lambda \eta^{\mu\nu})(q\cdot p_2) 
 -k^\nu q^\lambda p_2^\mu + k^\lambda q^\mu p_2^\nu 
  \Big ] \Big \}  
 \\[3pt]
 &\times \frac{- i \eta^{\mu\sigma}}{k^2 - m^2_W + i m_W \Gamma_W} 
 \bar{u}_L(k_1) \Big (  i \frac{g}{\sqrt{2}} \gamma^\sigma  \Big ) v_L (k_2 )~
   \\[3pt]
  =&\  \left (  i \frac{g}{\sqrt{2}} \right )^2 (i e) \frac{(- i)^2}{q^2 - m^2_W} \frac{1}{k^2 - m_W^2 + i m_W \Gamma_W}
 \   \epsilon^\lambda  j_q^\nu j_l^\mu \, V^{\lambda\nu\mu}~.
\end{split}
\end{equation}
where $q = p_2 - k = k_3 - p_1$.
\begin{equation}
\begin{split}
  i \epsilon\cdot\mathcal{M}_b =&\  \epsilon_\mu (p_2) \bar{u}_L(k_3) \Big (  i \frac{g}{\sqrt{2}} \gamma^\rho  \Big ) u_L (p_1 ) 
  \frac{- i \eta^{\rho\sigma}}{q^2 - m^2_W} 
  \\[3pt]
   & \times \bar{u}_L(k_1) \Big (  i \frac{g}{\sqrt{2}} \gamma^\sigma  \Big ) \frac{i(\slashed{p_2} - \slashed{k_2})}{(p_2-k_2)^2} 
  \left (  - i e \gamma^\mu  \right ) v_L (k_2 )~
  \\[3pt]
  =&\  \left (  i \frac{g}{\sqrt{2}} \right )^2 (-i e) \frac{(- i) i}{q^2 - m^2_W} \frac{1}{(p_2 - k_2)^2}
   \bar{u}_L(k_1) \slashed{j_q} (\slashed{p_2}-\slashed{k_2}) \slashed{\epsilon} v_L (k_2 )~,
\end{split}
\end{equation}
where $q = k_3 - p_1$.
\begin{equation}
\begin{split}
  i \epsilon\cdot\mathcal{M}_c =&\  \epsilon_\mu (p_2) \bar{u}_L(k_3)  \left ( - \frac{i}{3} e \gamma^\mu  \right )  \frac{i(\slashed{k_3} - \slashed{p_2})}{(k_3-p_2)^2} \Big (  i \frac{g}{\sqrt{2}} \gamma^\rho  \Big ) u_L (p_1 )
  \\[3pt]
   & \times  \frac{- i \eta^{\rho\sigma}}{k^2 - m^2_W + i m_W \Gamma_W}  \bar{u}_L(k_1) 
   \Big (  i \frac{g}{\sqrt{2}} \gamma^\sigma  \Big ) v_L (k_2 )~
   \\[3pt]
   =&\  \left (  i \frac{g}{\sqrt{2}} \right )^2 \left ( -\frac{i}{3} e \right ) \frac{(- i) i}{k^2 - m^2_W + i m_W \Gamma_W} \frac{1}{(k_3 - p_2)^2}
   \bar{u}_L(k_3) \slashed{\epsilon} (\slashed{k_3}-\slashed{p_2}) \slashed{j_l} u_L (p_1 )~,
\\[5pt]
  i \epsilon\cdot\mathcal{M}_d =&\  \epsilon_\mu (p_2) \bar{u}_L(k_3)  \Big (  i \frac{g}{\sqrt{2}} \gamma^\rho  \Big )  \frac{i(\slashed{p_1} + \slashed{p_2})}{(p_1 + p_2)^2} \left ( \frac{2i}{3} e \gamma^\mu  \right ) u_L (p_1 )
  \\[3pt]
   & \times  \frac{- i \eta^{\rho\sigma}}{k^2 - m^2_W + i m_W \Gamma_W}  \bar{u}_L(k_1) 
   \Big (  i \frac{g}{\sqrt{2}} \gamma^\sigma  \Big ) v_L (k_2 )~
   \\[3pt]
   =&\  \left (  i \frac{g}{\sqrt{2}} \right )^2  \left ( \frac{2i}{3} e \right ) \frac{(- i) i}{k^2 - m^2_W + i m_W \Gamma_W} \frac{1}{(p_1 + p_2)^2}
   \bar{u}_L(k_3) \slashed{j_l} (\slashed{p_1} + \slashed{p_2}) \slashed{\epsilon} u_L (p_1 )
\end{split}
\end{equation}
where $j_q^\mu = \bar{u}_L(k_3) \gamma^\mu  u_L (p_1 )$ and $j_l^\mu = \bar{u}_L(k_1) \gamma^\mu  v_L (k_2 )$.

\subsection{Phase space integration}
The partonic cross section of $2\rightarrow 3$ process in our coordinate system is obtained by the following phase space integration,
\begin{equation}
{\hat\sigma} = \frac{1}{512 \pi^4} \int_{1/2}^1 dz (1-z) \int_{-1}^1 d \cos\theta \int_{-1}^1 d\cos\psi \int_0^{2\pi} d\phi \left | \overline{\mathcal{M}} \right |^2~,
\end{equation}
where $\left | \overline{\mathcal{M}} \right |^2$ is the summed and averaged amplitude-squared over polarizations of the initial partons and $\left | \overline{\mathcal{M}} \right |$ has a negative mass dimension of one.

\subsection{Interference between SM and BSM amplitudes for coupling $\lambda_z$}
\label{sec:app:int}

\begin{figure}[tph]
\begin{center}
\includegraphics[width=0.45\textwidth]{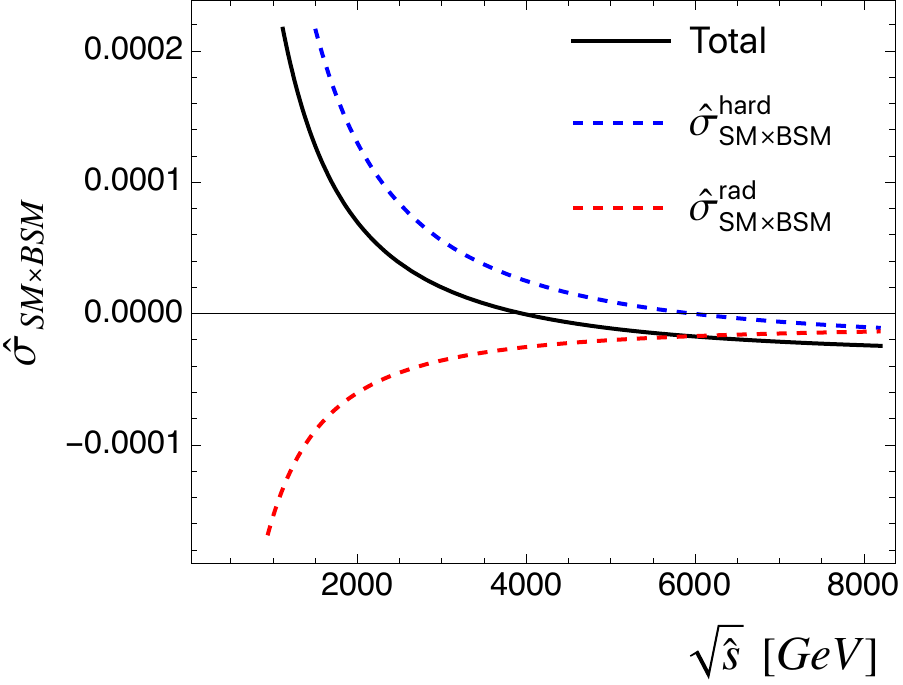}\quad
\includegraphics[width=0.45\textwidth]{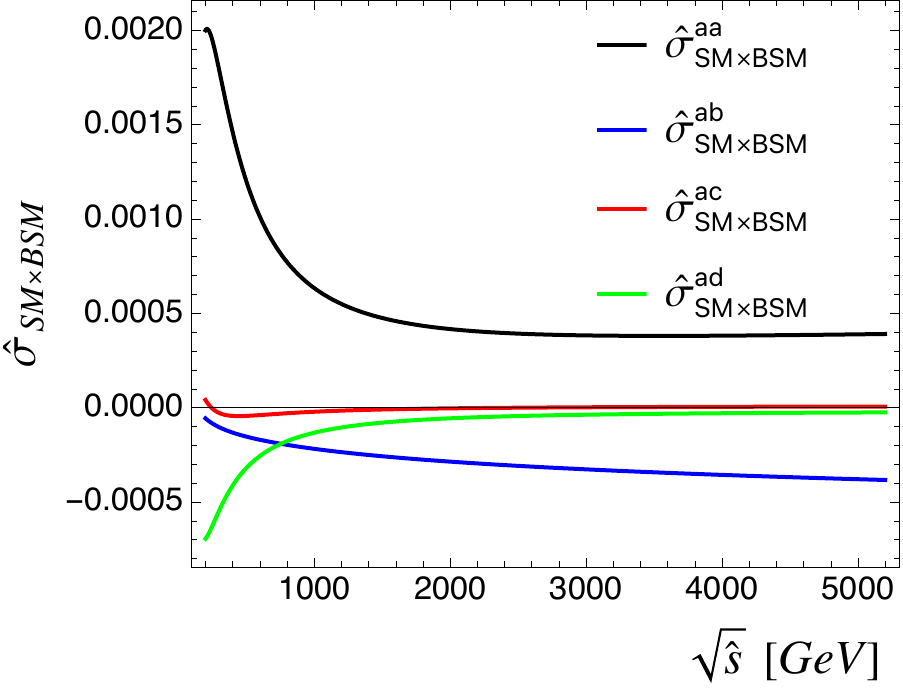}
\caption{\small The partonic inclusive cross section in an arbitrary rate for the interference between the SM and BSM,  $\hat\sigma_{\text{SM}\times \text{BSM}}(u_L\gamma_L \rightarrow d\nu e^+)$, integrated over the entire phase space.}
\label{fig:int:toy:amp}
\end{center}
\end{figure}

\begin{figure}[tph]
\begin{center}
\includegraphics[width=0.45\textwidth]{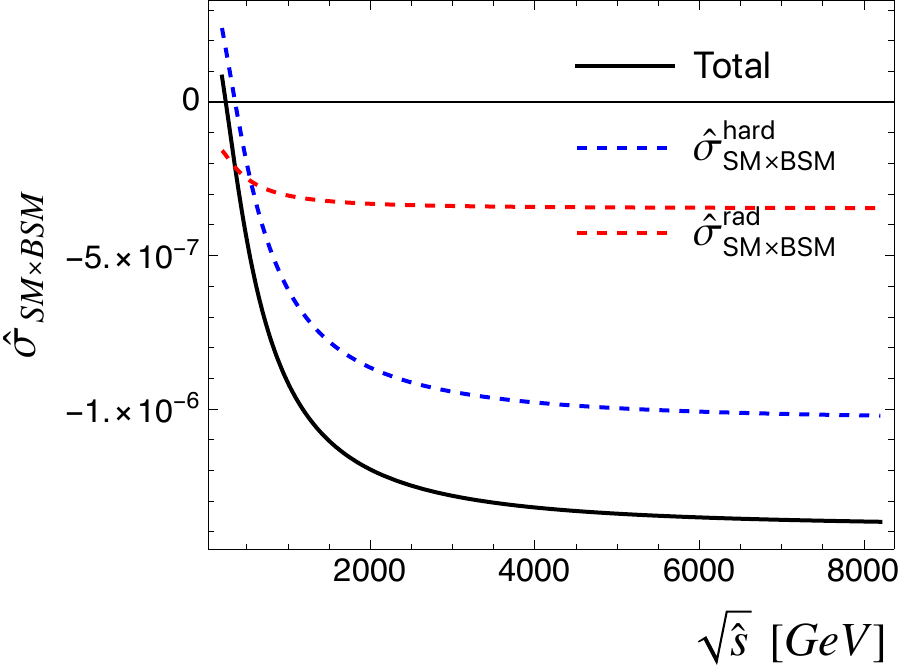}\quad
\includegraphics[width=0.45\textwidth]{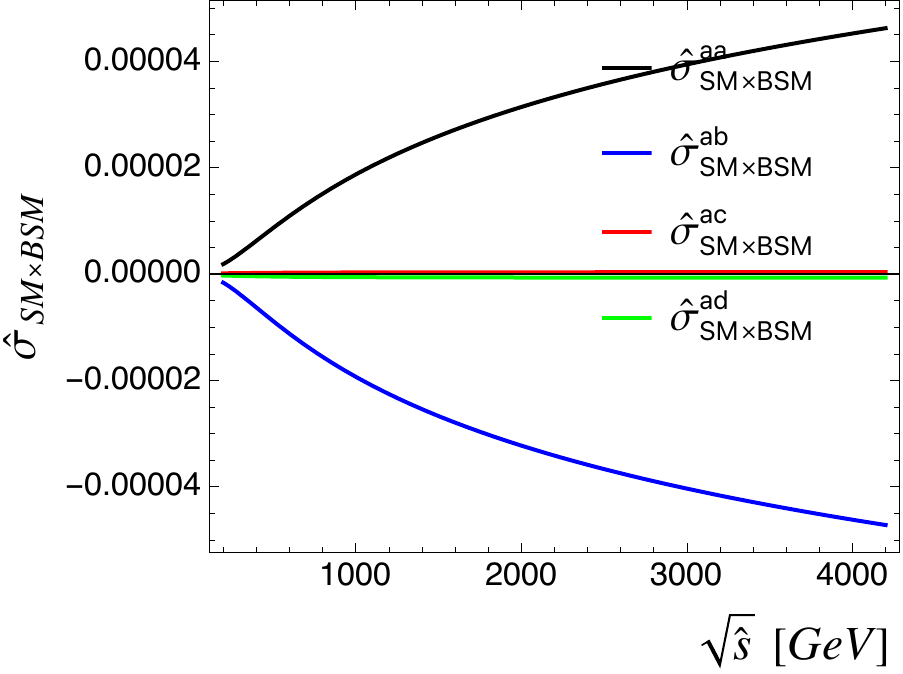}
\caption{\small The partonic inclusive cross section in an arbitrary rate for the interference between the SM and BSM, $\hat\sigma_{\text{SM}\times \text{BSM}}(u_L\gamma_L \rightarrow d\nu e^+)$, integrated over the restricted phase space $z = [1- \varepsilon,\, 1]$ where $\varepsilon = 0.1$ was chosen.}
\label{fig:int:toy:amp:ewa}
\end{center}
\end{figure}

In our $2\rightarrow 3$ toy process, diagrams $a$ and $b$ in Fig.~\ref{fig:aTGC:2to3:diagrams} are those of interest that probe the hard subprocess and diagrams $c$ and $d$ belong to the radiation type where $W$ decaying to $e^+\nu$ is attached to either incoming or outgoing quark line. 
Restricting only to the interference, we split the contribution into two categories.
\begin{equation}
\hat{\sigma}^{hard}_{\text{SM}\times \text{BSM}} \equiv \hat{\sigma}^{aa}_{\text{SM}\times \text{BSM}} +
\hat{\sigma}^{ab}_{\text{SM}\times \text{BSM}}~,\quad
\hat{\sigma}^{rad}_{\text{SM}\times \text{BSM}} \equiv \hat{\sigma}^{ac}_{\text{SM}\times \text{BSM}} +
\hat{\sigma}^{ad}_{\text{SM}\times \text{BSM}}~,
\end{equation}
where $\hat{\sigma}^{ij}_{\text{SM}\times \text{BSM}}$ refers to the partonic cross section from the product of two diagrams $i$ and $j$ in Fig.~\ref{fig:aTGC:2to3:diagrams}. The relative difference between two categories is purely due to the SM as the $\lambda_z$ dependence comes from the common diagram $a$.
The left panel of Fig.~\ref{fig:int:toy:amp} shows that $\hat{\sigma}^{hard}_{\text{SM}\times \text{BSM}}$ and $\hat{\sigma}^{rad}_{\text{SM}\times \text{BSM}}$ are comparable. While the magnitude of each $\hat{\sigma}^{aa}_{\text{SM}\times \text{BSM}}$ and $\hat{\sigma}^{ab}_{\text{SM}\times \text{BSM}}$ is bigger than both $\hat{\sigma}^{ac}_{\text{SM}\times \text{BSM}}$ and $\hat{\sigma}^{ad}_{\text{SM}\times \text{BSM}}$, there is a cancellation between two contributions from the hard subprocess, dictated by the gauge symmetry. It should be an artifact due to the gauge choice in the photon polarization. One may choose a particular gauge for the photon polarization to suppress the contribution from the radiation type diagrams. The observed property is more pronounced when the phase space is restricted to $z = [1 - \varepsilon,\, 1]$ with $\varepsilon = 0.1$. 
As is clearly seen in Fig.~\ref{fig:int:toy:amp:ewa}, an individual contribution from the hard subprocess becomes much bigger than those involving the radiation type diagrams, and the cancellation is more dramatic. The gauge dependence may not be a problem in the $2\rightarrow 4$ process where all gauge bosons including the photon are attached to the fermion currents. 

Another interesting observation is that the sign of interference is $\sqrt{\hat{s}}$-dependent. For instance, in Fig.~\ref{fig:int:toy:amp}, the interference stays positive until around $\sqrt{\hat{s}} \sim 4$ TeV whereas, in the situation corresponding to Fig.~\ref{fig:int:toy:amp:ewa}, the interference becomes negative well before TeV.

\section{Beyond the effective $W$ approximation in $qV \rightarrow q' \nu \ell$}
\label{app:sec:ewa}

It will be interesting to understand our result in the context of the EWA. In this section, we carefully compare the derivation of the EWA presented in~\cite{Borel:2012by} in the axial gauge with our amplitudes computed in the unitary gauge. In the unitary gauge, it is difficult to clearly separate the contributions from sub-amplitudes with definite polarizations. However, despite the different gauge choices, a meaningful comparison with~\cite{Borel:2012by} can be done.
We start our discussion by presenting our full amplitude decomposed into Fourier modes in $e^{i m \phi}$ (with $m$ as an integer of either sign or zero) only for the left-handed polarization of the photon as an illustration
\footnote{\label{app:for:bac}The diagram $b$ in Fig.~\ref{fig:aTGC:2to3:diagrams} takes the form
\begin{equation}
  \epsilon_L \cdot \mathcal{M}_b = \frac{\sum_{n} c_n e^{i n \phi}}{-\alpha + \beta \cos\phi} = \sum_m C_m e^{im \phi}~,
\end{equation}
where $\alpha, \beta >0$ and the summation over $n$ in the numerator stops at a finite $n$. Using the residue theorem to obtain $C_m$, the Fourier decomposition is given by
\begin{equation}\label{app:eq:Fourier}
  \epsilon_L \cdot \mathcal{M}_b = - \sum_m \left (  \sum_n c_n \frac{ \left ( \alpha - \sqrt{\alpha^2 - \beta^2} \right )^{|n-m|} }{\beta^{|n-m|} \sqrt{\alpha^2 - \beta^2}}  \right ) e^{i m \phi}~,
\end{equation}
where $m$ runs over $[-\infty, \, \infty]$. $ \sqrt{\alpha^2 - \beta^2} $ in Eq.~(\ref{app:eq:Fourier}) in terms of $\theta$ after the substitution is given by
\begin{equation}
 \sqrt{\alpha^2 - \beta^2} = \left | - 1 + z + z \cos\psi + \cos\theta (z - \cos\theta + z\cos\psi) \right |~,
\end{equation}
where $- 1 + z + z \cos\psi + \cos\theta (z - \cos\theta + z\cos\psi) >0$  in the forward quark limit $\theta \rightarrow 0$ (the opposite sign for the backward quark limit). We will refer to the forward quark region by phase space satisfying $- 1 + z + z \cos\psi + \cos\theta (z - \cos\theta + z\cos\psi) >0$ and the backward quark by those with the opposite sign.
}

\subsection{Full amplitude of $qV \rightarrow q' \nu \ell$}

All the amplitudes of $qV \rightarrow q' \nu \ell$ below, focusing only on the coupling $\lambda_z$, are multiplied by an overall phase factor $e^{-i\phi}$ (without loss of generality) for better comparison with literature. The amplitude is decomposed into the SM and BSM ones. After substituting $\theta \rightarrow \pi - \theta$ to parametrize the forward quark region in terms of angle $\theta$ (the forward quark corresponds to $\theta \sim 0$ after the substitution), our evaluation of the amplitudes for the SM and BSM for the coupling $\lambda_z$ are given by (showing only terms relevant for the forward quark, see footnote~\ref{app:for:bac})
\begin{equation}\label{app:eq:amp23:exact}
\begin{split}
\epsilon_L \cdot \mathcal{M}_{BSM} &= 
\lambda_z \frac{eg^2}{4 m_W^2} \frac{\hat{s}^{5/2} \sqrt{(2z-1)(1-z)} \sin\displaystyle\frac{\theta}{2}\, e^{-i \phi}}{\left [ (2z-1) \hat{s} - m_W^2 \right ] \left [ m_W^2 + \hat{s} (1-z) (1-\cos\theta ) \right ]}
\\[4pt]
&\quad \times \Big [ 2 \sqrt{2z-1}\, \sin\psi \cos\theta - (1-\cos\psi)\sin\theta\, e^{-i \phi}
\\[4pt]
&\hspace{2cm} + (2z-1) ( 1+ \cos\psi) \sin\theta\,  e^{i \phi} \Big ]~,
\\[7pt]
\epsilon_L \cdot \mathcal{M}_{SM} &= - e g^2 \frac{1}{m_W^2 + \hat{s} (1-z) (1 - \cos\theta)} 
\left [ \hat{s}^{3/2} \sqrt{\frac{1-z}{2z-1}} (1 + \cos\psi) \sec\frac{\theta}{2} \right .
\\[4pt]
& \times \frac{4(1-z)(2z-1) (1-\cos\theta) - 2 (5-4z) \displaystyle\frac{m_W^2}{\hat{s}}  }{6 \left [ (2z-1) \hat{s} - m_W^2 \right ] }
\\[4pt]
&+ \hat{s}^{1/2} \frac{(1-z)^{3/2}}{2z-1}  \sin\psi \sec^3\frac{\theta}{2}\sin\theta \, e^{i \phi}
\\[4pt]
& + \hat{s}^{1/2} \left (\frac{1-z}{2z - 1} \right )^{3/2} \frac{1}{2} (1-\cos\psi) \sec^5\frac{\theta}{2} \sin^2\theta\, e^{2i\phi}
\\[4pt]
& \left .+ \hat{s}^{1/2} \frac{(1-z)^{3/2}}{(2z-1)^2} \frac{1}{4} (1-\cos\psi)^2 \csc\psi \sec^7\frac{\theta}{2} \sin^3\theta\, e^{3i\phi} + \cdots \right ]~,
\end{split}
\end{equation}
where coefficients of Fourier modes are exact without any approximation, importantly, it works for $\theta \sim \mathcal{O}(1)$, and $\Gamma_W$ was neglected since here we focus on the off-shell $W$ decaying to $\ell \nu_\ell$. The series expansion in $e^{\pm i m \phi}$ for a large $m$ without being truncated arises due to the $\phi$-dependence in the denominator of diagram $b$ in Fig.~\ref{fig:aTGC:2to3:diagrams}.
In the forward limit of the quark, namely $\theta \ll 1$, the amplitudes for the SM and BSM will be approximated in power series of small $\theta$:
\begin{equation}\label{app:eq:amp:varepsilon:exp}
\begin{split}
  \epsilon\cdot\mathcal{M} &= \tilde{\theta} \left ( \mathcal{M}^{(0,0)}_+ + \mathcal{M}^{(1,0)}_+ \tilde{\theta}  
  + \mathcal{M}^{(0,1)}_+ \tilde{\theta}^*  + \cdots \right ) 
  \\[4pt]
  &\quad + \tilde{\theta}^* \left ( \mathcal{M}^{(0,0)}_- + \mathcal{M}^{(1,0)}_- \tilde{\theta}  
  + \mathcal{M}^{(0,1)}_- \tilde{\theta}^*  + \cdots \right )  + \cdots~,
\end{split}
\end{equation}
where a subscript $\pm$ is to distinguish two groups of terms multiplied by an overall $\tilde{\theta}$ and $\tilde{\theta}^*$ outside parenthesis. 
The contributions suppressed by $\frac{m_W^2}{E^2}$ (and, in general, those from the longitudinal polarizations although they do not appear in our computation as the external fermions are taken to be massless) are denoted by $\cdots$ in Eq.~(\ref{app:eq:amp:varepsilon:exp}).
It is important to notice that the factor $e^{\pm i m\phi}$ is always accompanied with $\theta^n$ for $m \leq n$ in the expansion in terms of $\theta \ll 1$ (due to a possible product of $\tilde{\theta} \equiv \theta e^{- i\phi}$ and $\tilde{\theta}^*$). $\tilde{\theta}$ and $\tilde{\theta}^*$ can be used as a way to compare with the derivation in~\cite{Borel:2012by} (this property will be clear below) as they are correlated with the specific polarization of $W$ radiated off the quark current (thus we can extract the information about the sub-amplitude with a specific polarization). 
By looking into the analytic evaluation of the SM and BSM amplitudes in Eq.~(\ref{app:eq:amp23:exact}), we definitely see that
\begin{equation}
\begin{split}
   \mathcal{M}^{(0,0)}_{-,SM} & \neq 0 
   \quad \mathcal{M}^{(0,0)}_{-,BSM} = 0~,
   \\[4pt]
   \mathcal{M}^{(0,0)}_{+,SM} &= 0
   \quad \mathcal{M}^{(0,0)}_{+,BSM} \neq 0~,
\end{split}
\end{equation}
which is the reflection of the helicity selection rule.

\subsection{What gets lost in effective $W$ approximation}

The derivation of the EWA in $2\rightarrow 3$ process in~\cite{Borel:2012by} has been carried out in the axial gauge where all radiation type diagrams were supposed to be sub-dominant. The derivation starts with the full matrix element expanded in power series of the virtuality of the gauge boson $V\equiv m^2 - q^2$ radiated off the forward quark line, while assuming $xE \sim (1-x) E$, $\delta_m = m/E$, $\delta_\perp = p_\perp/E \ll 1$ ($E$ as the scale of the hard subprocess, $m$ the gauge boson mass, $p_\perp$ the transverse momentum of the quark). Quoting Eq. (37) of~\cite{Borel:2012by} in their notation, the full amplitude takes the form
\begin{equation}\label{eq:EWA}
\begin{split}
 \mathcal{A_\text{total}} &= - \frac{i}{V^2} \sum_{h=\pm 1} \left [ J^\mu (\varepsilon_\mu^h)^* \right ] \left [\varepsilon_\nu^h \mathcal{A}^\nu_\text{hard} \right ]
 \\[3pt]
 &\hspace{2cm} - \frac{i}{V^2} \left [ J^\mu (\varepsilon_\mu^0)^* \right ] \left [ \left ( 1- \frac{V^2}{m^2} \right ) \varepsilon^0_\nu \mathcal{A}^\nu_\text{hard} \right ] \left ( 1 + \mathcal{O}(\delta_\perp^2 + \delta_m^2) \right )~,
\end{split}
\end{equation}
where the splitting amplitudes for the transverse polarizations were given by (similarly for the longitudinal polarization)
\begin{equation}
  -\frac{i}{V^2} \left [ J^\mu (\varepsilon_\mu^\pm)^* \right ] 
  = 2C \frac{p_\perp e^{\pm i\phi}}{V^2} g_\pm (x) \left ( 1 + \mathcal{O}(\delta_\perp^2 + \delta_m^2) \right )~,
\end{equation}
where $g_\pm (x)$ is the splitting function for transverse polarizations.
In our toy process $u\gamma \rightarrow d\nu e^+$ only with $\lambda_z$, the full amplitude would include terms in their language
\begin{equation}
 \propto \left [ J^\mu (\varepsilon_\mu^-)^* \right ]  \left [\varepsilon_\nu^- \mathcal{A}^\nu_{SM} \right ] 
 + \left [ J^\mu (\varepsilon_\mu^+)^* \right ]  \left [\varepsilon_\nu^+ \mathcal{A}^\nu_{BSM} \right ]~,
\end{equation}
where the polarization is that of $W$ radiated off the quark current. Note that the total helicity of the sub-amplitude of $W\gamma \rightarrow \ell \nu_\ell$ for the SM and BSM with the insertion of $\text{tr}(W_{\mu\nu}^3)$ are different while the total helicity of the full amplitude can match. 
Following the notation of~\cite{Borel:2012by} and its procedure, the sub-amplitudes are expanded in $\frac{\tilde{p}_\perp}{E}$ and $\frac{\tilde{p}^*_\perp}{E}$ (Eq. (55) of~\cite{Borel:2012by}),
\begin{equation}
\begin{split}
  \mathcal{A}_\pm  \equiv g_\pm (x) \left [\varepsilon_\nu^\pm \mathcal{A}^\nu \right ] 
  &= \mathcal{A}^{(0,0)}_\pm  + \mathcal{A}^{(1,0)}_\pm \frac{\tilde{p}_\perp}{E} + \mathcal{A}^{(0,1)}_\pm \frac{\tilde{p}^*_\perp}{E}
\\[3pt]
  &\quad +  \mathcal{A}^{(1,1)}_\pm \frac{\tilde{p}_\perp \tilde{p}^*_\perp}{E^2}
  + \mathcal{A}^{(2,0)}_\pm \frac{\tilde{p}^2_\perp}{E^2} + \mathcal{A}^{(0,2)}_\pm \frac{\tilde{p}^{*2}_\perp}{E^2}
   + \cdots~,
\end{split}  
\end{equation}
where $\tilde{p}_\perp \equiv p_\perp^1 - i p_\perp^2 \equiv p_\perp e^{-i\phi}$ and $\tilde{p}^*_\perp = p_\perp e^{i\phi}$ is the complex conjugate. 
Since $p_T(q)=(1-z)\sqrt{\hat{s}}\sin\theta$ and $m^2_{e\nu} = (2z-1)\hat{s}$ in our toy process, taking $m_{e\nu} \sim E$ for $z\sim \mathcal{O}(1)$ as an energy of the hard subprocess, we have the relation $p_\perp \sim E\, \theta$ with $\theta \ll 1$ from which we can relate $\tilde{p}_\perp$ and $\tilde{\theta}$.
The total amplitude is rewritten as, keeping only transverse polarizations to simplify the discussion,
\begin{equation}
\begin{split}
 \mathcal{A_\text{total}} &= \frac{2C}{V^2} \left [ \tilde{p}_\perp \mathcal{A}_+ +  \tilde{p}^*_\perp \mathcal{A}_- \right ]
 \\[3pt]
 &=\frac{2C}{V^2} \Big [ \tilde{p}_\perp \left ( \mathcal{A}^{(0,0)}_+  + \mathcal{A}^{(1,0)}_+ \frac{\tilde{p}_\perp}{E} + \mathcal{A}^{(0,1)}_+ \frac{\tilde{p}^*_\perp}{E} + \cdots \right ) 
 \\[3pt]
 &\hspace{2cm} +  \tilde{p}^*_\perp \left ( \mathcal{A}^{(0,0)}_-  + \mathcal{A}^{(1,0)}_- \frac{\tilde{p}_\perp}{E} + \mathcal{A}^{(0,1)}_- \frac{\tilde{p}^*_\perp}{E}  + \cdots \right ) \Big ]~,
\end{split}
\end{equation}
where we also kept higher-order terms which are important in our situation unlike the case of the derivation in~\cite{Borel:2012by} which keeps only the leading terms $\mathcal{A}^{(0,0)}_\pm$.
Upon squaring the full amplitude, we would expect interference terms to the total cross section such as
\begin{equation}\label{app:eq:inter}
  \propto \frac{(\tilde{p}_\perp \tilde{p}^*_\perp)^2}{E^2} \left ( \mathcal{A}^{(1,0)*}_{-,SM}  \mathcal{A}^{(0,1)}_{+,BSM} + h.c. \right )
  + \cdots
\end{equation}
whereas the leading SM and quadratic terms of the BSM are given by
\begin{equation}\label{app:eq:sm:bsmsq}
\begin{split}
  &\propto (\tilde{p}_\perp \tilde{p}^*_\perp) \left | \mathcal{A}^{(0,0)}_{-,SM} \right |^2
+ (\tilde{p}_\perp \tilde{p}^*_\perp) \left | \mathcal{A}^{(0,0)}_{+,BSM} \right |^2  + \cdots ~,
\end{split}
\end{equation}
where only leading $\phi$-independent terms (that can survive in the inclusive cross section) were written in Eqs. (\ref{app:eq:inter}) and (\ref{app:eq:sm:bsmsq}) for an illustration.
Therefore, the leading contributions to the interference and quadratic terms in the inclusive cross section scale as, showing only leading $\phi$-independent terms, 
\begin{equation}\label{eq:ewa:int:quad}
 \frac{ \left |\mathcal{A}_\text{total} \right |^2_{SM\times BSM}}{ \left | \mathcal{A}_\text{total} \right |^2_{SM}} \propto \lambda_z \theta^2 \frac{E^2}{\Lambda^2}~,\quad
 \frac{ \left | \mathcal{A}_\text{total} \right |^2_{BSM^2}}{\left | \mathcal{A}_\text{total} \right |^2_{SM}} \propto \lambda_z^2 \frac{E^4}{\Lambda^4}~.
\end{equation}
The interference appears suppressed by $\theta^2$ in case of $\theta \ll 1$, compared to the typical energy-growing behavior. Note that there could be also terms suppressed by $\delta^2_m = \frac{m^2}{E^2}$ in the interference in Eq.~(\ref{eq:ewa:int:quad}). However, importantly, our exact result in Eq. (\ref{app:eq:amp23:exact}) extend to the situation with a sizeable $\theta$ which can be thought of kind of the resummation.
If only leading terms $\mathcal{A}^{(0,0)}_\pm$ are taken as in~\cite{Borel:2012by}, the interference between the SM and BSM would have only $\cos(2\phi)$ term which vanishes upon the integration over $\phi$. The explicit analytic evaluation and numerical confirmation of our toy process reveals a sizable phase space beyond the relevant regime of the EWA that contributes to the interference in the inclusive cross section. We suspect that our situation belongs to an exceptional case mentioned in~\cite{Borel:2012by}.

\section{Detail of BDT Analysis}
\label{app:sec:BDT}

%
\begin{figure}[tph]
\begin{center}
\includegraphics[width=0.50\textwidth]{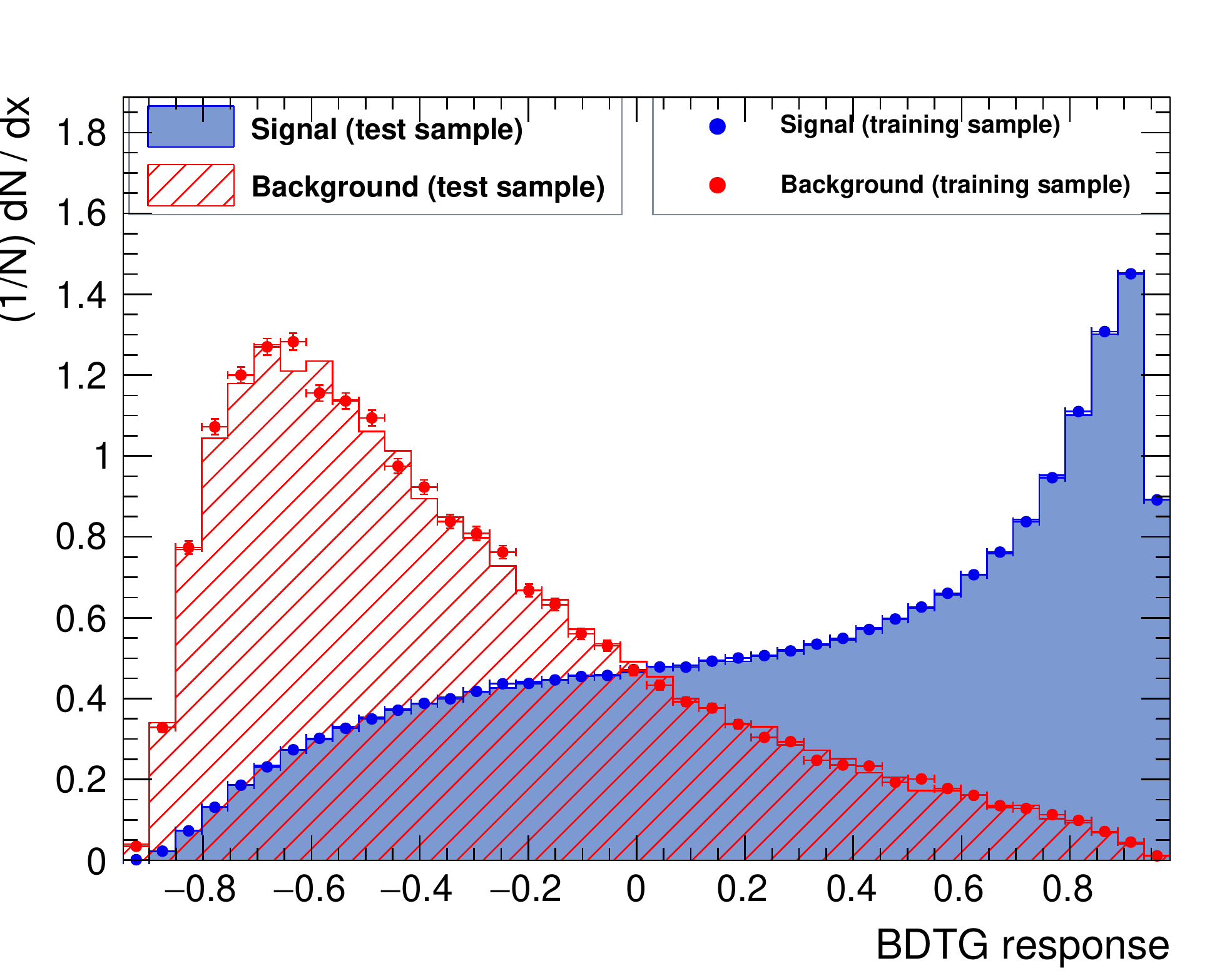}
\caption{\small Our validation of the BDT analysis with the variable set in Eq.~(\ref{eq:bdt:set}), using the gradient boosting algorithm in {\tt TMVA} package.}
\label{fig:BDTG:output}
\end{center}
\end{figure}
For the purpose of the training and testing, we made separate inclusive EW $\ell\ell jj$ and QCD Drell-Yan samples over the entire $m_{\ell\ell}$ range whereas the samples (for the same processes) for the actual BDT analysis were generated in multiple $m_{\ell\ell}$ bins to guarantee the smoothness with enough statistics up to the high invariant mass tail. The ratio of samples for the training and testing to those for the actual analysis is 1 to 4.
For $t\bar{t}$+jets samples, we used 30\% for the training and testing and the remaining 70\% for the analysis. We trained and tested over the EW $\ell\ell jj$ in the SM as a signal and the remaining as the background using the gradient boosting algorithm (called BDTG) provided in {\tt TMVA} package. Our validation of the BDT analysis is illustrated in Fig.~\ref{fig:BDTG:output} which shows the clear separation of the EW $\ell\ell$ + jets events from the QCD Drell-Yan and top pair backgrounds. 

\begin{figure}[tph]
\begin{center}
\includegraphics[width=0.48\textwidth]{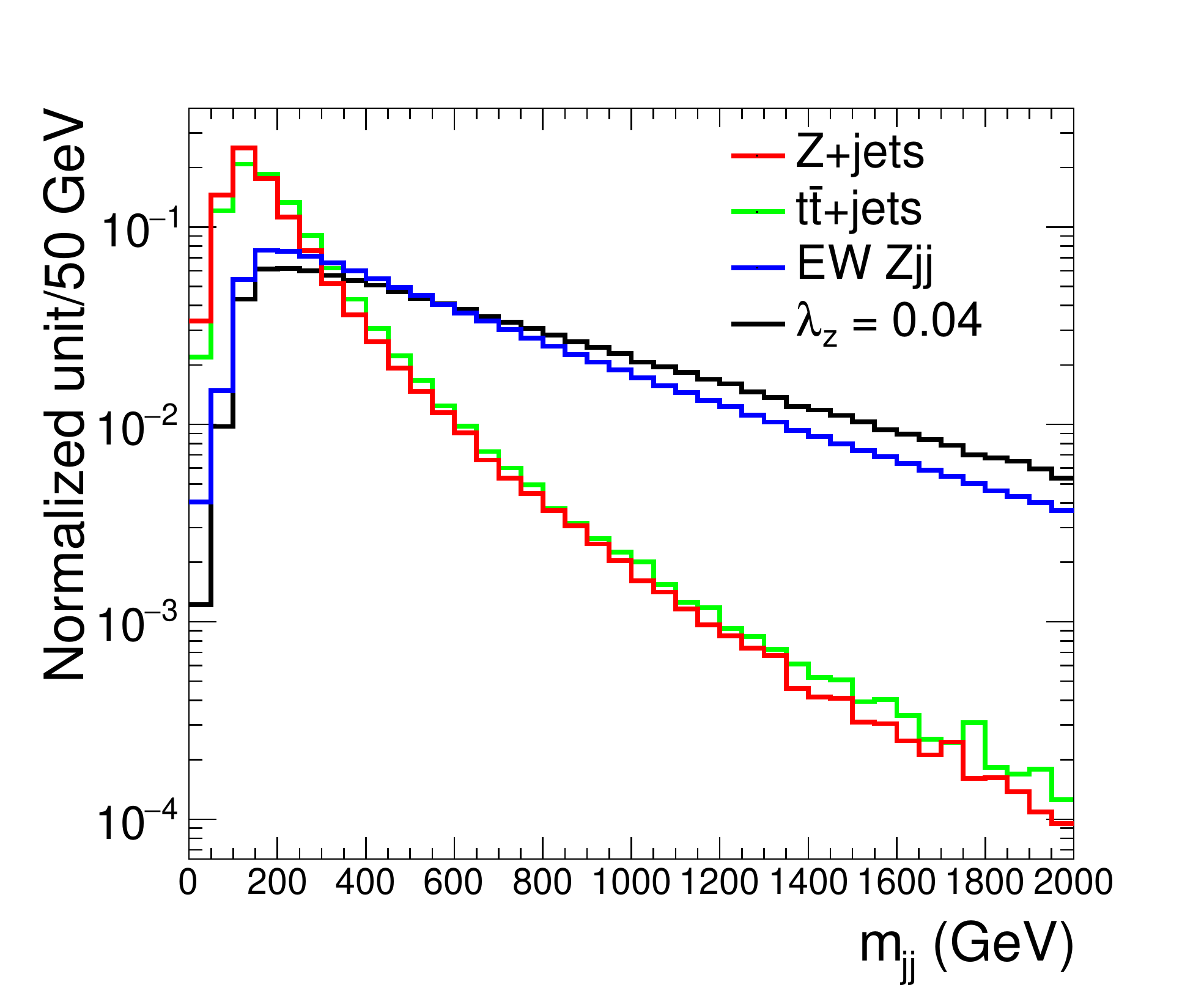}
\includegraphics[width=0.48\textwidth]{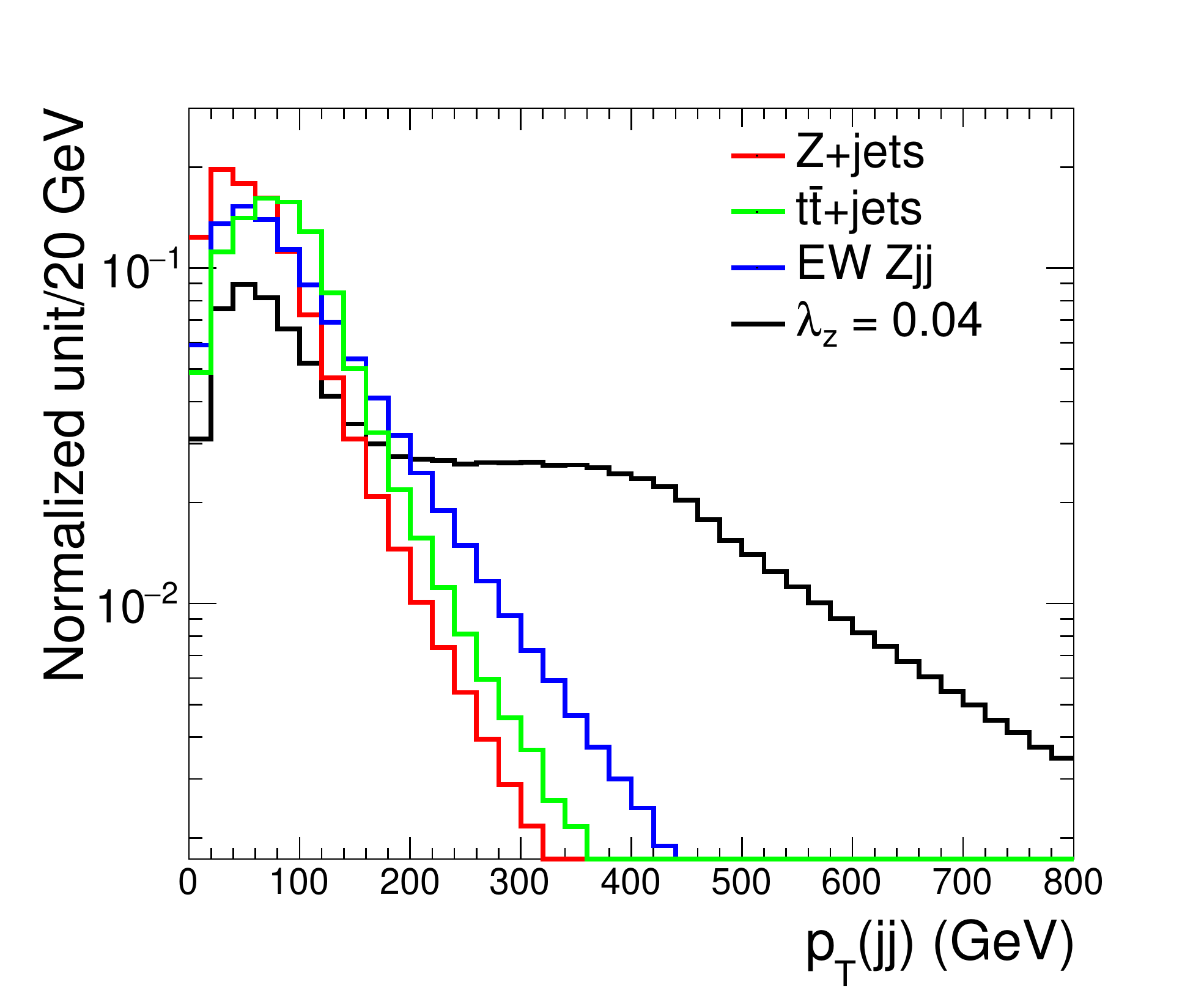}
\\
\includegraphics[width=0.48\textwidth]{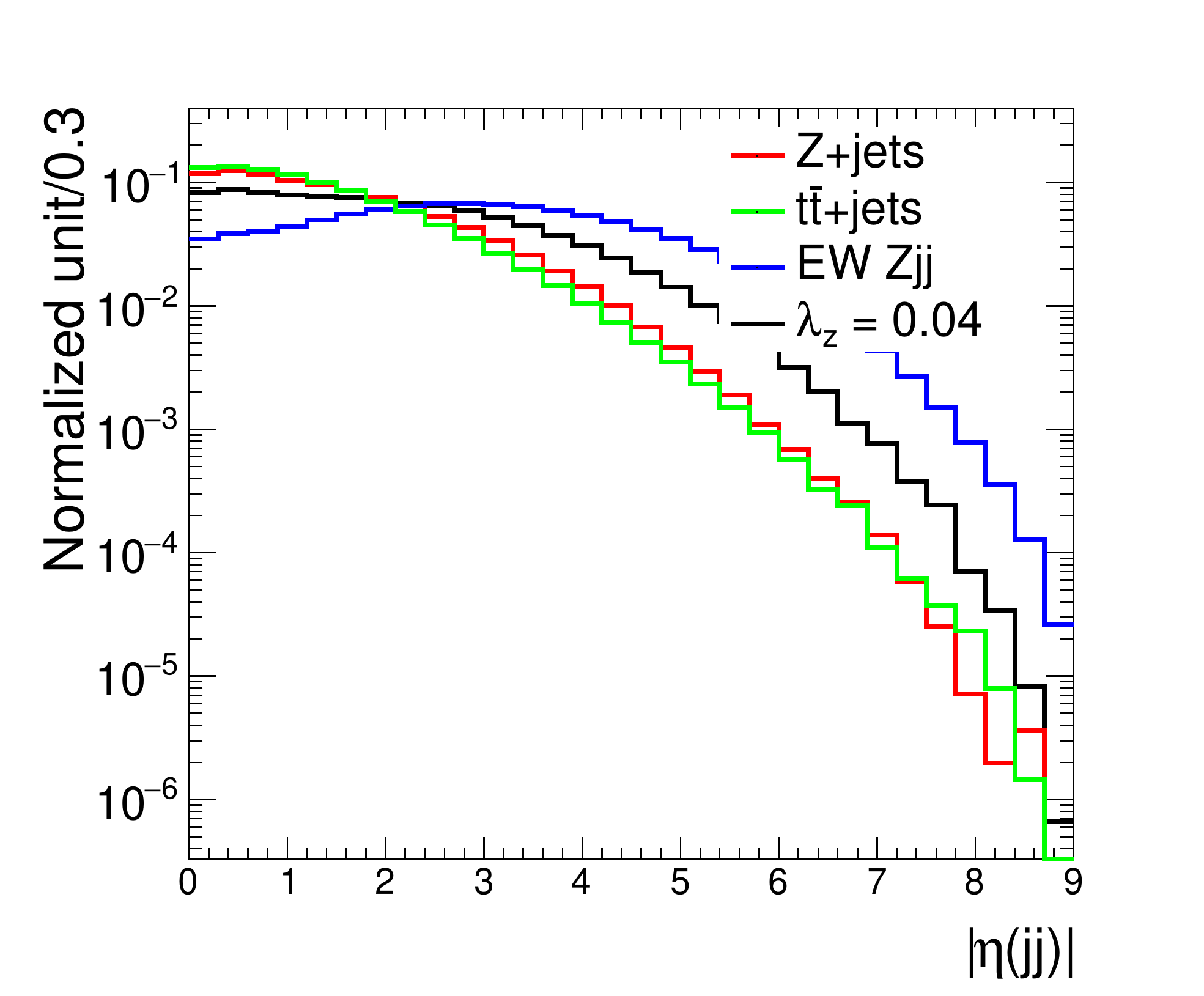}
\\
\includegraphics[width=0.48\textwidth]{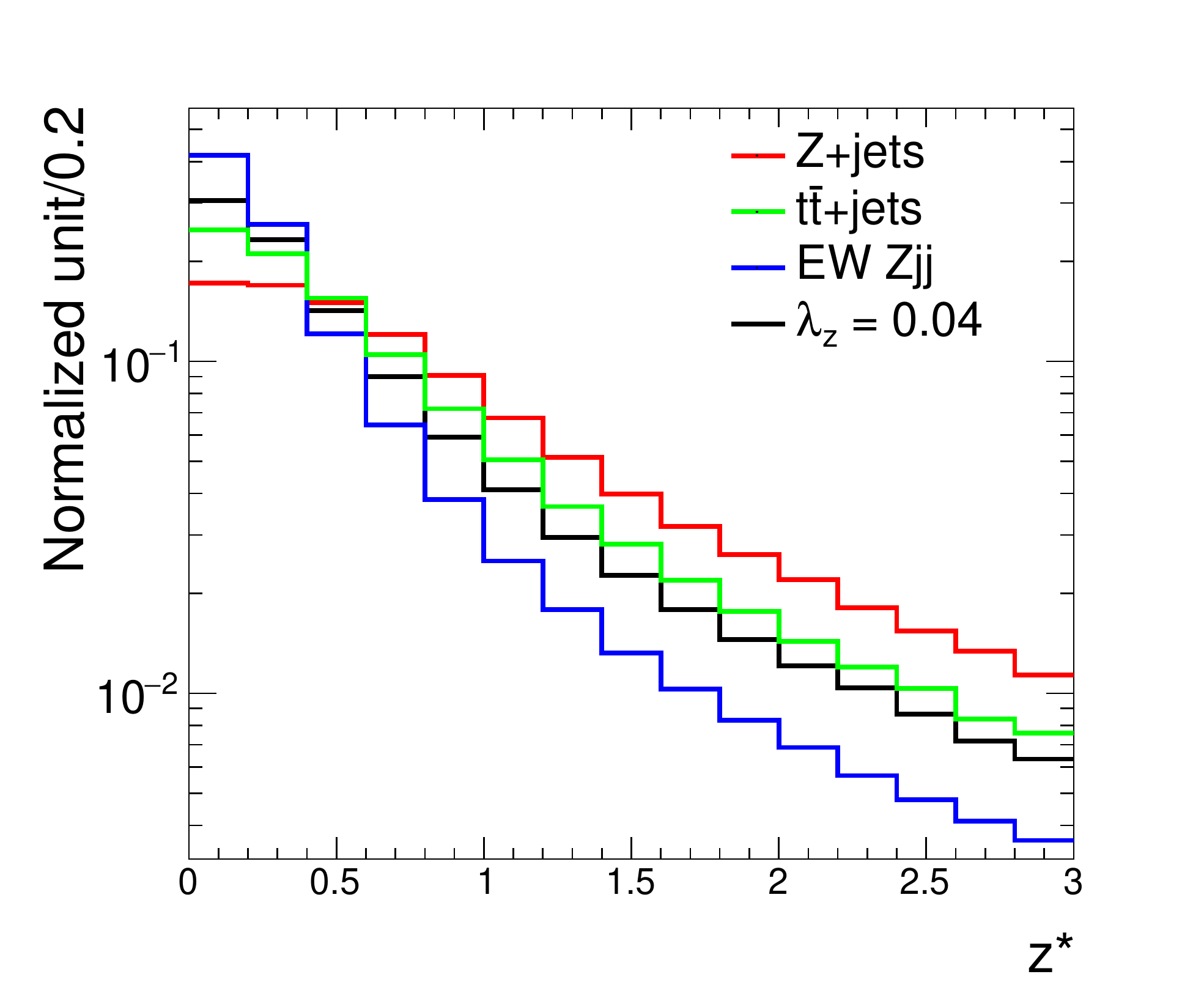}
\includegraphics[width=0.48\textwidth]{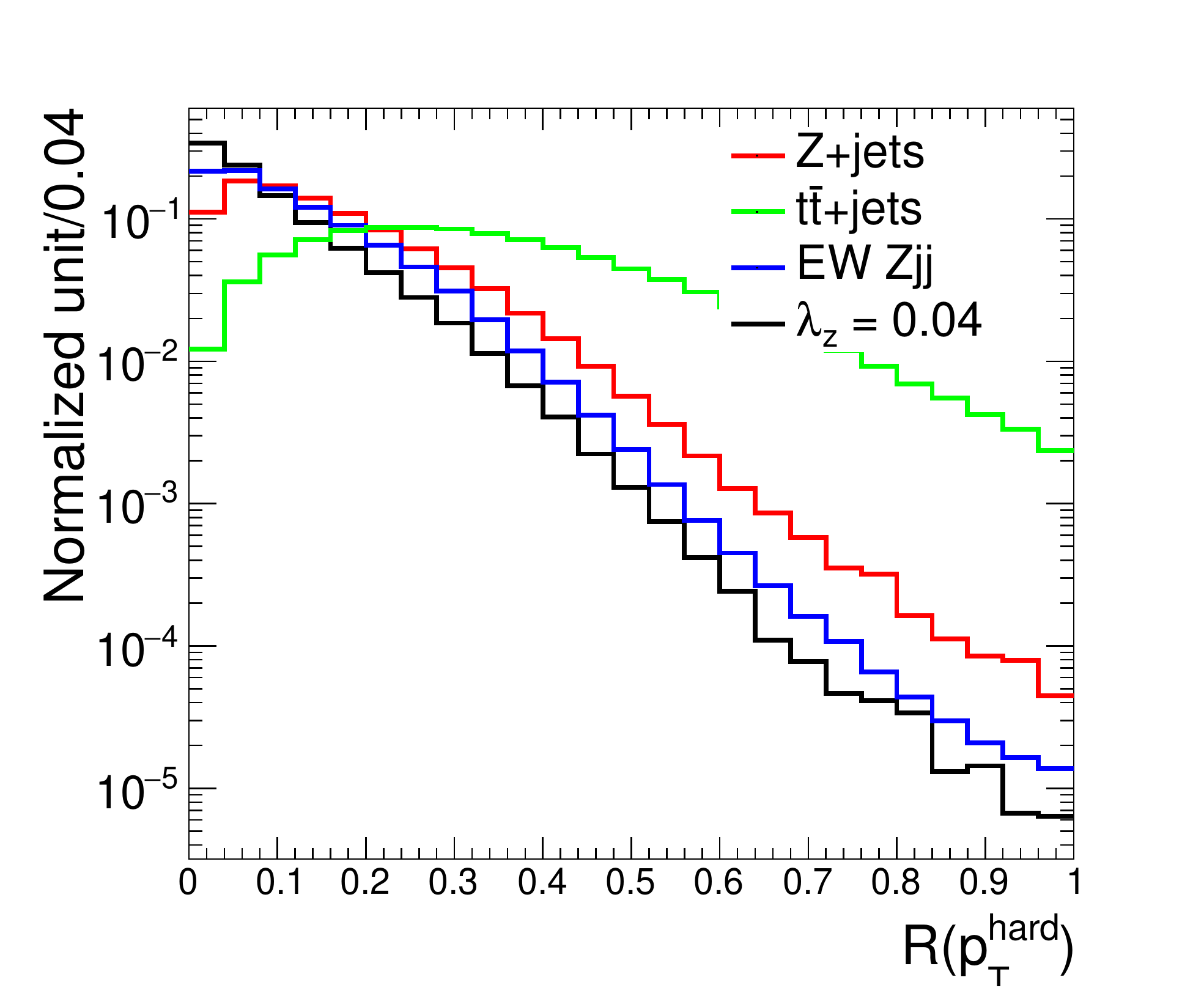}
\caption{\small The normalized distribution of BDT variables for the EFT signal for $\lambda_z = 0.04$ and backgrounds after imposing $p_T(j_1) >50$ GeV, $p_T(j_2) >30$ GeV, $p_T(\ell_1) >30$ GeV, $p_T(\ell_2) >20$ GeV, $|\eta(j)| <4.5$ , $|\eta(\ell)| < 2.5$. Recall that EW $Zjj$ and EFT samples were generated with $m_{jj} > 120$ GeV at the generation level.}
\label{fig:bdtset}
\end{center}
\end{figure}

Training and testing by taking EFT benchmark points as signals and the remaining as backgrounds may help in boosting the discrimination of the EFT signals from the background, and VBFhardness may play a role in that situation. We also have not included any top-related variables, including $b$-jets, which may be important in the binned analysis of $m_{\ell\ell}$ as top backgrounds remain significant up to a higher energy tail (see right panel of Fig.~\ref{fig:diff:ptandmll:benchmark}).  
The distributions for part of the BDT variables, given in Eq.~(\ref{eq:bdt:set}), after imposing $p_T$ and $\eta$ cuts on jets and leptons are illustrated in Fig.~\ref{fig:bdtset} where we also added one selected EFT benchmark point for $\lambda_z = 0.04$ as an illustration.

\newpage

{\small
\bibliography{lit}{}
\bibliographystyle{JHEP}}

\end{document}